\documentclass[prd,twocolumn,showpacs,nofootinbib, aps, superscriptaddress]{revtex4-1}
\usepackage{bm}
\usepackage{graphicx}
\usepackage{multirow}
\usepackage{amsmath}
\usepackage{amssymb}
\usepackage{tabulary}
\usepackage{hyperref}
\usepackage{color}
\usepackage{xspace}
\usepackage{verbatim}

\newcommand{\mwimp}{$m_\chi$\xspace}
\newcommand{\vmin}{\ifmmode v_{\mathrm{min}} \else $v_{\mathrm{min}}$ \fi\xspace}
\newcommand{\vesc}{\ifmmode v_{\mathrm{esc}} \else $v_{\mathrm{esc}}$ \fi\xspace}
\newcommand{\sigmapsi}{\relax \ifmmode \sigma_{\mathrm{p}}^{\mathrm{SI}}\else $\sigma_{\mathrm{p}}^{\mathrm{SI}}$\fi\xspace}
\newcommand{\sigmaNsi}{\ifmmode \sigma_{\mathrm{N}}^{\mathrm{SI}} \else $\sigma_{\mathrm{N}}^{\mathrm{SI}}$\fi\xspace}
\newcommand{\sigmapsd}{\relax \ifmmode \sigma_{\mathrm{p}}^{\mathrm{SD}}\else $\sigma_{\mathrm{p}}^{\mathrm{SD}}$\fi\xspace}
\newcommand{\sigmansd}{\ifmmode \sigma_{\mathrm{n}}^{\mathrm{SD}}\else $\sigma_{\mathrm{n}}^{\mathrm{SD}}$\fi\xspace}
\newcommand{\kms}{\ifmmode \textrm{ km s}^{-1}\else $\textrm{ km s}^{-1}$\fi\xspace}

\newcommand{\dbd}[2]{\ifmmode \frac{\textrm{d}#1}{\textrm{d}#2}\else $\textrm{d}#1/\textrm{d}#2$\fi\xspace}

\begin{document}

\hfill SACLAY-t14/179

\title{Probing WIMP particle physics and astrophysics with direct detection and neutrino telescope data}

\author{Bradley J. Kavanagh}
\email{bradley.kavanagh@cea.fr}
\affiliation{School of Physics \& Astronomy, University of Nottingham, University Park, Nottingham, NG7 2RD, UK}
\affiliation{Institut de physique th\'eorique, Universit\'e Paris Saclay, CNRS, CEA, F-91191 Gif-sur-Yvette, France}

\author{Mattia Fornasa}
\email{fornasam@gmail.com}
\affiliation{School of Physics \& Astronomy, University of Nottingham, University Park, Nottingham, NG7 2RD, UK}
\affiliation{GRAPPA Institute, University of Amsterdam, Science Park 904, 1098 XH Amsterdam, The Netherlands}

\author{Anne M. Green}
\email{anne.green@nottingham.ac.uk}
\affiliation{School of Physics \& Astronomy, University of Nottingham, University Park, Nottingham, NG7 2RD, UK}

\preprint{SACLAY-t14/179}

\begin{abstract}
With positive signals from multiple direct detection experiments it will, in principle, be possible to measure the mass and cross sections of weakly-interacting massive particle (WIMP) dark matter. Recent work has shown that, with a polynomial parameterisation of the WIMP speed distribution, it is possible to make an unbiased measurement of the WIMP mass, without making any astrophysical assumptions. However, direct detection experiments are not sensitive to low-speed WIMPs and, therefore, any model-independent approach will lead to a bias in the cross section. This problem can be solved with the addition of measurements of the flux of neutrinos from the Sun. This is because the flux of neutrinos produced from the annihilation of WIMPs which have been gravitationally captured in the Sun is sensitive to low-speed WIMPs. Using mock data from next-generation direct detection experiments and from the IceCube neutrino telescope, we show that the complementary information from IceCube on low-speed WIMPs breaks the degeneracy between the cross section and the speed distribution. This allows unbiased determinations of the WIMP mass and spin-independent and spin-dependent cross sections to be made, and the speed distribution to be reconstructed. We use two parameterisations of the speed distribution: binned and polynomial. While the polynomial parameterisation can encompass a wider range of speed distributions, this leads to larger uncertainties in the particle physics parameters. 
\end{abstract}

\maketitle

\section{Introduction}
\label{sec:introduction}
\makeatletter{}
Experiments aiming at detecting dark matter (DM) {\it directly} rely on 
measuring the signatures left by DM particles when they interact with the 
nuclei of a detector \cite{Cerdeno:2010jj,Strigari:2013iaa}. This technique 
was devised to search for a specific class of DM candidates: Weakly 
Interacting Massive Particles (WIMPs)~\cite{Goodman:1984dc}. WIMPs 
characteristically have a mass of the order of GeV to TeV, and their elastic 
scatterings off target nuclei induce recoils with energy on the order of keV. 

The number and energies of the recoil events can in principle be used to infer 
the properties of the DM particle, e.g. its mass and scattering cross 
sections. However, with a single experiment, this requires knowledge of the 
local WIMP velocity distribution, $f(\mathbf{v})$ 
(e.g.~Ref.~\cite{Peter:2011}). Data analyses usually use the so-called 
Standard Halo Model (SHM), where the velocity distribution is assumed to have 
the following simple form (in the Galactic rest frame):
\begin{equation}
f_\textrm{Gal}(\textbf{v}) = \frac{1}{(2\pi\sigma_v^2)^{3/2}}
\exp\left(-\frac{\textbf{v}^2}{2\sigma_v^2}\right) \,.
\label{eqn:SHM}
\end{equation}
This corresponds to an isothermal, spherical DM halo with density profile 
$\rho(r) \propto r^{-2}$, in equilibrium, in which case the dispersion $\sigma_v$ 
is related to the local circular speed $v_c \approx 220 \kms$ 
\cite{Feast:1997,Bovy:2012} by $\sigma_v = v_c/\sqrt{2}$. The SHM velocity 
distribution is usually truncated manually at the Galactic escape speed, 
which we take to be $\vesc = 544 \kms$, consistent with the estimates from 
the RAVE survey \cite{RAVE:2007,RAVE:2014}.

Despite its simplicity and extensive use, it is unlikely that the SHM 
provides a good description of the DM velocity distribution. More realistic 
models have been proposed that allow for a triaxial DM halo \cite{Evans:2000},
a more general density profile \cite{Widrow:2000} and/or an anisotropic 
velocity distribution \cite{Bozorgnia:2013pua}. Furthermore, if the DM
speed distribution is reconstructed self-consistently from the potential of 
the Milky Way \cite{Bhattacharjee:2012,Fornasa:2013} the resulting distribution 
deviates from the Maxwellian distribution of the SHM in Eq.~(\ref{eqn:SHM}).

Distribution functions extracted from $N$-body simulations also show 
deviations from the SHM \cite{Fairbairn:2008gz,Vogelsberger:2009,Kuhlen:2010,
Mao:2012}. In particular, DM substructures (e.g. streams) may lead to `spikes' 
in the speed distribution, while DM which has not yet completely phase-mixed
(so-called `debris flows') gives broad features \cite{Kuhlen:2012}. 
Simulations including baryonic physics suggest the possibility of a dark 
disk, produced by the DM tidally stripped from subhalos that are 
preferentially dragged into the stellar disk during the late stages of halo 
assembly \cite{Read:2009,Read:2010}. The resulting dark disk corotates with 
approximately the same speed as the stellar disk, but with a smaller velocity 
dispersion $\sigma_v \sim 50 \kms$. This dark disk may contribute an 
additional $20-100\%$ of the density of the halo (depending on the merger 
history of the Milky Way), although more recent simulations 
\cite{Pillepich:2014} indicate a smaller density, $\sim 10\%$.

From the previous discussion it is clear that the velocity distribution is 
still a quite uncertain quantity, despite its fundamental importance in 
the interpretation of direct detection data \cite{Green:2011bv,Green:2010gw,
McCabe:2010zh}. Furthermore, probing the speed distribution would provide
information on the structure and evolution of the Milky Way.

Various model independent techniques have recently been introduced for
analysing direct detection data without rigid assumptions about the WIMP 
speed distribution, with the goal of obtaining unbiased constraints on the 
WIMP particle properties, e.g.~\cite{Drees:2007hr,Strigari:2009zb,Fox:2010bz,
Peter:2009ak,Peter:2011, Feldstein:2014}. See Ref.~\cite{Peter:2013aha} for a 
review. In particular, Refs.~\cite{Peter:2009ak,Peter:2011} suggested 
employing an empirical parameterisation of the velocity distribution, and 
using data from multiple experiments to constrain its parameters along with 
the WIMP mass and cross section. Care must still be taken with the choice of 
parameterisation in order to avoid a biased determination of the WIMP mass 
\cite{Peter:2011,Kavanagh:2012}. We found that particular functional forms for 
the \textit{logarithm} of $f(\mathbf{v})$, for instance Legendre and Chebyshev 
polynomials, allow an unbiased reconstruction of the WIMP mass 
\cite{Kavanagh:2013a,Kavanagh:2013b,Peter:2013aha}. However, if only direct 
detection data is used, data analysis is inevitably hindered by the lack of 
sensitivity to low-speed WIMPs that produce recoil energies below the 
experimental energy thresholds. As experiments are blind to low-speed WIMPs, 
they can only detect some unknown fraction of the WIMPs. This then translates 
into a biased reconstruction of the scattering cross section.

Here we provide a solution to this problem using (simulated) future 
measurements from neutrino telescopes of the flux of neutrinos from the Sun. 
WIMPs scattering off the nucleons in the Sun can lose enough energy to get
captured in its gravitational potential \cite{Press:1985,Silk:1985,
Gaisser:1986,Krauss:1986,Srednicki:1987,Griest:1987}. They accumulate there 
until their density is high enough to annihilate, producing (among other 
particles) neutrinos that can travel to us and be detected by neutrino 
telescopes, such as IceCube~\cite{Icecube}. The expected number of neutrinos 
depends on the WIMP capture rate in the Sun, which is sensitive to the 
velocity distribution of WIMPs {\it below} a certain value.

A xenon-based direct detection experiment with ${\cal O}({\rm keV})$ energy
threshold is sensitive to WIMPs with speeds above 500~$\kms$ for light WIMPs 
with mass of order a few GeV, or above tens of~$\kms$ for heavier WIMPs. On 
the other hand, the maximum Solar capture speed (above which WIMPs are too 
fast to be captured) is larger than this for WIMP masses between 10 GeV and 1 
TeV. Neutrino telescopes are therefore sensitive to the entire low-speed WIMP
population which lies below the direct detection energy threshold. This 
indicates that combining direct detection and neutrino telescope data will 
allow us to probe the entire speed distribution and improve the accuracy of 
the constraints obtainable on both the WIMP mass and interaction cross section.

The complementarity of direct detection and neutrino telescope experiments has 
been studied previously in Ref.~\cite{Arina:2013}. In that paper, 
astrophysical uncertainties were included by marginalising over parameters of 
the SHM,  and by comparing with the results obtained assuming speed 
distributions from N-body simulations. In the current work, we account for 
such uncertainties using two general parametrisations of $f(v)$ described 
above, and investigate how well $f(v)$ can be reconstructed from data. Due to 
the degeneracy between the WIMP mass and speed distribution, such a general 
approach requires complementary information from several direct detection 
experiments, rather than a single experiment, as considered in 
Ref.~\cite{Arina:2013}.

In the following sections we estimate the sensitivity of this general approach
by means of Bayesian inference. We choose a set of well-motivated benchmarks 
for the mass and cross sections of the WIMP, as well as for its speed 
distribution. For each of these benchmarks we simulate the data expected in 
next-generation direct detection experiments and in a neutrino telescope. 
This data is encoded in a likelihood function, with which we scan over a 
parameter space that includes both particle physics quantities (e.g. the WIMP 
mass and scattering cross sections) and astrophysical ones (e.g. the 
coefficients entering in our parametrisation of the speed distribution). This 
technique allows us to estimate the precision with which future experiments 
will be able to reconstruct these parameters.

The paper is organized as follows. In Sec. \ref{sec:formalism} we summarize
the formalism for the computation of recoil events in a direct detection
experiment, as well as the expected signal in a neutrino telescope. 
In Sec. \ref{sec:reconstruction} we introduce the benchmark models considered 
and the parameterisations of the speed distribution. We also describe the 
sampling technique. In Sec. \ref{sec:DDonly} we present our results based on 
direct detection data only, while in Sec. \ref{sec:DDwithIC} we also include 
the information from a neutrino telescope. Finally, we discuss our results 
in Sec. \ref{sec:discussion} and summarize our main conclusions in 
Sec. \ref{sec:conclusions}.

\section{Dark matter event rate formalism}
\label{sec:formalism}
\makeatletter{}
\subsection{Direct detection}
The differential event rate per unit time and detector mass for nuclear 
recoils of energy $E_{\rm R}$ in a direct detection experiment is given by 
\cite{Jungman:1995}
\begin{equation}
\label{eq:dRdE}
\dbd{R}{E_R} = \frac{\rho_0}{m_\chi m_{\rm N}} 
\int_{\vmin}^{\infty} v f_1(v) \dbd{\sigma}{E_{\rm R}} \, \mathrm{d}v\,.
\end{equation}
The local DM mass density is denoted by $\rho_0$, the WIMP mass by $m_\chi$ 
and the target nuclear mass by $m_{\rm N}$. The prefactor 
$\rho_0/(m_\chi m_{\rm N})$ is the number of WIMPs per unit volume multiplied by 
the number of target nuclei per unit detector mass. The integral in 
Eq.~(\ref{eq:dRdE}) is of the differential cross section weighted by the 
one-dimensional WIMP speed distribution $f_1(v)$ (see later).

The WIMP velocity distribution in the Earth's frame is related to that in the 
Galactic frame by a Galilean transformation: 
$f(\mathbf{v})=f_\textrm{Gal}(\textbf{v} - \textbf{v}_{\rm lag})$, where 
$\textbf{v}_{\rm lag} \approx 230 \, {\rm km \, s^{-1}}$ is the velocity of the 
Earth with respect to the Galactic rest frame \cite{Lee:2013xxa,McCabe:2014}. 
This includes a contribution from the velocity of the Sun with respect to the 
Galactic frame as well as a contribution from the Earth's motion as it orbits 
the Sun. The dependence on the Earth's orbit implies that the velocity 
distribution will be time-varying, producing an annual modulation in the 
event rate \cite{Freese:1988}. However, this modulation is expected to be 
$\lesssim 10\%$ and we consider here only the time averaged event rate. The 
one-dimensional speed distribution is obtained by integrating over all 
directions in the Earth frame
\begin{equation}
f_1(v) = \oint f_\textrm{Gal}(\textbf{v} - \textbf{v}_\textrm{lag}) v^2 \, 
\mathrm{d}\Omega_\textbf{v} \equiv v^2 f(v)\,.
\end{equation}
The function $f(v)$ is the directionally-averaged velocity distribution and 
is the quantity which we parametrise (and subsequently reconstruct) in order 
to account for astrophysical uncertainties.

The lower limit of the integral in Eq.~(\ref{eq:dRdE}) is the minimum WIMP 
speed that can excite a recoil of energy $E_{\rm R}$:
\begin{equation}
\label{eq:vmin}
\vmin = \sqrt{\frac{m_{\rm N} E_{\rm R}}{2 \mu_{\chi {\rm N}}^2}} \,,
\end{equation}
where $\mu_{\chi {\rm N}}$ is the reduced mass of the WIMP-nucleon system.

The differential cross section is typically divided into spin-dependent (SD) 
and spin-independent (SI) contributions:
\begin{equation}
\dbd{\sigma}{E_{\rm R}} = \dbd{\sigma_{{\rm SD}}}{E_{\rm R}} +
\dbd{\sigma_{{\rm SI}}}{E_{\rm R}}\,.
\end{equation}
The SI contribution can be written as
\begin{equation}
\dbd{\sigma_{{\rm SI}}}{E_{\rm R}} = 
\frac{m_{\rm N}\sigmapsi}{2\mu_{\chi {\rm p}}^2 v^2} 
A^2 F_{{\rm SI}}^2(E_{\rm R})\,,
\end{equation}
where we have assumed that the coupling to protons and neutrons is equal 
($f_{\rm p}=f_{\rm n}$). In this case, the SI contribution scales with the square 
of the mass number $A$ of the target nucleus. The interaction strength is 
controlled by \sigmapsi, the WIMP-proton SI cross section, and by
$\mu_{\chi {\rm p}} = m_\chi m_{\rm p}/(m_\chi + m_{\rm p})$, the reduced mass of the 
WIMP-proton system. The loss of coherence due to the finite size of the 
nucleus is captured in the form factor $F_{{\rm SI}}^2(E_{\rm R})$, which is 
obtained from the Fourier transform of the nucleon distribution in the nucleus. 
We take the form factor to have the Helm form \cite{Helm:1956}
\begin{equation}
F_{{\rm SI}}^2(E_{\rm R}) = \left[\frac{3j_1(qR_1)}{qR_1}\right]^2 
\mathrm{e}^{-q^2s^2}\,,
\end{equation}
where $j_1(x)$ is a spherical Bessel function of the first kind and 
$q=\sqrt{2 m_{\rm N} E_{\rm R}}$ is the momentum transfer. We use nuclear 
parameters from Ref. \cite{Lewin:1996}, based on fits to muon spectroscopy 
data \cite{Fricke:1995}:
\begin{align}
R_1 & = \sqrt{c^2 + \frac{7}{3}\pi^2a^2 - 5s^2} \,, \\
c & = 1.23A^{1/3} - 0.60 \mbox{ fm} \,, \\
a & = 0.52 \mbox{ fm} \,, \\
s & = 0.9 \mbox{ fm} \,.
\end{align}
Muon spectroscopy probes the \textit{charge} distribution in the nucleus. 
However, detailed Hartree-Fock calculations indicate that the charge 
distribution can be used as a good proxy for the nucleon distribution 
(especially in the case $f_{\rm p}=f_{\rm n}$). It has been shown that using the 
approximate Helm form factor introduces an error of less than $\sim$5\% in 
the total event rate \cite{Duda:2007,Co:2012}.

The standard expression for the SD contribution is
\begin{equation}
\label{eq:dSdE_SD}
\dbd{\sigma_{{\rm SD}}}{E_{\rm R}} = 
\frac{m_{\rm N}\sigmapsd}{2\mu_{\chi {\rm p}}^2 v^2} \frac{4(J+1)}{3J} 
\left( \langle S_{\rm p} \rangle + 
\frac{a_{\rm n}}{a_{\rm p}} \langle S_{\rm n} \rangle \right)^2 
F_{{\rm SD}}^2(E_{\rm R})\,.
\end{equation}
As before, \sigmapsd denotes the WIMP-proton SD cross section and we have 
assumed that the coupling to protons and neutrons is equal ($a_n=a_p$). The 
total nuclear spin of the target is denoted $J$, while the expectation values 
of the proton and neutron spin operators are given by 
$\langle S_{\rm p} \rangle$ and $\langle S_{\rm n} \rangle$ respectively. The SD 
form factor can be written as
\begin{equation}
F_{{\rm SD}}^2(E_{\rm R}) = \frac{S(E_{\rm R})}{S(0)}\,,
\end{equation}
where $S(E_{\rm R})$ describes the energy dependence of the recoil rate due to 
the fact that the nucleus is not composed of a single spin but rather a 
collection of spin-1/2 nucleons. This response function $S(E_{\rm R})$ is 
usually decomposed in terms of spin structure functions \cite{Cannoni:2011}:
\begin{equation}
S(E_{\rm R}) = a_0^2S_{00}(E_{\rm R}) + a_0a_1S_{01}(E_{\rm R}) +
a_1^2S_{11}(E_{\rm R})\,,
\end{equation}
where $a_0 = a_{\rm p} + a_{\rm n}$ is the isoscalar coupling and 
$a_1= a_{\rm p} - a_{\rm n}$ is the isovector coupling. Under the assumption 
$a_{\rm n}=a_{\rm p}$, then  $a_0 = 2a_{\rm p}$ and $a_1 = 0$, and only the 
isoscalar structure function $S_{00}(E_{\rm R})$ will be relevant for our 
analysis. 

The functional form for $S_{ij}$ can be calculated from shell models for the 
nucleus \cite{Ressell:1997}. However, competing models (such as the Odd Group 
Model \cite{Engel:1989}, Interacting Boson Fermion Model \cite{Iachello:1991} 
and Independent Single Particle Shell Model \cite{Ellis:1988}, among others) 
may lead to different spin structure functions, generating a significant 
uncertainty in the value of the SD cross section. This issue was explored by 
Ref.~\cite{Cerdeno:2012}, who developed a parametrisation for the spin 
structure functions in terms of the parameter $u = (qb)^2/2$, where 
\begin{equation}
b = \sqrt{\frac{41.467}{(45.0 A^{-1/3} - 25.0 A^{-2/3})}} \mbox{ fm} \,,
\end{equation} 
is the oscillator size parameter \cite{Warburton:1990,Ressell:1997}. Their 
parametrisation takes the form
\begin{equation}
\label{eq:SDparametrization}
S_{ij} = N ((1-\beta)\mathrm{e}^{-\alpha u} + \beta)\,.
\end{equation}

The values we use for the parameters $(N, \alpha, \beta)$ are 
$(0.0595, 3.75, 0.0096)$ for ${}^{129}$Xe, $(0.035, 3.925, 0.12)$ for 
${}^{131}$Xe and $(0.195, 4.25, 0.07)$ for ${}^{73}$Ge. These values were 
chosen to approximately reproduce the median values obtained from a range 
of spin structure function calculations \cite{Ressell:1993,Dimitrov:1995,
Ressell:1997,Menendez:2012}. We keep the values of these parameters fixed in 
our analysis in order to focus on the impact of astrophysical uncertainties. 
We note that argon has zero spin and therefore has no SD interaction with 
WIMPs.

The proton and neutron spins $\langle S_{{\rm p},{\rm n}} \rangle$ can be 
rewritten in terms of the total nuclear spin and the spin structure functions 
(as in Ref.~\cite{Cannoni:2013}). Using this to rewrite 
Eq.~(\ref{eq:dSdE_SD}), gives the following expression (in the case $a_p=a_n$):
\begin{equation}
\dbd{\sigma_{\rm SD}}{E_R} = 
\frac{8\pi m_N\sigmapsd}{3\mu_{\chi p}^2 v^2}  \frac{S_{00}(E_R)}{(2J+1)}\,.
\end{equation}

Both the SI and SD differential cross sections are inversely proportional to
the WIMP velocity squared. Factoring out all the terms that do not depend on
$v$ in Eq.~(\ref{eq:dRdE}), the integral over the WIMP speed is normally 
written as $\eta(v_{\rm min})$:
\begin{equation}
\label{velocityintegral}
\eta(v_{\rm min}) = \int_{v_{\rm min}}^\infty \frac{f_{1}(v)}{v}
\mathrm{d}v \,,
\end{equation}
and we will subsequently refer to this quantity as the velocity integral.

\subsection{Neutrino telescopes}
The WIMP capture rate per unit shell volume for a shell at distance $r$ from 
the center of the Sun, due to species $i$ is given by 
\cite{Gould:1987,Gould:1992}
\begin{equation}
\dbd{C_i}{V} = \int_{0}^{v_{\textrm{max}}} \textrm{d}v \,
\frac{f_1(v)}{v} \, w \, \Omega^{-}_{w_\textrm{esc},i}(w)\,,
\end{equation}
where $v$ is the asymptotic WIMP speed\footnote{In the literature, the 
asymptotic WIMP speed is typically written as $u$. Here, we denote it as $v$ 
for consistency with the notation of the direct detection formalism.}, 
$w(r) = \sqrt{v^2 + w_{\rm esc}(r)^2}$ and $w_{\rm esc}(r)$ is the local escape 
speed at radius $r$ inside the Sun\footnote{For compactness, we subsequently 
suppress the radial dependence when denoting $w(r)$ and $w_{\rm esc}(r)$.}. The 
rate per unit time at which a single WIMP travelling at speed $w$ is scattered 
down to a speed less than $w_{\rm esc}$, due to the interaction with species 
$i$, is $\Omega^{-}_{w_{\rm esc},i}(w)$. Finally, the upper limit of integration is 
given by
\begin{equation}
\label{eq:vmax}
v_\mathrm{max} = \frac{\sqrt{4m_\chi m_{{\rm N}_i}}}{m_\chi - m_{{\rm N}_i}} w_{\rm esc}\,,
\end{equation}
where $m_{{\rm N_i}}$ is the mass of the nucleus of species $i$. Above 
$v_{\mathrm{max}}$, WIMPs cannot lose enough energy in a recoil to drop below the 
local escape speed and, therefore, they are not captured by the Sun.

The scatter rate $\Omega^{-}_{w_{\rm esc},i}$ from a species with number density 
$n_{{\rm N_i}}$ can be written as:
\begin{equation}
\label{eq:omega}
\Omega^{-}_{w_{\rm esc},i}(w) = w \, n_{{\rm N}_i} 
\int_{E_v}^{E_\mathrm{max}} \dbd{\sigma}{E_{\rm R}} \,
\mathrm{d}E_{\rm R}\,,
\end{equation}
where $E_{\rm R}$ is the energy lost by the scattering WIMP. The limits of 
integration run from the minimum energy loss required to reduce the WIMP 
speed below $w_{\rm esc}$,
\begin{equation}
E_v = \frac{m_\chi}{2}\left(w^2 - w_{\rm esc}^2\right) = \frac{m_\chi}{2}v^2\,,
\end{equation}
to the maximum possible energy loss in the collision,
\begin{equation}
E_\mathrm{max} = \frac{2\mu_{\chi {\rm N}_i}^2}{m_{{\rm N}_i}}w^2\,.
\end{equation}

As in the direct detection case, we can decompose the differential cross 
section into SI and SD components. While all of the constituent elements of 
the Sun are sensitive to SI interactions, only spin-1/2 hydrogen is sensitive 
to SD scattering. The differential cross section is therefore given by
\begin{equation}
\dbd{\sigma}{E_{\rm R}} = \frac{m_{{\rm N}_i}}{2\mu_{\chi {\rm p}}^2 v^2} \times
\begin{cases}
\sigmapsi + \sigmapsd & \textrm{ for } A = 1 \,, \\
\sigmapsi A_i^2 F_i^2(E_{\rm R}) & \textrm{ for } A > 1 \,.
\end{cases}
\end{equation}
No form factor is needed for hydrogen, which consists of only a single 
nucleon. For the remaining nuclei, we approximate the form factor as 
\cite{Gould:1987}
\begin{align}
F^2_i(E_{\rm R}) & = \exp{\left(- \frac{E_{\rm R}}{E_i}\right)} \,, \\
E_i & = \frac{3}{2m_{{\rm N}_i} R_i^2} \,.
\end{align}
This allows Eq.~(\ref{eq:omega}) to be calculated analytically and introduces 
an error in the total capture rate of only a few percent.

\begin{figure}[t]
  \centering
  \includegraphics[trim=0.8cm 0.9cm 0cm 0cm, width=0.5\textwidth]{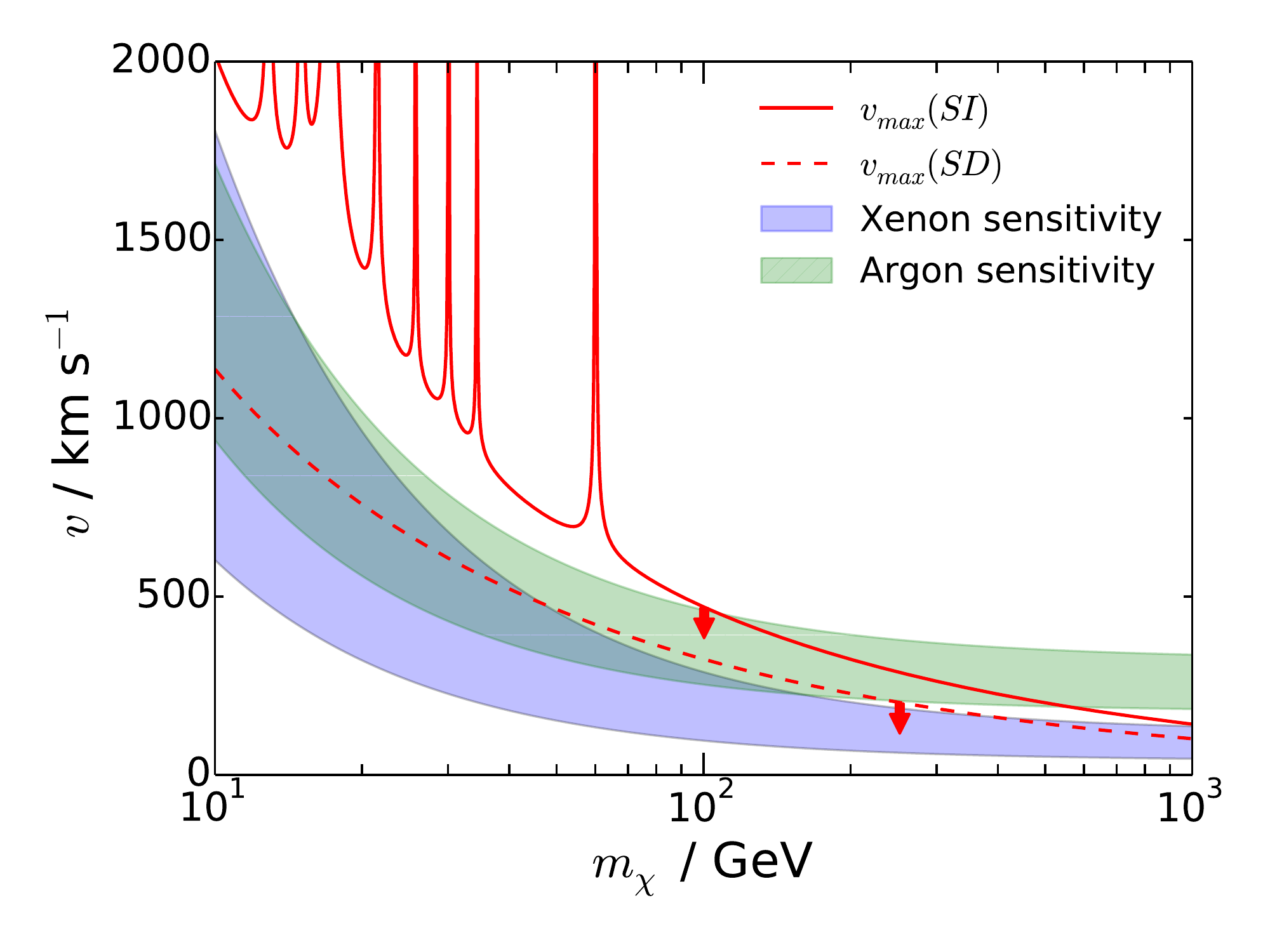}
  \caption{The ranges of WIMP velocity that Solar capture and direct
    detection experiments are sensitive to, as a function of the WIMP
    mass. The blue band shows the range of speeds to which a xenon-based 
    detector with an energy window of $[5,45]$ keV is sensitive. The green 
    hatched band shows the corresponding range of speeds for an 
    argon-based detector with an energy window of $[30,100]$ keV. The solid 
    (dashed) red lines shows the maximum speed to which Solar WIMP capture 
    is sensitive for SI (SD) interactions. See the text for further details.}
  \label{fig:speedoverlap}
\end{figure}

Fig. \ref{fig:speedoverlap} shows the maximum Solar capture speed 
$v_\textrm{max}$ given by Eq.~(\ref{eq:vmax}) (averaged over the Solar radius). 
We consider separately the SD contribution (dashed red line) from hydrogen and 
the SI contribution (solid red line) averaged over all elements in the Sun. 
The former goes from a value slightly larger than 1000~\kms for a 10 GeV WIMP 
down to $\sim$~100 \kms for a mass of 1 TeV, while the latter is larger by a 
factor of approximately two. The sharp peaks in the SI curve for $v_\textrm{max}$
are resonances due to mass matching between the WIMP and one of the nuclei in 
the Sun. In these cases, energy transfer during recoils can be very efficient 
and WIMPs with high speeds can be captured.  

The two red lines should be compared with the blue (green hatched) band, 
which shows the velocity window to which a xenon-based (argon-based) 
experiment is sensitive (assuming the energy thresholds described in 
Sec. \ref{sec:reconstruction} and Tab. \ref{tab:Experiments}). We note that, 
over the whole range of masses considered, the maximum Solar capture speed is
always larger (both for SI and SD interactions) than the lower edge of the 
blue band. This means that, as anticipated in the introduction, neutrino 
telescopes are sensitive to all of the low-speed tail of the velocity 
distribution that is inaccessible to direct detection experiments. 

WIMPs which are captured can annihilate in the Sun to Standard Model 
particles. Over long timescales, equilibrium is reached between the capture 
and annihilation rates. In such a regime, the annihilation rate $\Gamma_A$ is
equal to half the capture rate, independent of the unknown annihilation cross 
section \cite{Griest:1987}. For large enough scattering cross sections 
($\sigmapsd \gtrsim 10^{-43} \textrm{ cm}^2$), capture is expected to be 
efficient enough for equilibrium to be reached \cite{Peter:2009}. The 
assumption of equilibrium is, therefore, reasonable for the benchmarks 
considered in this work.

The majority of Standard Model particles produced by WIMP annihilations cannot 
escape the Sun. However, some of these particles may decay to neutrinos or 
neutrinos may be produced directly in the annihilation. Neutrinos can reach the 
Earth and be detected by neutrino telescope experiments. In this work, we 
focus on the IceCube experiment \cite{Aartsen:2013b}, which measures the 
\v{C}erenkov radiation produced by high energy particles travelling through 
ice. IceCube aims at isolating the contribution of muons produced by muon 
neutrinos interacting in the Earth or its atmosphere. The amount of 
\v{C}erenkov light detected, combined with the shape of the \v{C}ereknow
cascade, allows the energy and direction of the initial neutrino to be 
reconstructed.

The spectrum of neutrinos arriving at IceCube is given by
\begin{equation}
\label{eqn:neutrino_spectrum}
\dbd{N_\nu}{E_\nu} = \frac{\Gamma_A}{4\pi D^2}\sum_f B_f
\dbd{N_\nu^{f}}{E_\nu}\,,
\end{equation}
where $D$ is the distance from the Sun to the detector and the sum is over 
all annihilation final states $f$, weighted by the branching ratios $B_{f}$. 
The factor $\mathrm{d}N_\nu^{f} / \mathrm{d}E_\nu$ is the neutrino spectrum 
produced by final state $f$, taking into account the propagation of neutrinos 
as they travel from the Sun to the detector \cite{Blennow:2008,Baratella:2014}.
The branching ratios depend on the specific WIMP under consideration. For 
simplicity, it is typically assumed (as we do here) that the WIMPs annihilate 
into a single channel. For the computation of Eq.~(\ref{eqn:neutrino_spectrum}) 
we use a modified version of the publicly available DarkSUSY code 
\cite{Gondolo:2004,DarkSUSYweb}, that also accounts for the telescope 
efficiency (see also Sec. \ref{sec:reconstruction}).

\section{Benchmarks and parameter reconstruction}
\label{sec:reconstruction}
\makeatletter{}
In order to determine how well the WIMP parameters can be recovered, we 
generate mock data sets for IceCube and three hypothetical direct detection 
experiments. 

Table~\ref{tab:Experiments} displays the parameters we use for the three 
direct detection experiments. They are chosen to broadly mimic next-generation 
detectors that are currently in development. Each experiment is described by 
the energy window it is sensitive to and the total exposure, which is the 
product of the fiducial detector mass, the exposure time and the experimental 
and operating efficiencies (which we implicitly assume to be constant). We 
also include a gaussian energy resolution of 
$\sigma_E = 1 \textrm{ keV}$~\footnote{The precise value of $\sigma_E$ is not 
expected to strongly affect parameter reconstruction, unless 
$\sigma_E > E_\mathrm{th}$ \cite{Green:2007rb}.} and a flat background rate of 
$10^{-7}$ events/kg/day/keV.

We choose three experiments using different target nuclei as it has been
shown that the employment of multiple targets significantly enhances the 
accuracy of the reconstruction of the WIMP mass and cross sections 
\cite{Pato:2010zk,Cerdeno:2013gqa,Cerdeno:2014uga}. Furthermore, if the WIMP 
velocity distribution is not known, multiple targets are a necessity 
\cite{Kavanagh:2012}. For more details on the reconstruction performance of different ensembles of target materials, we refer to reader the Ref.~\cite{Peter:2013aha}. We note that our modelling of the 
detectors is rather unsophisticated. More realistic modelling would include, 
for instance, energy-dependent efficiency. However, the detector modelling we 
employ here is sufficient to estimate the precision with which the WIMP 
parameters can be recovered. 

We divide the energy range of each experiment into bins and generate Asimov 
data \cite{Cowan:2013} by setting the observed number of events in each bin 
equal to the expected number of events. While this cannot correspond to a 
physical realisation of data as the observed number of events will be 
non-integer, it allows us to disentangle the effects of Poissonian 
fluctuations from the properties of the parametrisations under study. 
Including the effect of Poissonian fluctuations would require the generation 
of a large number of realisations for each benchmark. The precision in the 
reconstruction of the WIMP parameters will, in general, be different for each 
realisation. This leads to the concept of {\it coverage}, i.e. how many times 
the benchmark value is contained in the credible interval estimating the 
uncertainty in the reconstruction (c.f. Ref.~\cite{Strege:2012kv}). We leave 
this for future work, noting here that Ref. \cite{Kavanagh:2013b} showed that 
the polynomial parameterisation we use (Sec.~\ref{subsec-param}) provides 
almost exact coverage for the reconstruction of the WIMP mass (at least in the 
case of \mwimp = 50 GeV).

\begin{table}[t]
\setlength{\extrarowheight}{3pt}
\begin{center}
\begin{tabular}{cm{2cm}m{2cm}m{2cm}}
 \hline\hline
Experiment & Energy Range (keV) & Exposure (ton-yr) & Energy bin width (keV) \\
\hline
Xenon & 5-45 & 1.0 & 2.0 \\
Argon &  30-100 & 1.0 & 2.0 \\
Germanium & 10-100 & 0.3 & 2.0 \\
\hline\hline
\end{tabular}
\end{center}
\caption{Summary of the parameters describing the mock direct detection 
  experiments. All experiments have a constant energy resolution of 
  $\sigma_E = 1 \textrm{ keV}$ and a flat background rate of $10^{-7}$ 
  events/kg/day/keV. The energy windows are chosen to be similar to those 
  proposed in Refs. \cite{Aprile:2010,Bauer:2013} and \cite{Grandi:2005} for
  xenon, germanium and argon respectively. We assume natural isotopic 
  abundances of the target materials.}
\label{tab:Experiments}
\end{table}

For the mock neutrino telescope data, we consider the IceCube 86-string 
configuration. We follow Ref.~\cite{Arina:2013} and use an exposure time of 
900 days (corresponding to five 180-days austral Winter observing seasons) and 
an angular cut around the Solar position of $\phi_\textrm{cut} = 3\,^{\circ}$. 
This value is chosen to reflect the typical angular resolution of the IceCube 
detector \cite{Danninger:2012} and has previously been shown to be the optimal 
angular cut over a range of DM masses \cite{Silverwood:2013}. This results in 
approximately 217 background events over the full exposure. As with the direct 
detection experiments, we set the observed number of events equal to the 
expected number of signal plus background events. We use only the observed 
number of events as data and not their individual energies. While event-level 
likelihood methods have previously been developed \cite{Scott:2012} for the 
use of IceCube 22-string data \cite{Abbasi:2009}, a similar analysis has not 
yet been performed for IceCube-86. In particular, the probability 
distributions for the number of lit digital optical modules as a function of 
neutrino energy are not yet available for IceCube-86. Nonetheless, using only 
the number of observed events is a first step towards the characterisation of 
the WIMP speed distribution with neutrino telescopes.

\subsection{Benchmarks}
The four benchmark models we use to generate mock data sets are summarised 
in Table~\ref{tab:benchmarks}, along with the number of events produced in 
each experiment. In all cases, we use a SI WIMP-proton cross section of 
$\sigmapsi = 10^{-45} \textrm{ cm}^2$ and a SD cross section of 
$\sigmapsd = 2 \times 10^{-40} \textrm{ cm}^2$, both of which are close to the 
current best exclusion limits \cite{Akerib:2014,Aprile:2013}. 

\begin{table*}[t]
\setlength{\extrarowheight}{3pt}
\begin{center}
\begin{tabular}{c|ccc|ccccccc}
\hline\hline
Benchmark & $m_\chi \textrm{ (GeV)}$ & Speed dist. & Annihilation channel & $N_{\mathrm{Xe}}(\mbox{SI})$ & $N_{\mathrm{Xe}}(\mbox{SD})$  & $N_{\mathrm{Ar}}(\mbox{SI})$ & $N_{\mathrm{Ar}}(\mbox{SD})$ & $N_{\textrm{Ge}}(\mbox{SI})$ & $N_{\textrm{Ge}}(\mbox{SD})$ & $N_{\textrm{IC}}$ \\
\hline
A & 100 & SHM & $W^{+}W^{-}$ & 154.9 & 262.7 & 16.1 & 0 & 25.4 & 18.7 & 43.3 \\
B & 100 & SHM+DD & $W^{+}W^{-}$ & 167.1 & 283.9 & 16.2 & 0 & 25.7 & 18.9 & 242.9 \\
C & 30  & SHM & $\nu_\mu \bar{\nu}_\mu$ & 175.1 & 301.1 & 6.2 & 0 & 20.5 & 16.1 & 13.2 \\
D & 30  & SHM+DD & $\nu_\mu \bar{\nu}_\mu$ & 175.0 & 300.9 & 5.8 & 0 & 20.4 & 16.0 & 40.2 \\
\hline\hline
\end{tabular}
\end{center}
\caption{Summary of the WIMP benchmarks. The first section shows the
 WIMP mass, speed distribution and annihilation channel. SHM refers to the
 Standard Halo Model, with a speed distribution described by 
 Eq.~(\ref{eqn:SHM}) with $v_\textrm{lag} = 230 \kms$ and $\sigma_v = 163 \kms$. 
 In the case of SHM+DD, we also include the contribution of a dark disk,
 described by an additional term as in  Eq.~(\ref{eqn:SHM}), but with 
 $v_\textrm{lag} = 50 \kms$ and $\sigma_v = 50 \kms$. For all benchmarks, we 
 consider only isospin-conserving interactions (i.e. $f_{\rm p} = f_{\rm n}$ and 
 $a_{\rm p} = a_{\rm n}$) and assume $\sigmapsi = 10^{-45} \textrm{ cm}^2$ and 
 $\sigmapsd = 2 \times 10^{-40} \textrm{ cm}^2$. The second section gives the 
 number of events produced in the xenon-, argon- and germanium-based direct 
 detection experiments (with the number of recoils induced by SI and SD 
 interactions listed separately) and IceCube.}
\label{tab:benchmarks}
\end{table*}

The WIMPs in benchmarks A and B have an intermediate mass of 100 GeV and the
production of neutrinos originates from annihilations into $W^{+}W^{-}$. This
is a similar configuration to benchmark B used by Ref.~\cite{Arina:2013}. For 
benchmarks C and D we decrease the mass to 30 GeV, which allows a more 
accurate reconstruction of the WIMP mass (see Secs. \ref{sec:DDonly} and 
\ref{sec:DDwithIC}). The IceCube detector (with DeepCore) is sensitive to 
WIMPs with masses down to about 20 GeV \cite{Aartsen:2012kia}. For benchmarks 
C and D we assume that annihilations take place directly into 
$\nu_\mu \bar{\nu}_\mu$. 

Other annihilation channels may produce fewer neutrinos and, thus, reduce the impact of IceCube in the reconstruction of the particle physics nature of DM and its $f(v)$. However, note that, according to Ref.~\cite{Cirelli:2010xx}, the only annihilation channels not decaying with a significant probability into neutrinos are electrons, gluons and gamma rays. The scan of non-minimal supersymmetry performed in Ref.~\cite{Roszkowski:2014iqa} showed that these are subdominant channels.

Benchmarks A and C assume a SHM speed distribution as described in 
Sec. \ref{sec:formalism}, with $v_\textrm{lag} = 230 \kms$ 
\cite{Schonrich:2012,Bovy:2012} and $\sigma_v = 163 \kms$. Benchmarks B and D 
also include a moderate dark disk with a population of low-speed WIMPs which 
contribute an additional 30\% to the local DM density. We assume that the dark 
disk velocity distribution is also given by Eq.~(\ref{eqn:SHM}), with 
$v_\textrm{lag} = 50 \kms$ and $\sigma_v = 50 \kms$ \cite{Bruch:2008rx}.
As shown in Ref.~\cite{Choi:2013}, the capture rate in the Sun is not affected 
by variations in the shape of $f(v)$ (such as the differences between 
distribution functions extracted from different $N$-body simulations). However, 
significant enhancement of the capture rate can occur if there is a dark disk 
\cite{Bruch:2008rx}, and our benchmarks have been chosen in order to 
investigate this scenario. Finally, we assume a fixed value for the SHM local 
DM density of $\rho_0 = 0.3 \textrm{ GeV cm}^{-3}$. There is an uncertainty in 
this value of around a factor of 2 (see e.g.~Refs.~\cite{Catena:2010,
Weber:2010,Zhang:2013,Nesti:2013,Read:2014qva}). However this is degenerate 
with the cross sections.

In this work, we assume a common speed distribution $f(v)$ experienced by 
both Earth-based experiments and by the Sun. In principle gravitational 
focussing and diffusion could lead to differences between the forms of $f(v)$ 
at the Earth and Sun \cite{Gould:1991,Peter:2009}. However, a recent study 
using Liouville's theorem showed that such effects must be balanced by inverse 
processes \cite{Sivertsson:2012}. We can, therefore, treat the WIMP 
population as being effectively free and consider only a single common form 
for $f(v)$.

\subsection{Parametrisations of the speed distribution}
\label{subsec-param}
We use the mock data generated for the benchmarks in Table~\ref{tab:benchmarks} 
to evaluate the likelihood employed in the Bayesian scans over $m_\chi$, 
$\sigmapsi$ and $\sigmapsd$. In order to study the synergy between direct 
detection experiments and neutrino telescopes in the reconstruction of the 
speed distribution, some of these scans will also include parameters which 
describe the form of $f(v)$. We consider two possible parametrisations:
\begin{itemize}
\item {\bf Binned parametrisation:} This parameterisation was introduced in 
Ref. \cite{Peter:2011} and it involves dividing $f(v)$ into $N$ bins of 
width $\Delta v$ with bin edges $\tilde{v}_i$ and parametrising the bin 
heights by ${g}_i$:
\begin{equation}
\label{eq:Speed:binned}
f(v) = \sum_{i = 1}^N \frac{3{g}_i \, W(v;\tilde{v}_i,\Delta v)}{(\tilde{v}_i + \Delta v)^{3} - \tilde{v}_i^3} \,,
\end{equation}
where the top-hat function, $W$, is defined as:
\begin{equation}
W(v;\tilde{v}_i,\Delta v) =
\begin{cases}
   1 &  v \in [\tilde{v}_i,\tilde{v}_i+\Delta v] \,, \\
   0  & \text{otherwise \,.}
  \end{cases} 
\end{equation}
The bin heights then satisfy the normalisation condition
\begin{equation}
\label{eq:normalisation}
\sum_{i=1}^{N} {g}_i = 1\,.
\end{equation}
We parametrise $f(v)$ up to some maximum speed $v_\textrm{max} = N\Delta v$, 
above which we set $f(v) = 0$. We choose $v_\textrm{max} = 1000 \kms$, 
conservatively larger than the escape velocity in the Earth frame, which is 
around $800 \kms$ \cite{RAVE:2007,RAVE:2014}. The binned parametrisation was 
studied in detail in Ref.~\cite{Kavanagh:2012}, where it was demonstrated 
that, when only direct detection data is used, this method results in a bias 
towards smaller WIMP masses. 

\item {\bf Polynomial parametrisation:} In Ref.~\cite{Kavanagh:2013a} we 
proposed that the natural logarithm of $f(v)$ be expanded in a series of 
polynomials in $v$, i.e.
\begin{equation}
f(v) = \exp\left[ \sum_{k = 0}^{N-1} a_k P_k(2v/v_\textrm{max} - 1)\right] \,.
\end{equation}
The first coefficent $a_0$ is fixed by requiring the speed distribution to be 
normalised to 1. As detailed in Ref.~\cite{Kavanagh:2013b}, various polynomial 
bases can be used. Chebyshev and Legendre polynomials allow an unbiased 
reconstruction of the WIMP mass across a wide range of astrophysical and 
particle physics benchmarks \cite{Kavanagh:2013a,Kavanagh:2013b}, and the 
scans are normally faster if Chebyshev polynomials are used. We, therefore, 
use a basis of $N$ Chebyshev polynomials $P_k$ weighted by the parameters 
$a_k$.
\end{itemize}

By studying two different speed parametrisations, we can examine how particle 
physics parameter reconstruction is affected by the choice of speed 
parametrisation. While the binned parametrisation may lead to a bias in the 
WIMP mass, it is straight-forward and provides a good approximation to 
smoothly varying speed distributions. As discussed in 
Ref.~\cite{Kavanagh:2012}, this bias is due in part to a lack of information 
about $f(v)$ at low speeds.  We therefore expect that the addition of IceCube 
data will reduce this bias. By comparison, the polynomial distribution is 
unbiased and allows for a wider range of shapes for $f(v)$, although some of 
these are rapidly varying and may not be physically well-motivated. For a large 
number of parameters, these two methods should converge and both could be 
used as a consistency check. 

\subsection{Parameter sampling}
\label{sec:sampling}
We perform parameter scans using a modified version of the publicly available 
\textsc{MultiNest 3.6} package \cite{Feroz:2007,Feroz:2008,Feroz:2013hea}. 
This allows us to map out the likelihood $\mathcal{L}(\mathbf{\Theta})$ for 
the model parameters $\mathbf{\Theta}$. We use $N_\textrm{live} = 20000$ live 
points and a tolerance of $10^{-4}$. The priors we use for the various 
parameters are displayed in Table~\ref{tab:priors}.

\begin{table}
\setlength{\extrarowheight}{3pt}
\begin{center}
\begin{tabular}{ccc}
\hline\hline
Parameter & Prior range & Prior type \\
\hline
$m_\chi$ (GeV) & $10\--1000$ & log-flat \\
$\sigmapsi \textrm{ (cm}^2\textrm{)}$ & $10^{-48} \-- 10^{-42}$ & log-flat \\
$\sigmapsd \textrm{ (cm}^2\textrm{)}$ & $10^{-43} \-- 10^{-37}$ & log-flat \\
Polynomial coefficients $\left\{a_k\right\}$ & $-20 \-- 20$ & linear-flat \\
Bin heights $\left\{g_i\right\}$ & $0\--1$ & simplex \\
\hline\hline
\end{tabular}
\end{center}
\caption{Summary of prior types and ranges. The `simplex' prior is described in Sec.~\ref{sec:sampling}.}
\label{tab:priors}
\end{table}

Due to the normalisation condition on the bin heights (given in 
Eq.~(\ref{eq:normalisation})) for the case of the binned parametrisation of 
$f(v)$, we must sample these parameters from the so-called `simplex' priors: 
i.e. we uniformly sample $\left\{g_2,...,g_N\right\}$ such that they sum 
to less than one and then fix $g_1$ as
\begin{equation}
g_1 = 1 - \sum_{i = 2}^N {g}_i \,.
\end{equation}
The ellipsoidal sampling performed by \textsc{MultiNest} becomes increasingly 
inefficient as the number of bins $N$ increases, since the volume of the 
parameter space for which Eq.~(\ref{eq:normalisation}) is satisfied becomes 
very small. We therefore use \textsc{MultiNest} in constant efficiency mode 
when using the binned parametrisation, with a target efficiency of 0.3. We use 
a total of 10 bins in this parameterisation (9 free parameters, with one fixed 
by normalisation), which should allow us to obtain a close approximation to 
the rapidly varying SHM+DD distribution. 

For the polynomial $\ln f(v)$ parametrisation, we use 6 basis polynomials (5 
free coefficients, with one fixed by normalisation). This is a smaller number 
of parameters than for the binned parameterisation because the polynomial 
coefficients are allowed to vary over a much wider range. The volume of the 
polynomial parameter space is therefore significantly larger than for the 
binned parametrisation and a much larger number of live points would be 
required to accurately map the likelihood using 10 parameters. As we will see, 
using 6 basis functions still allows a wide range of speed distributions to 
be explored and provides a good fit to the data (see also the discussion in 
Ref. \cite{Kavanagh:2013b}). With a larger numbers of events, it would be 
feasible to increase the number of basis functions and more precisely 
parametrise the form of the speed distribution.

The likelihood function we use for each experiment is:
\begin{equation}
\mathcal{L}(\mathbf{\Theta}) = \prod_{i = 1, N_{\textrm{bins}}} 
\frac{(N_{\rm e}^i)^{N_{\rm o}^i}}{N_{\rm o}^i!} e^{-N_{\rm e}^i}\,,
\end{equation}
where the energy window is divided into $N_{\textrm{bins}}$ bins. For each bin, 
$N_{\rm e}^i$ is the expected number events, for a given set of parameters 
$\mathbf{\Theta}$, and $N_{\rm o}^i$ is the observed number of events (i.e.~the 
mock data). The total likelihood is the product over the likelihoods for 
each experiment considered. 

Finally, it is often interesting to consider the probability distribution of a
subset of parameters from the full parameter space. We do this by 
{\it profiling}, so that the profile likelihood $\mathcal{L}_p(\theta_i)$ of 
the $i$-th parameter is obtained by maximising $\mathcal{L}$ over the other 
parameters:
\begin{equation}
\mathcal{L}_p(\theta_i) = 
\max\limits_{0,...,i-1,i+1,...,N} \mathcal{L}(\mathbf{\Theta})\,.
\end{equation}
We have checked that, for the data sets and likelihoods used here, the 
marginalised posterior distributions of the parameters of interest do not 
differ qualitatively from the profile likelihood. We therefore do not display 
the marginalised posterior distributions. Because we only make use of the 
likelihood (and not the posterior distribution) in parameter inference, 
we expect that the priors will not strongly impact the results. In addition, 
the use of Asimov data with a large number of energy bins means that we expect 
the confidence intervals obtained from the asymptotic properties of the 
profile likelihood to be valid.

\section{Direct detection data only}
\label{sec:DDonly}
\makeatletter{}\begin{figure*}[Htbh!]
\centering
\includegraphics[width=0.32\textwidth]{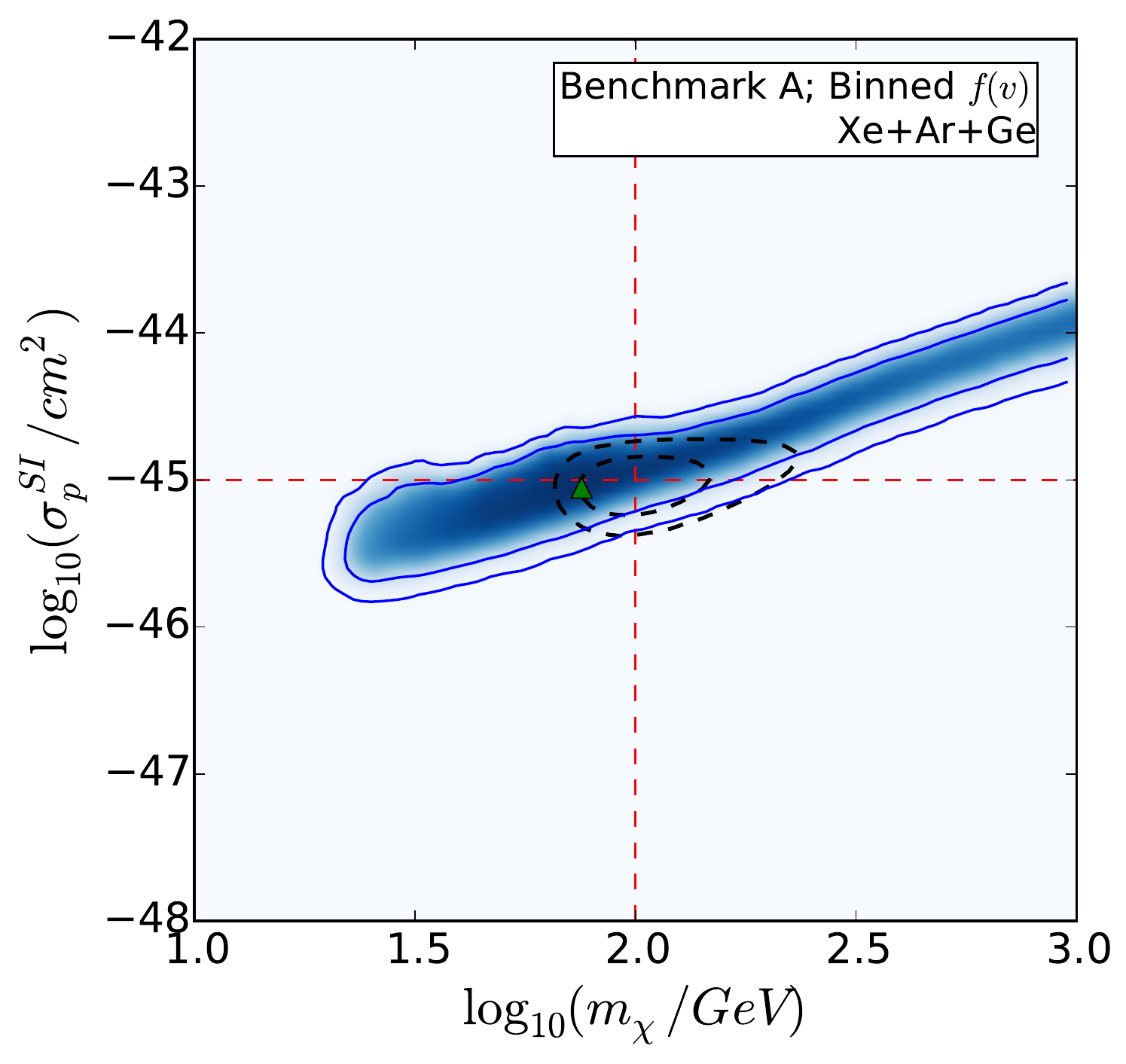}
\includegraphics[width=0.32\textwidth]{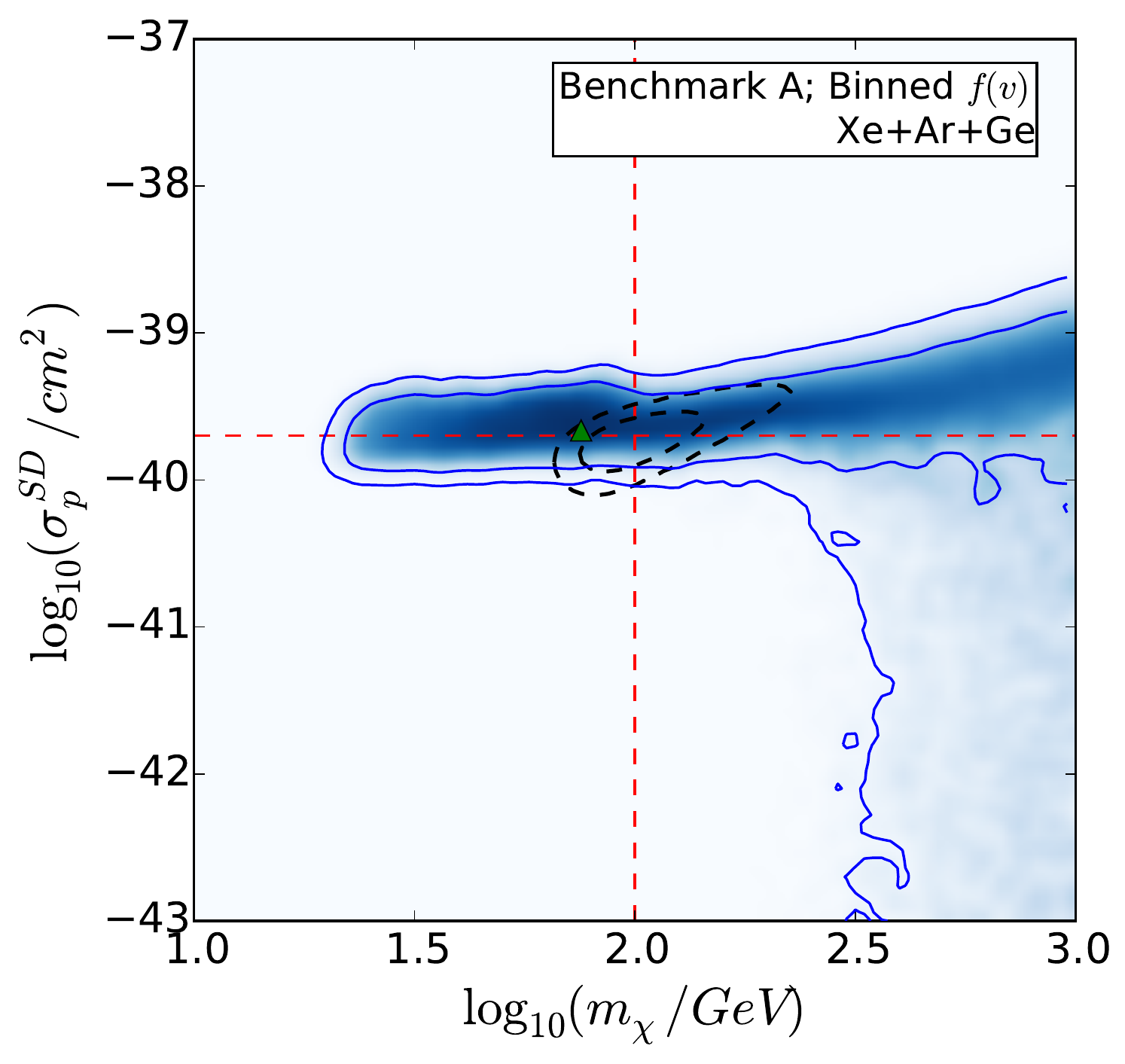}
\includegraphics[width=0.32\textwidth]{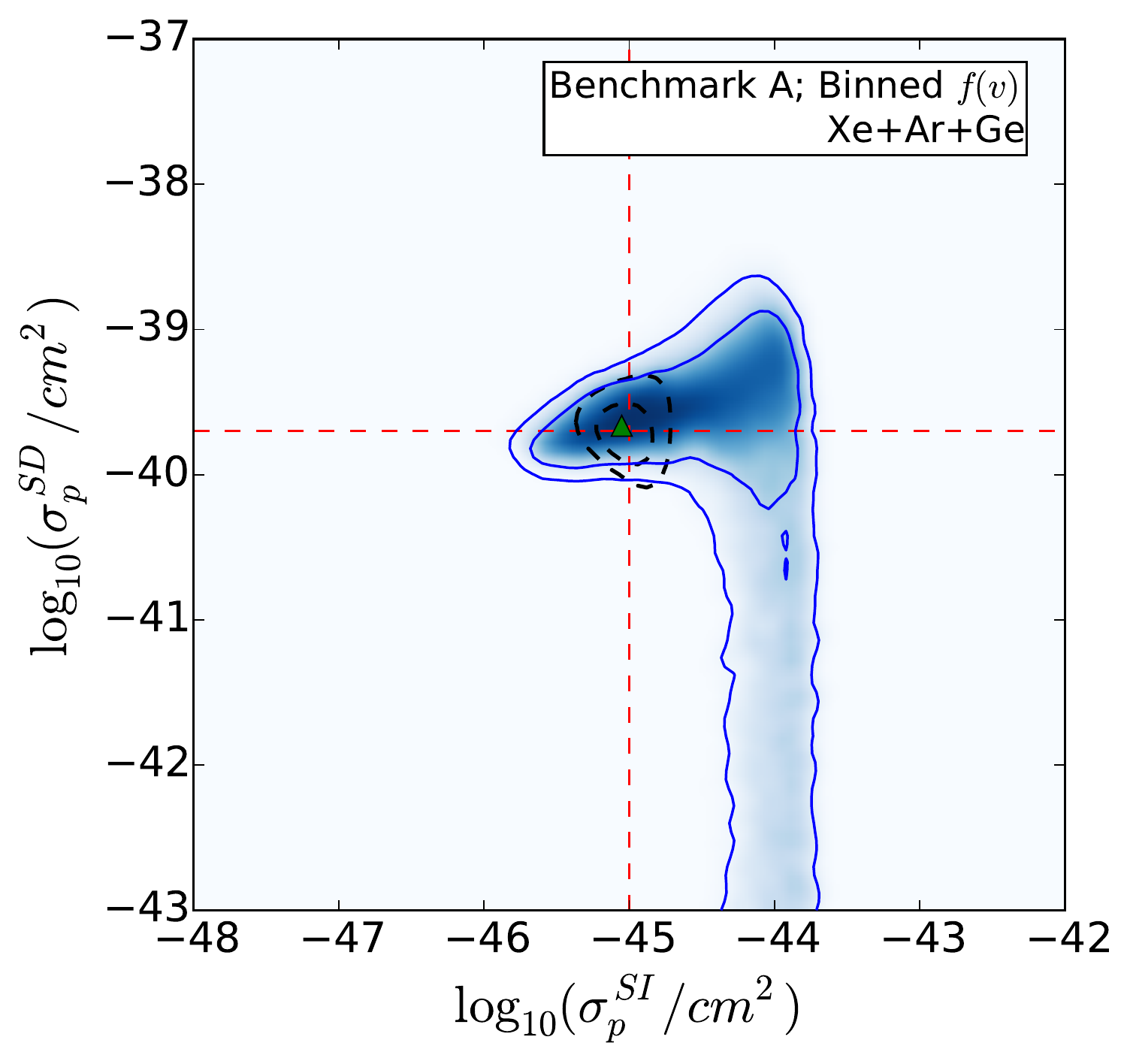}
\includegraphics[width=0.32\textwidth]{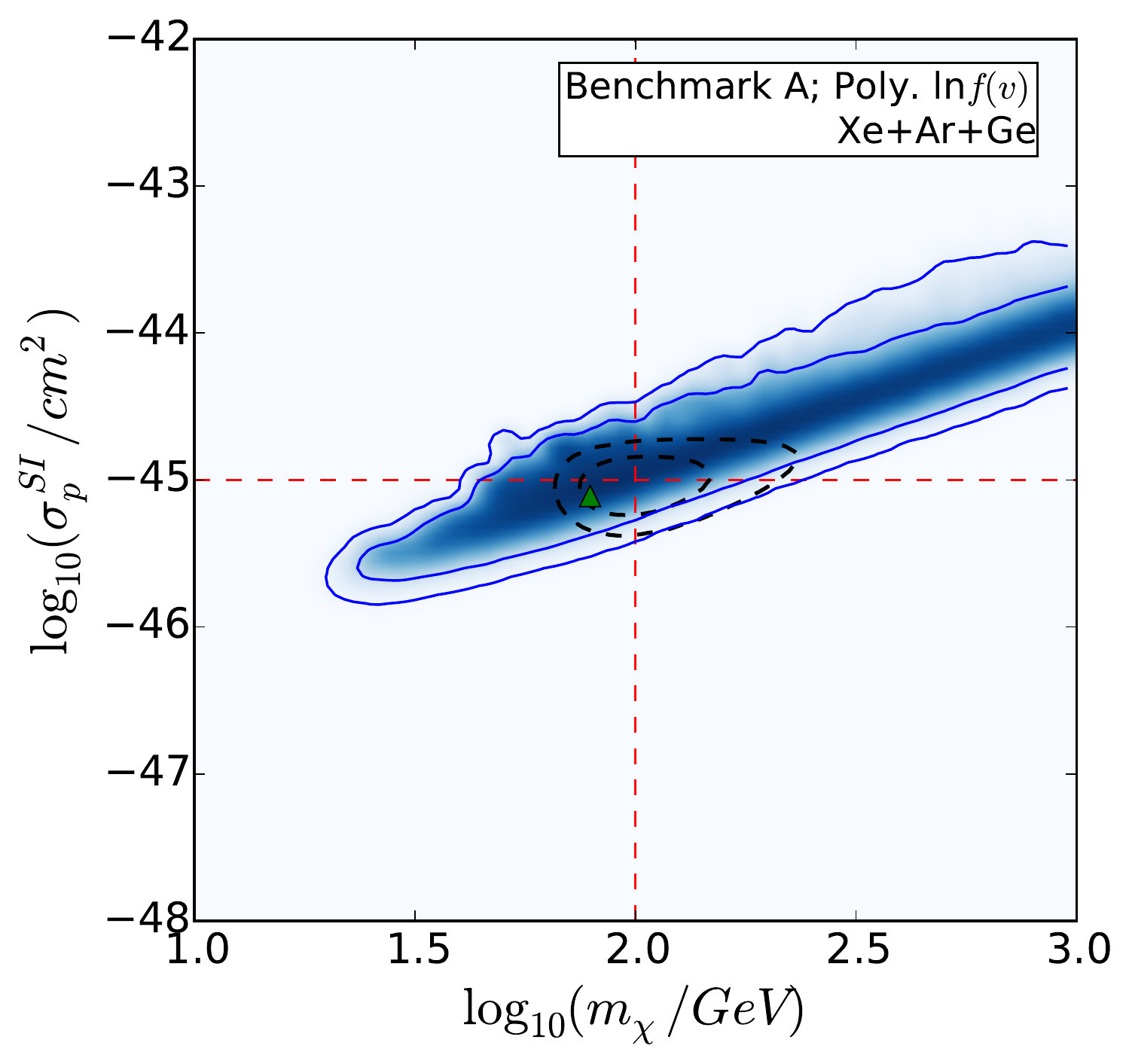}
\includegraphics[width=0.32\textwidth]{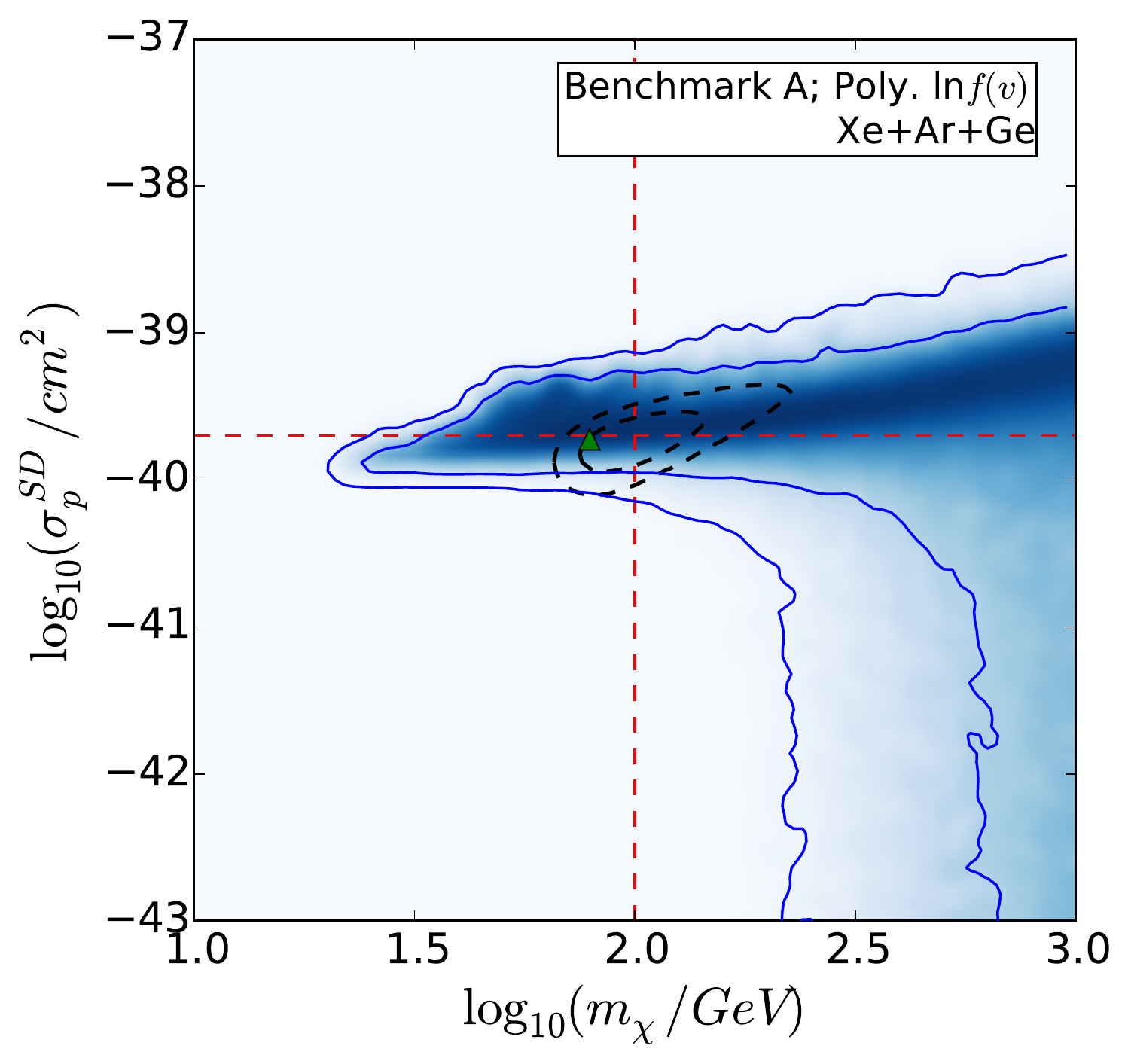}
\includegraphics[width=0.32\textwidth]{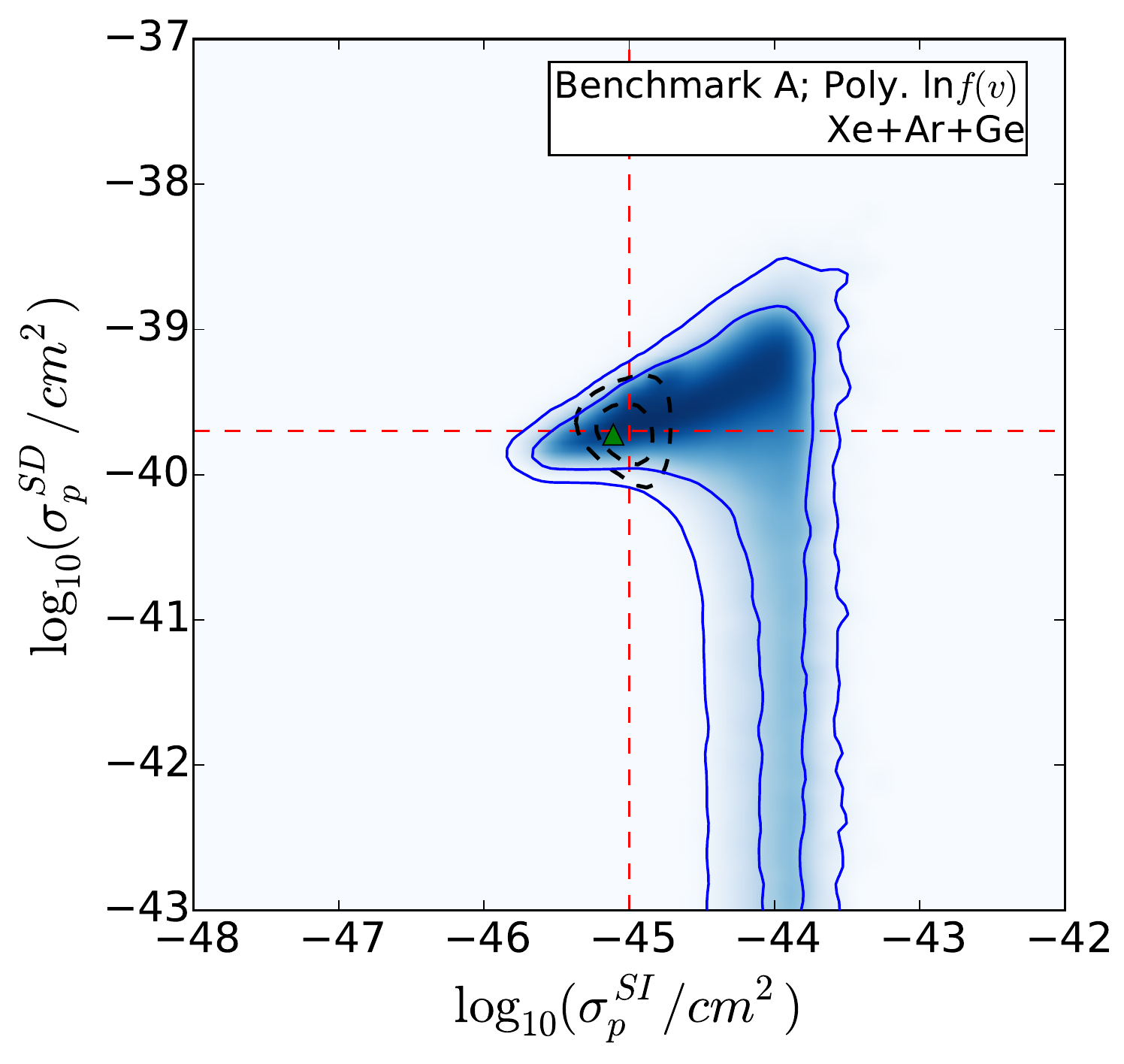}
\caption{2-dimensional profile likelihood for (\mwimp, \sigmapsi) (left column), (\mwimp, \sigmapsd) (central column) and (\sigmapsi, \sigmapsd) (right column), using only information from three direct detection experiments with Xenon, Germanium and Argon targets (see Tab. \ref{tab:Experiments}) for benchmark A. The shaded areas and solid blue contours are obtained from a scan of a parameter space that includes both the particle physics quantities (\mwimp, \sigmapsi and \sigmapsd) and the parameters defining the speed distribution. The panels in the top row are for a binned parametrisation, while those in the second row are for a polynomial parametrisation (see Sec. \ref{sec:reconstruction}). The dashed black contours are from a separate scan, performed with the speed distribution fixed to its input form (see Tab. \ref{tab:benchmarks}). In all cases, the inner (outer) contours enclose the 68\% (95\%) confidence level region. The green triangle marks the position of the best-fit point for the scan including the parameters of the speed distributions. The red dashed lines show the nominal values for \mwimp, \sigmapsi and \sigmapsd assumed for this benchmark.}
\label{fig:DDonly_benchmarkA}
\end{figure*}

In this section we present the results of the scans performed with only
direct detection data, leaving the discussion of the impact of IceCube data for
the following section.

\subsection{Benchmark A}
Figure~\ref{fig:DDonly_benchmarkA} shows the 2-dimensional profile likelihood
distributions for benchmark A, i.e. \mwimp=100 GeV, 
\sigmapsi= $10^{-45} \textrm{ cm}^2$, 
\sigmapsd= $2 \times 10^{-40} \textrm{ cm}^2$ and a SHM $f(v)$. Left panels are 
for (\mwimp, \sigmapsi), central ones for (\mwimp, \sigmapsd), while the ones 
on the right are for (\sigmapsi, \sigmapsd). Shaded regions and solid blue
contours are for scans carried out over particle physics quantities 
(i.e.~\mwimp, \sigmapsi and \sigmapsd) as well as the parameters entering in
the parametrisation of the speed distribution. The first row is for the binned
$f(v)$ and the second row for the polynomial expansion (see 
Sec. \ref{subsec-param}). In addition, the dashed black contours are from
a scan performed keeping the speed distribution fixed, i.e. assuming that the 
correct $f(v)$ is known and there are no astrophysical uncertainties.

In the case of a fixed speed distribution (dashed black lines), the 
reconstruction is very good: the mass is well constrained and closed contours 
are obtained for both the SI and SD cross section (right column). These constraints on the WIMP mass are similar to those obtained in Ref.~\cite{Arina:2013}  whose Benchmark B is the same as our Benchmark A (see the middle row of Fig.~1 of Ref.~\cite{Arina:2013}). However, the possible degeneracy between the two cross sections \cite{Cerdeno:2013gqa}, which is observed in Ref.~\cite{Arina:2013}, is 
broken by the fact that we use three different target experiments, one of 
which (argon) is only sensitive to SI interactions.

\begin{figure}[th!]
\centering
\includegraphics[width=0.45\textwidth]{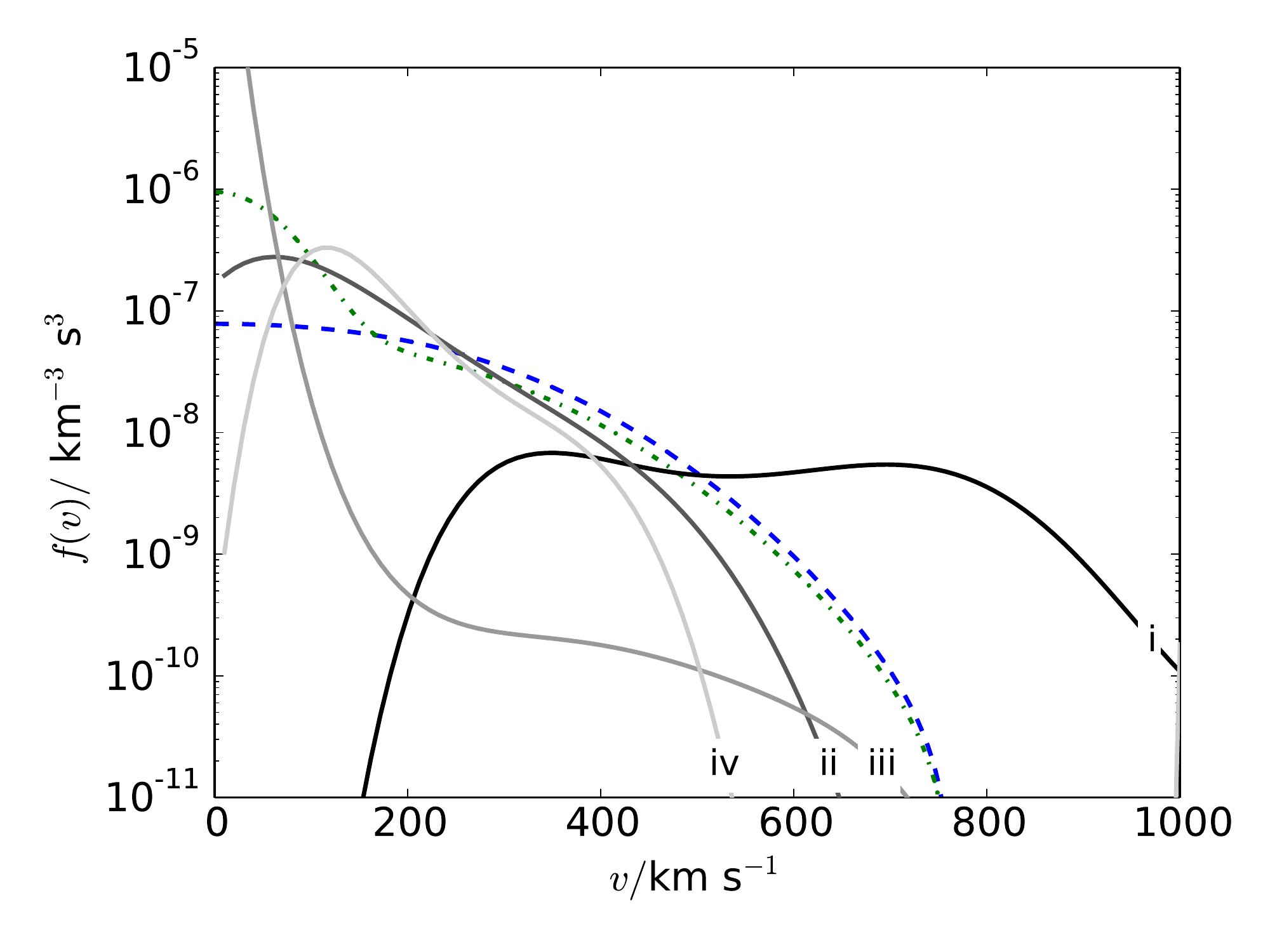}
\caption{Example shapes for the directionally-averaged velocity distribution $f(v)$. These are labeled i-iv and are referred to in the text of Sec.~\ref{sec:DDonly}, \ref{sec:DDwithIC} and \ref{sec:SpeedDist} to explain the different regions of parameter space which can be fit to the data. For comparison, the Standard Halo Model (SHM) and the SHM with a dark disk (SHM+DD) are shown as dashed blue and dot-dashed green lines respectively.}
\label{fig:SpeedExamples}
\end{figure}

When we allow for the realistic possibility of a variable speed distribution, 
the contours unsurprisingly increase in size. For both the binned and 
polynomial parametrisation, the contours have a similar shape, extending down 
to 20 -- 30 GeV in \mwimp. This is because both parametrisations can encompass 
distributions that are flatter than the SHM. Decreasing the WIMP mass steepens 
the recoil spectrum, allowing these flatter distributions to be compatible 
with the mock data. An example of such a speed distribution is the one 
labelled `i', in Fig.~\ref{fig:SpeedExamples}. In order to remain normalised, 
such distributions must be depleted at low speeds (i.e. below 200 
${\rm km \, s}^{-1}$), where the experiments are no longer sensitive (see 
Fig.~\ref{fig:speedoverlap}). 

Similarly, the data can also be well fit by higher WIMP masses. Increasing the 
WIMP mass moves the $v_{\rm min}$ intervals probed by the three direct detection 
experiments to lower values (see Fig. \ref{fig:speedoverlap}). However, in 
order to provide a good fit to the data, the relative number of recoil events 
in each experiment must remain roughly the same. With the SHM, this is not 
possible, since too few events would be produced in the xenon experiment. A 
velocity integral which is similar to the SHM in the region probed by 
germanium and argon, but steeper in the region probed by xenon can be used to 
compensate for this by increasing the number of xenon events. An example of a 
speed distribution which produces this effect is shown in 
Fig.~\ref{fig:SpeedExamples}, labeled `ii'. Such an $f(v)$ is possible with 
both the polynomial and binned parametrisation, and therefore larger WIMP 
masses are allowed. Furthermore as \mwimp is increased above the masses of the 
target nuclei, the shape of the recoil spectrum becomes almost independent of 
\mwimp. This effect is well understood \cite{Green:2007rb,Green:2008rd} and 
it contributes to the degeneracy between the mass and the cross section for 
large \mwimp. As a consequence of this effect, we also do not expect the 
results obtained here to change qualitatively if the upper limit of the prior 
were increased beyond 1000 GeV.

For large values of \mwimp the contours also extend down to small values of 
\sigmapsd. This means that in the (\sigmapsi, \sigmapsd) plane the contours 
are now open (lower right panel), as opposed to the closed contours obtained 
with fixed astrophysics. As explained in the previous paragraphs, the region 
at large mass and large SI and SD cross sections provides a good fit to the 
data with a velocity integral that is slightly steeper than the SHM. 
Decreasing $\sigmapsd$ means that fewer events will be produced in the xenon 
and germanium experiments, with no effect on the argon detector. The same 
relative numbers of events for the three targets can be maintained with a 
velocity integral that is even steeper at low speeds (where xenon and 
germanium are sensitive), but unchanged in the region probed by argon, between 
200 and 400 ${\rm km \, s}^{-1}$. This requires a shape for $f(v)$ which rises 
more rapidly at low speeds than example `ii' in Fig.~\ref{fig:SpeedExamples}. 
With decreasing $\sigmapsd$, a point is reached where all the events are 
explained by SI interactions and lowering the SD cross section further has no 
effect. Conversely, it is not possible to explain the data in terms of only 
SD interactions, as $\sigmapsi$ is constrained by the (small) number of events 
in the argon experiment which couples only via SI interactions. Therefore the 
contours do not extend to low values of \sigmapsi.

\begin{figure*}[thp!]
\centering
\includegraphics[width=0.31\textwidth]{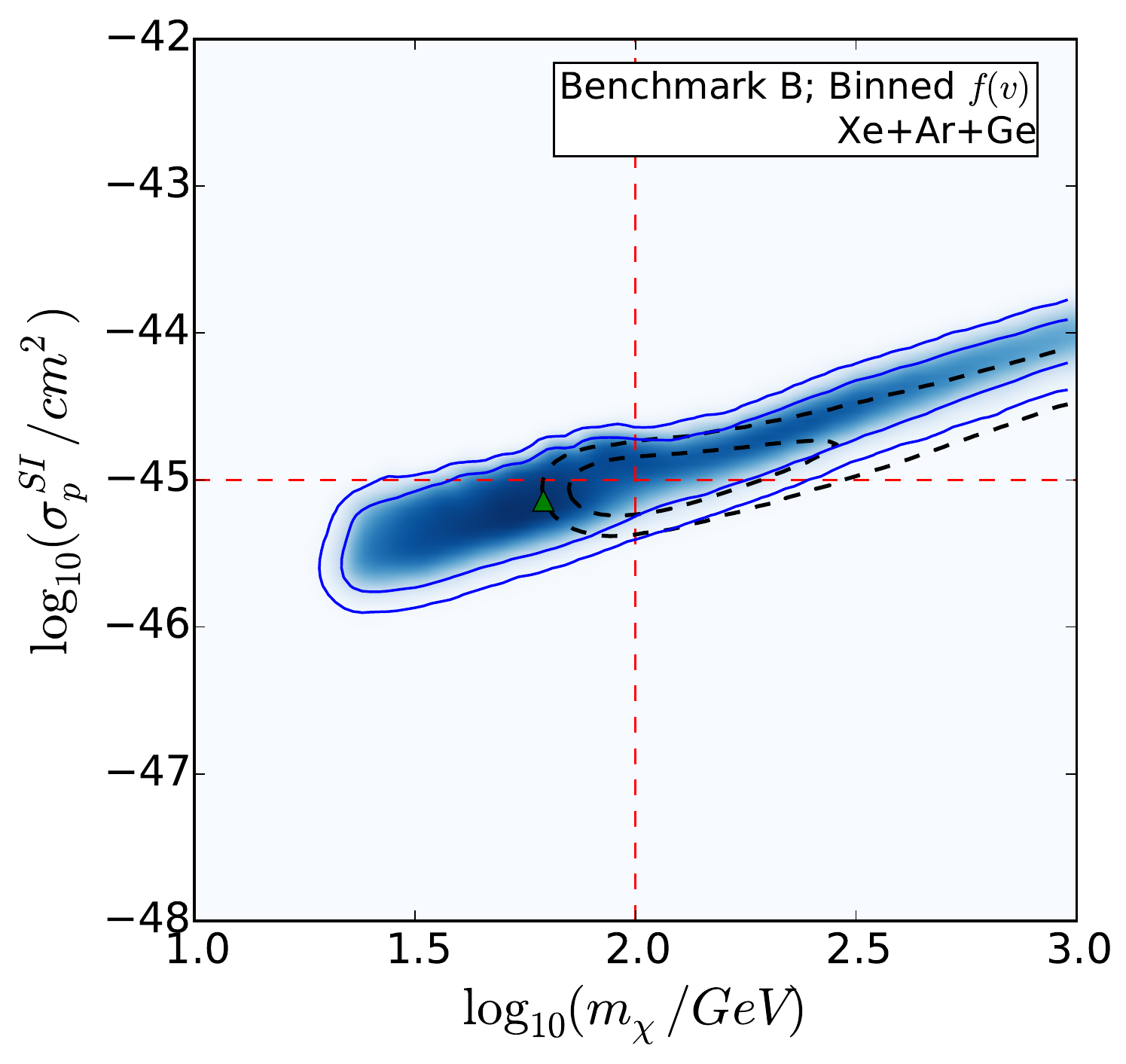}
\includegraphics[width=0.31\textwidth]{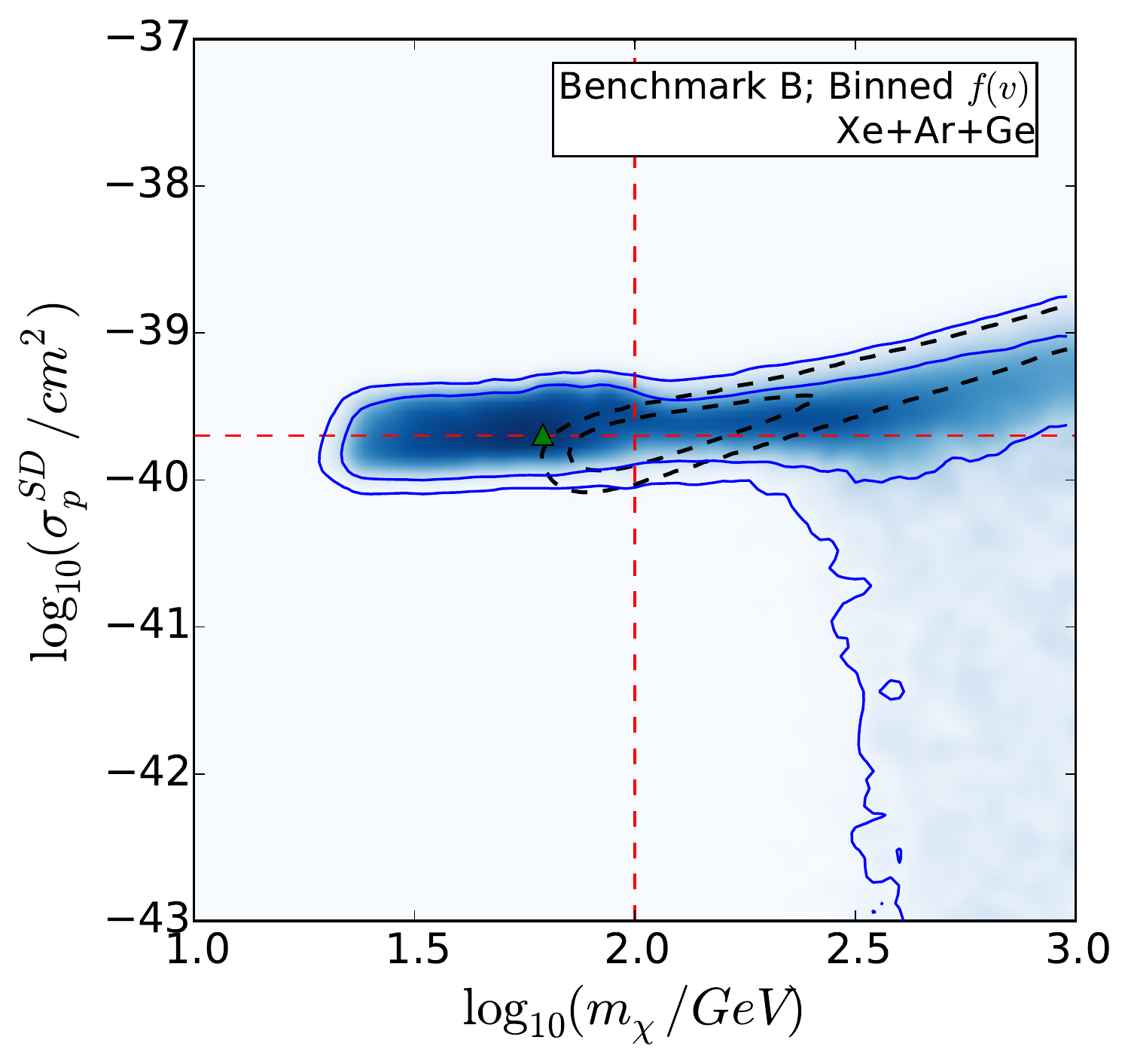}
\includegraphics[width=0.31\textwidth]{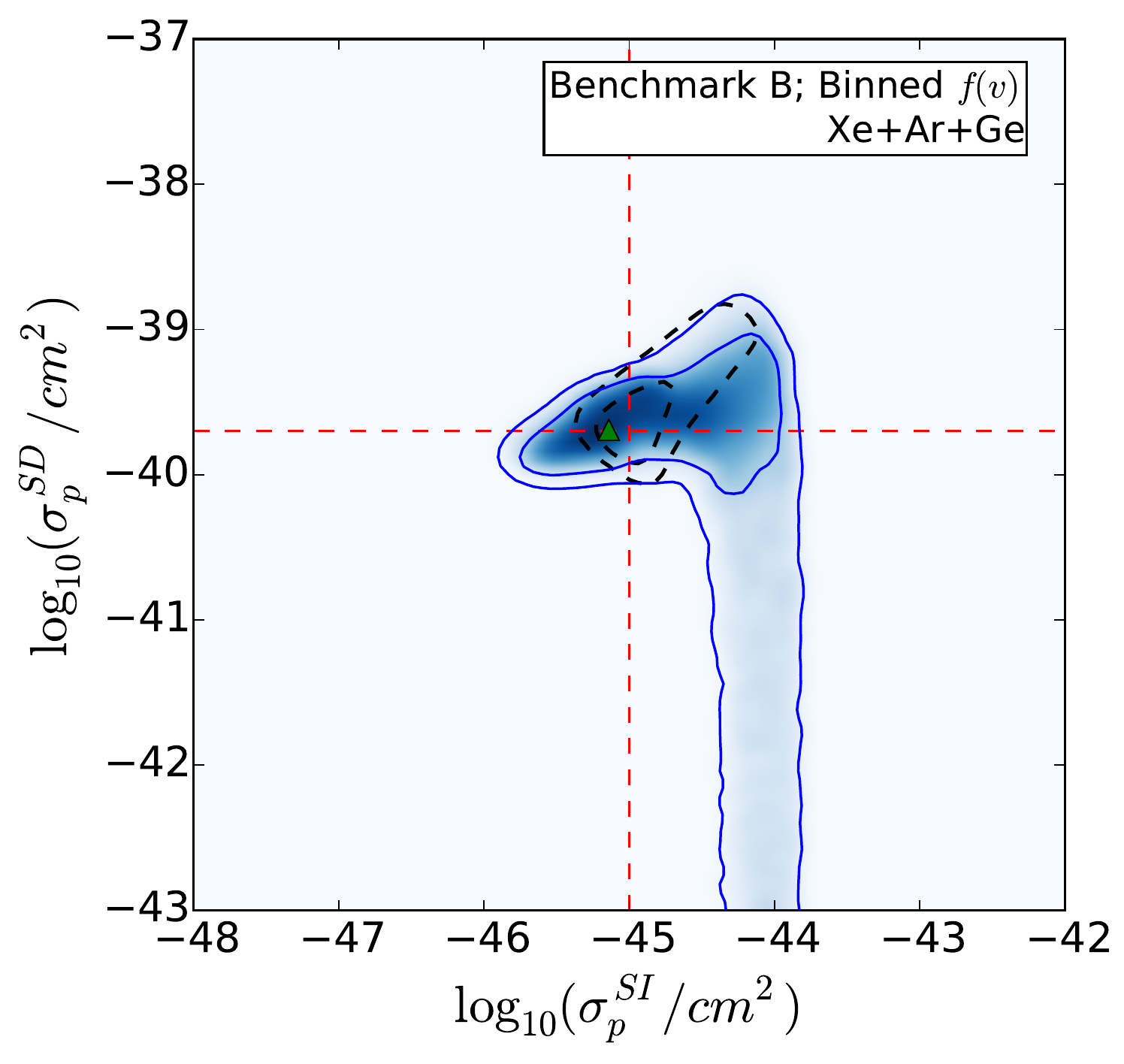}
\includegraphics[width=0.31\textwidth]{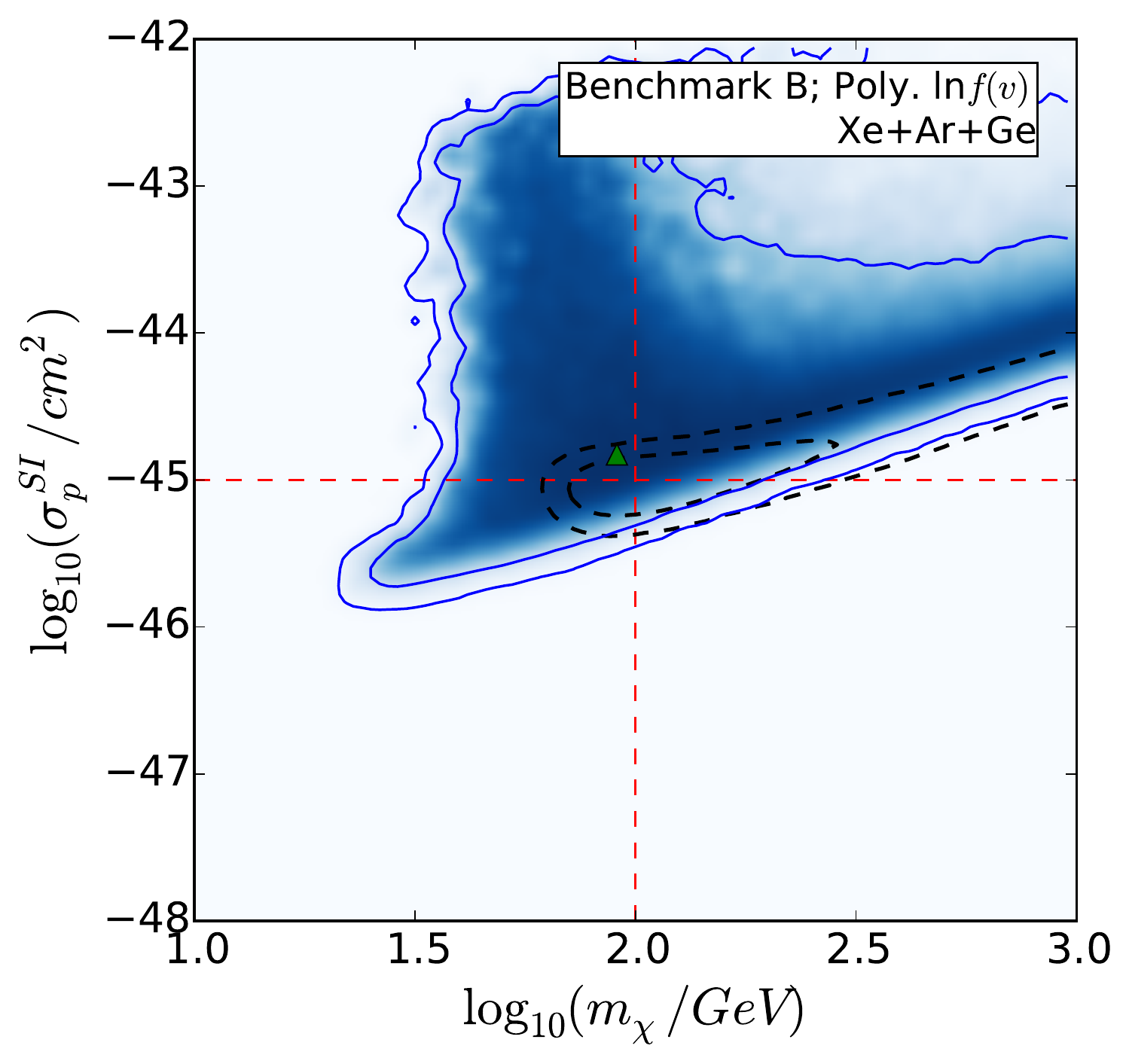}
\includegraphics[width=0.31\textwidth]{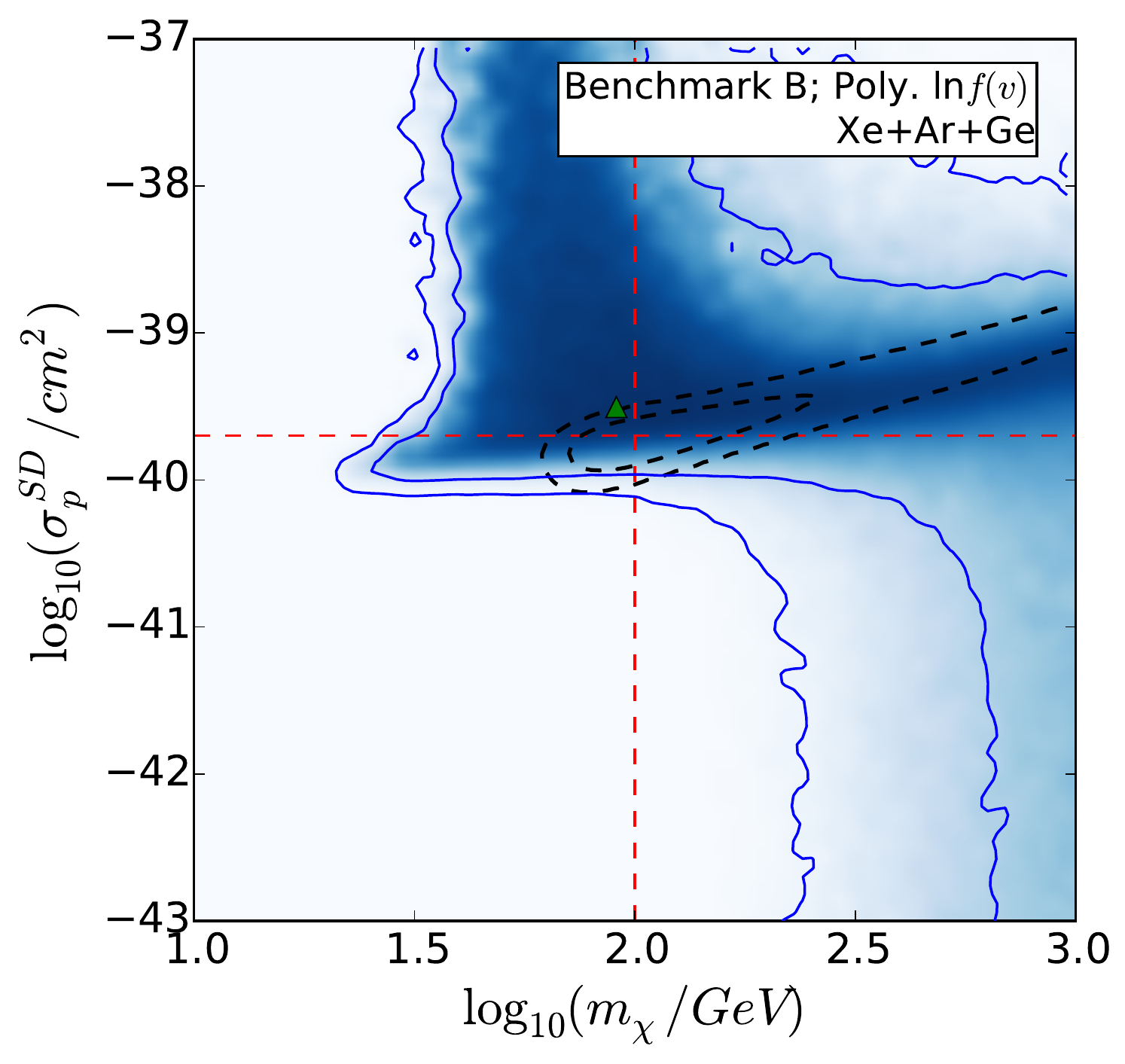}
\includegraphics[width=0.31\textwidth]{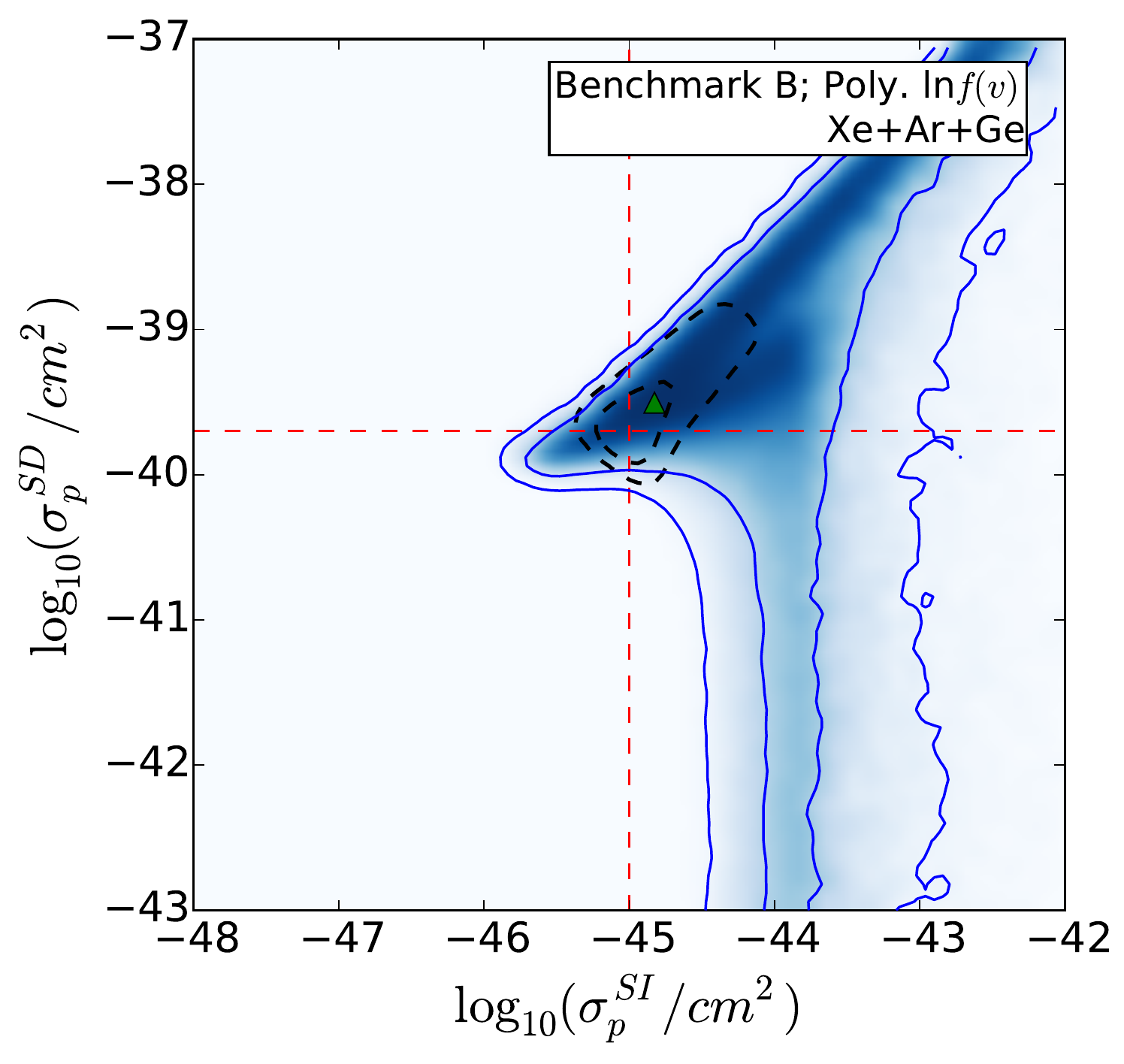}
\caption{Same as Fig. \ref{fig:DDonly_benchmarkA} but for benchmark B.}
\label{fig:DDonly_benchmarkB}
\end{figure*}

\begin{figure*}[tbhp!]
\centering
\includegraphics[width=0.31\textwidth]{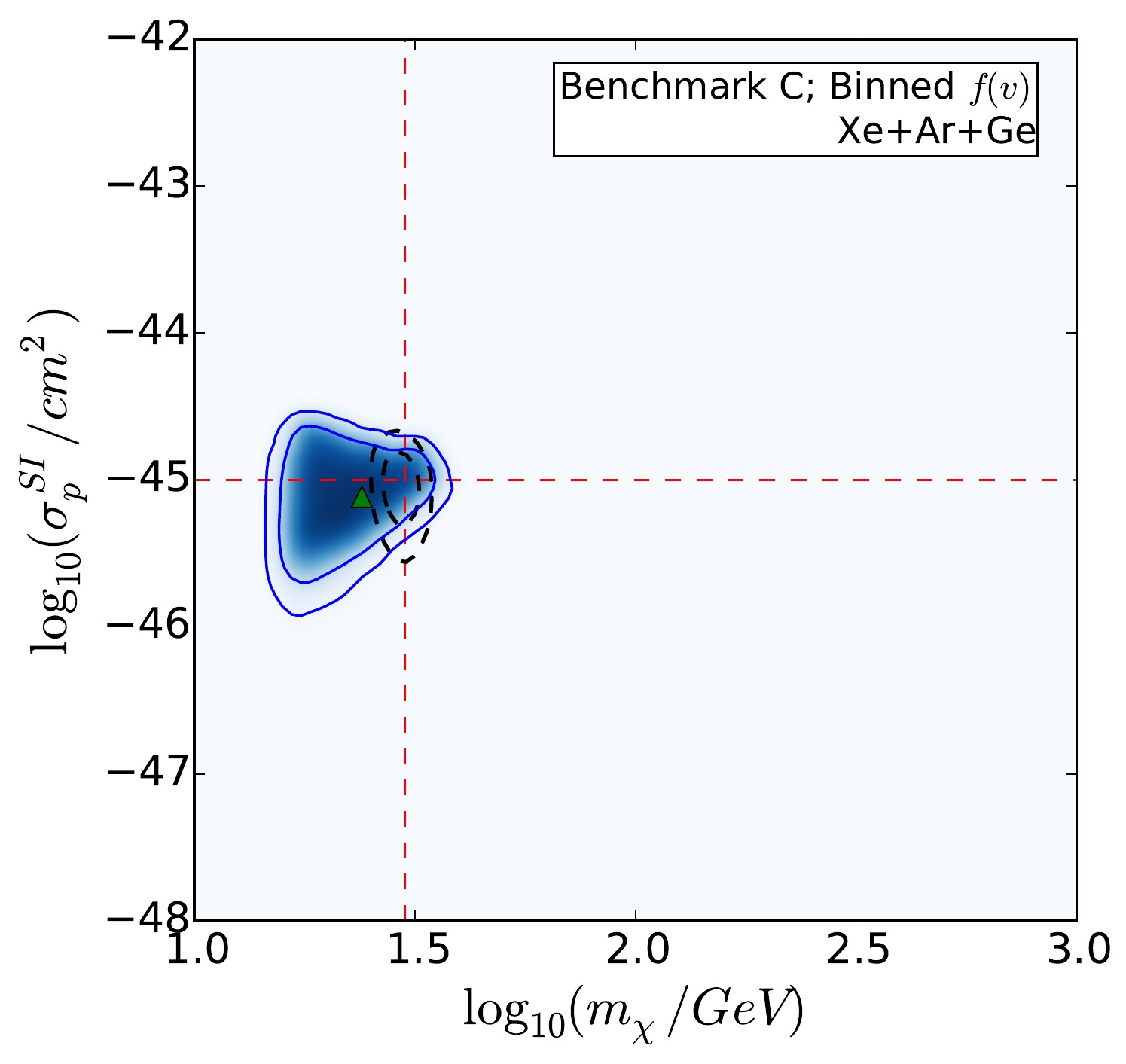}
\includegraphics[width=0.31\textwidth]{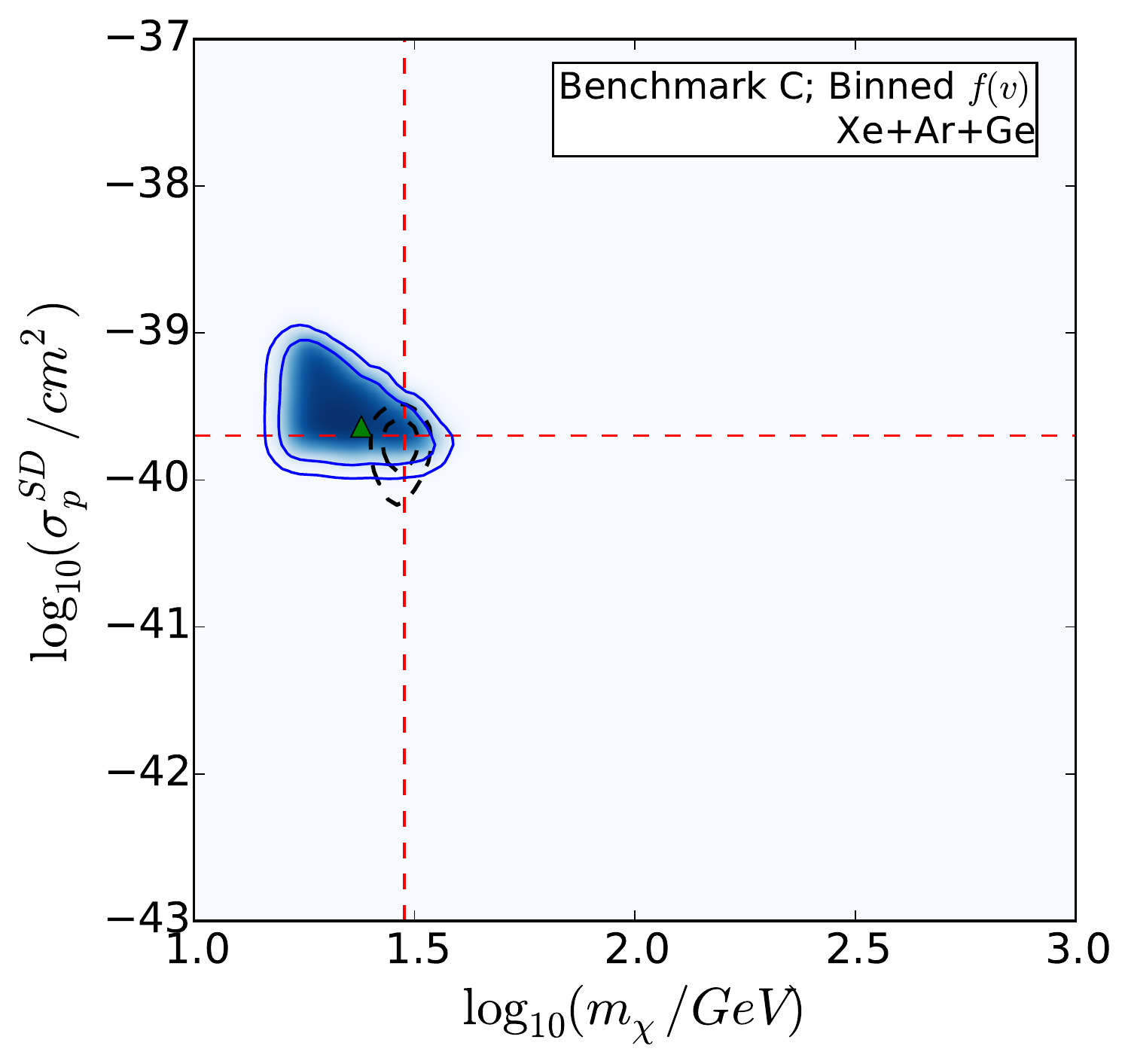}
\includegraphics[width=0.31\textwidth]{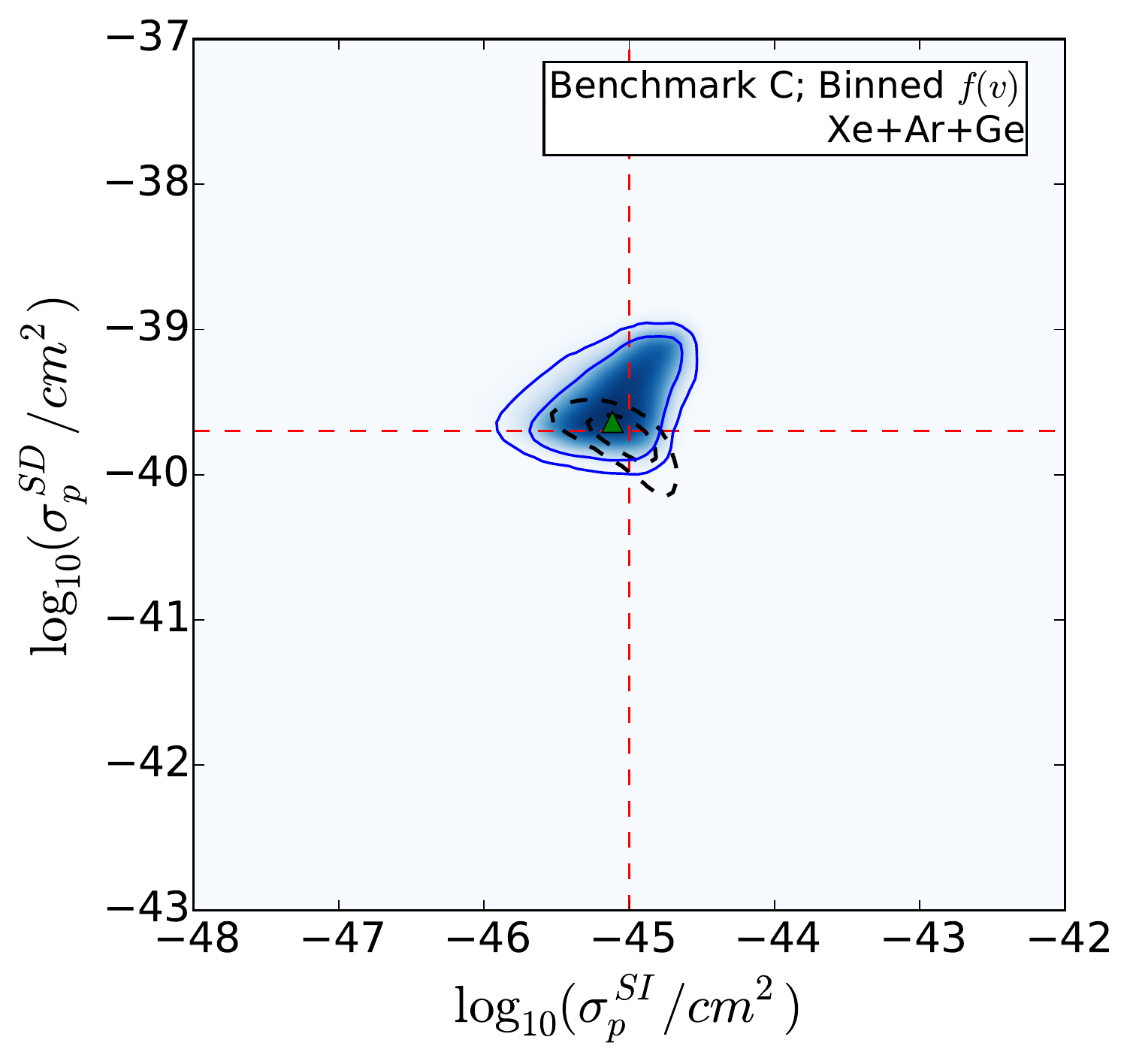}
\includegraphics[width=0.31\textwidth]{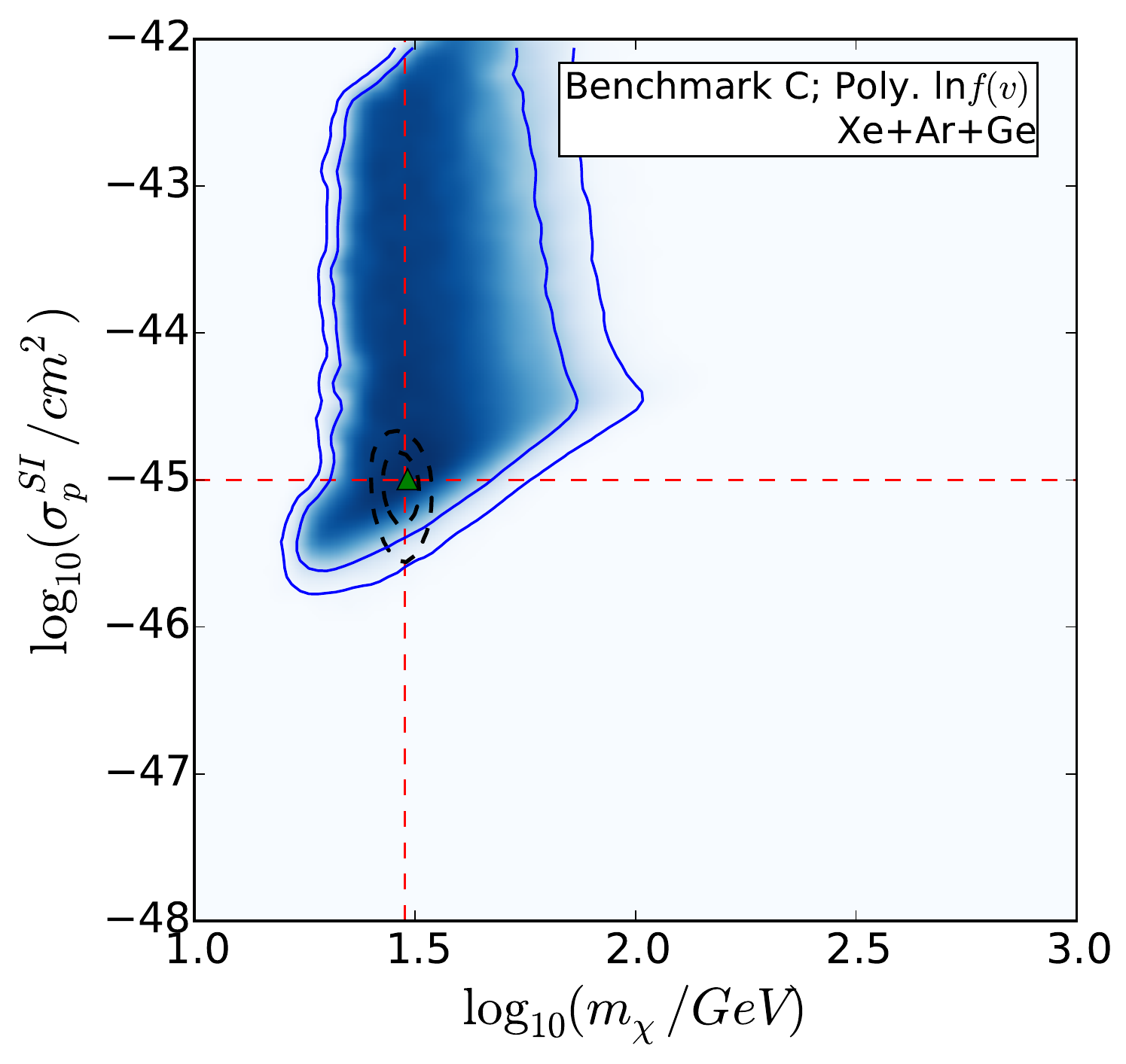}
\includegraphics[width=0.31\textwidth]{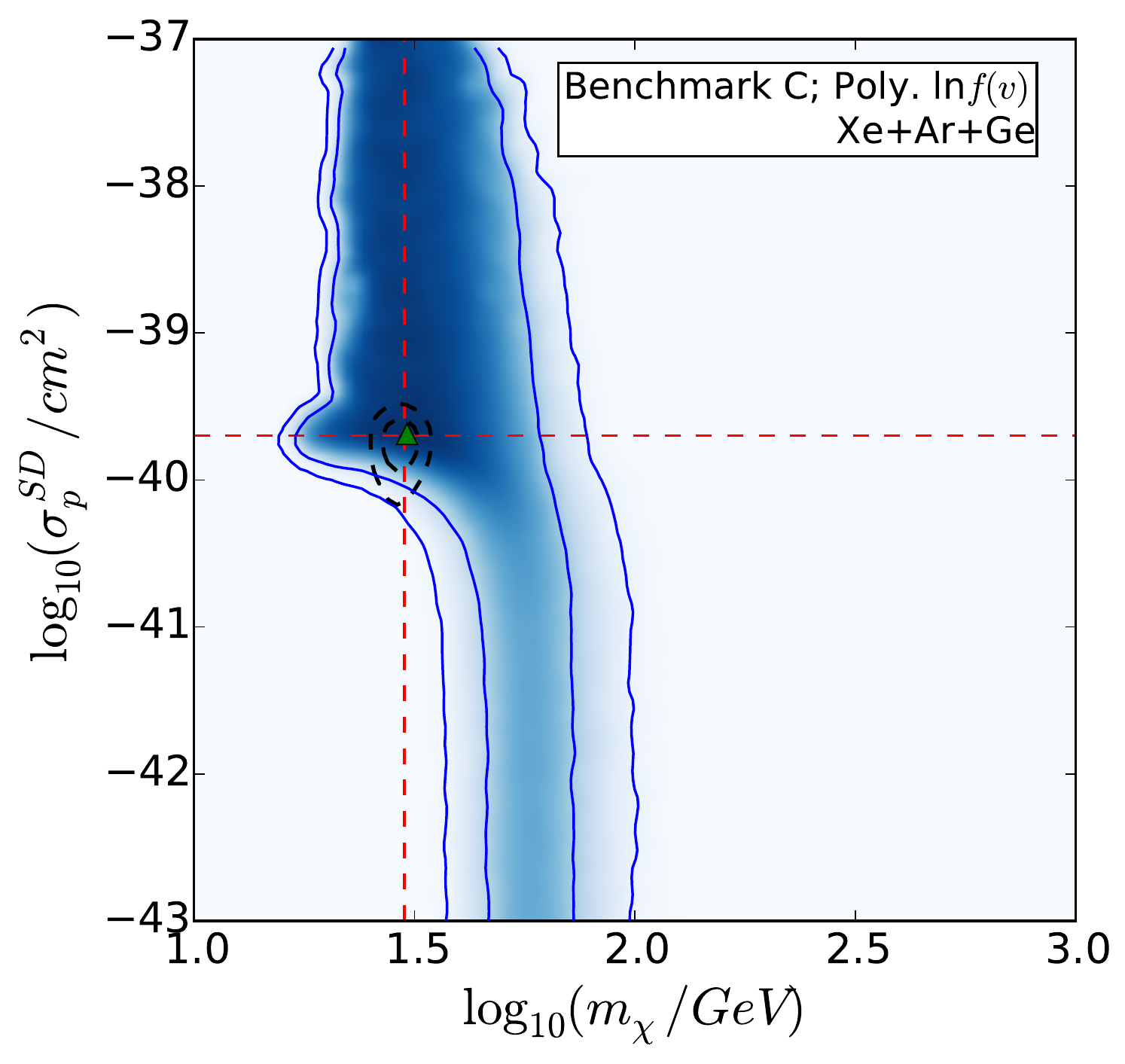}
\includegraphics[width=0.31\textwidth]{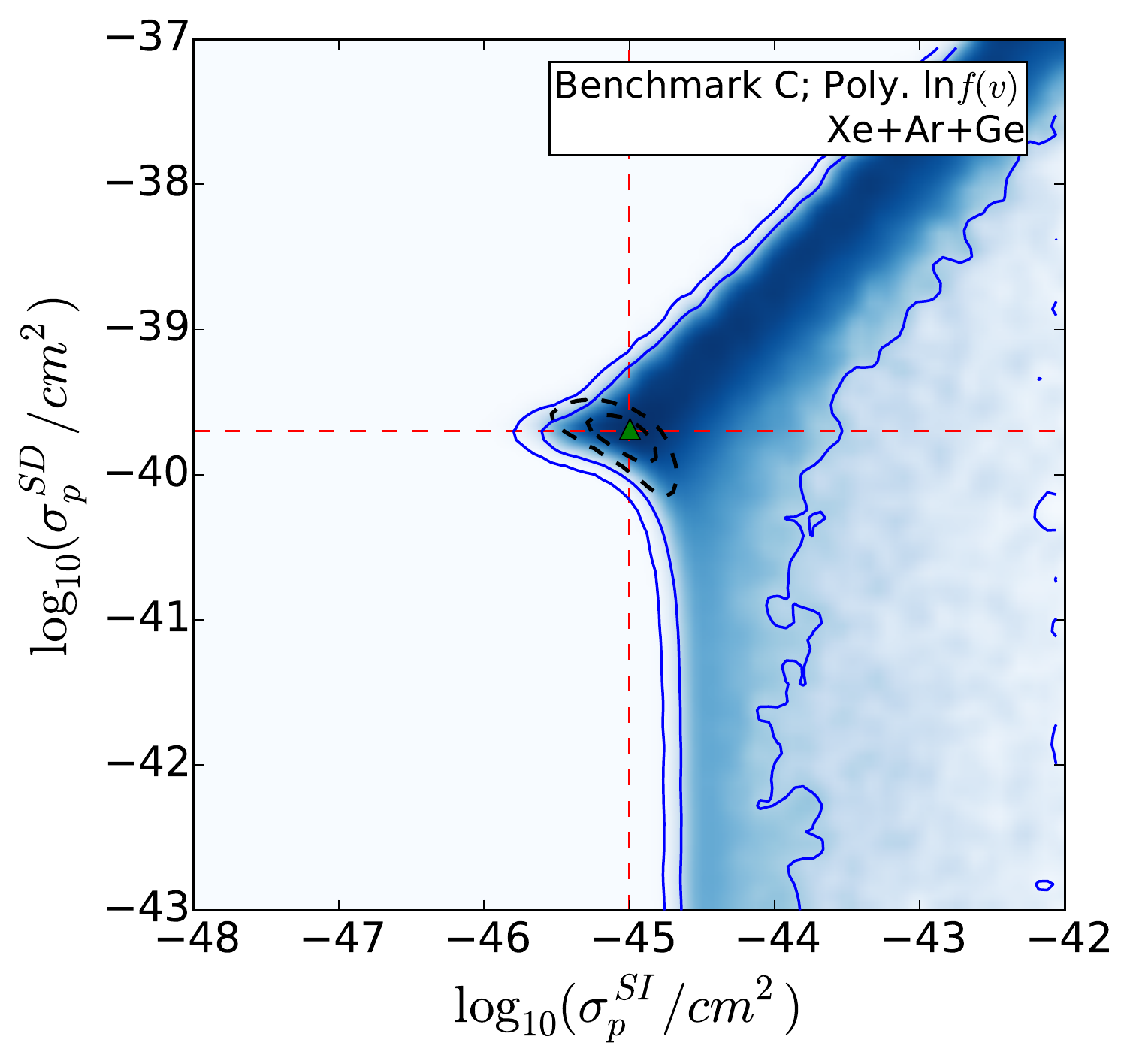}
\caption{Same as Fig. \ref{fig:DDonly_benchmarkA} but for benchmark C.}
\label{fig:DDonly_benchmarkC}
\end{figure*}

\subsection{Benchmark B}
Figure~\ref{fig:DDonly_benchmarkB} shows the 2-dimensional profile likelihood
distributions in the case of benchmark B, which has the same values of the 
particle physics parameters as benchmark A, but the input speed distribution 
includes a dark disk. The results for benchmark B with the speed distribution 
fixed to its input form (dashed black) are similar to those for benchmark A. 
However, the 95\% confidence contours now extend up to large WIMP masses 
(left and central panels). When the WIMP mass is increased, the relative 
number of events in the xenon experiment can be too small. However, in 
benchmark B (unlike benchmark A) this is counteracted by the steep velocity 
integral at low speeds, due to the presence of the dark disk.

Again, when we allow the speed distribution to vary, the contours are 
significantly wider. In the case of the binned speed distribution, the 
likelihood peaks at around $m_\chi \approx 50 \textrm{ GeV}$, compared to the 
input value of $100 \textrm{ GeV}$. A possible bias in the WIMP mass when 
using the binned distribution has been noted previously 
\cite{Peter:2011,Kavanagh:2012}, although in this case the effect is 
relatively minor and the input value lies within the 68\% contours. When the 
polynomial parametrisation is used, the best-fit point is closer to the input 
parameter values. However, there is a strong degeneracy between the mass and 
the cross sections, and consequently for both parameterisations the 
displacement of the best-fit point away from the input parameter values is 
much smaller than the uncertainties on the parameters.

A significant difference between the two parameterisations is that the 
contours for the polynomial parametrisation extend up to large values of 
\sigmapsi and \sigmapsd (this is most apparent in the lower-right panel of 
Fig.~\ref{fig:DDonly_benchmarkB}). This is a manifestation of the degeneracy 
described in Sec.~\ref{sec:introduction}. Direct detection experiments do not 
probe the low-speed WIMP population. Thus, a velocity integral which is 
compatible with the input one in the region probed by the experiments but 
sharply increasing towards low speeds can still produce a good fit, provided 
that the cross section is also increased to give the correct total number of 
events. An example of such a distribution is shown in 
Fig.~\ref{fig:SpeedExamples}, labeled `iii'. These rapidly varying 
distributions are more easily accommodated in the polynomial parametrisation 
than in the binned one, which explains why the contours do not extend to 
large cross sections in that case (top row).

This region at large cross sections for the polynomial parameterisation did 
not appear in the case of benchmark A. This is because the parameter space 
describing the shape of the speed distribution is very large and distribution 
functions which rise rapidly at low $v$ do not make up a large fraction of the 
parameter space and, therefore, may not be well explored. In the case of 
benchmark B (which has a dark disk component), the input $f(v)$ is already 
increasing towards low speeds. This means that such rapidly rising 
distributions are better explored and this degeneracy becomes clear. The 
degeneracy up to high cross sections would become manifest for benchmark A if 
significantly more live points were used in the parameter scan. Therefore, 
the boundaries of the contours in Fig.~\ref{fig:DDonly_benchmarkA} for 
benchmark A at large \sigmapsi and \sigmapsd should be considered as lower 
limits.

\begin{figure*}[tbh!]
\centering
\includegraphics[width=0.32\textwidth]{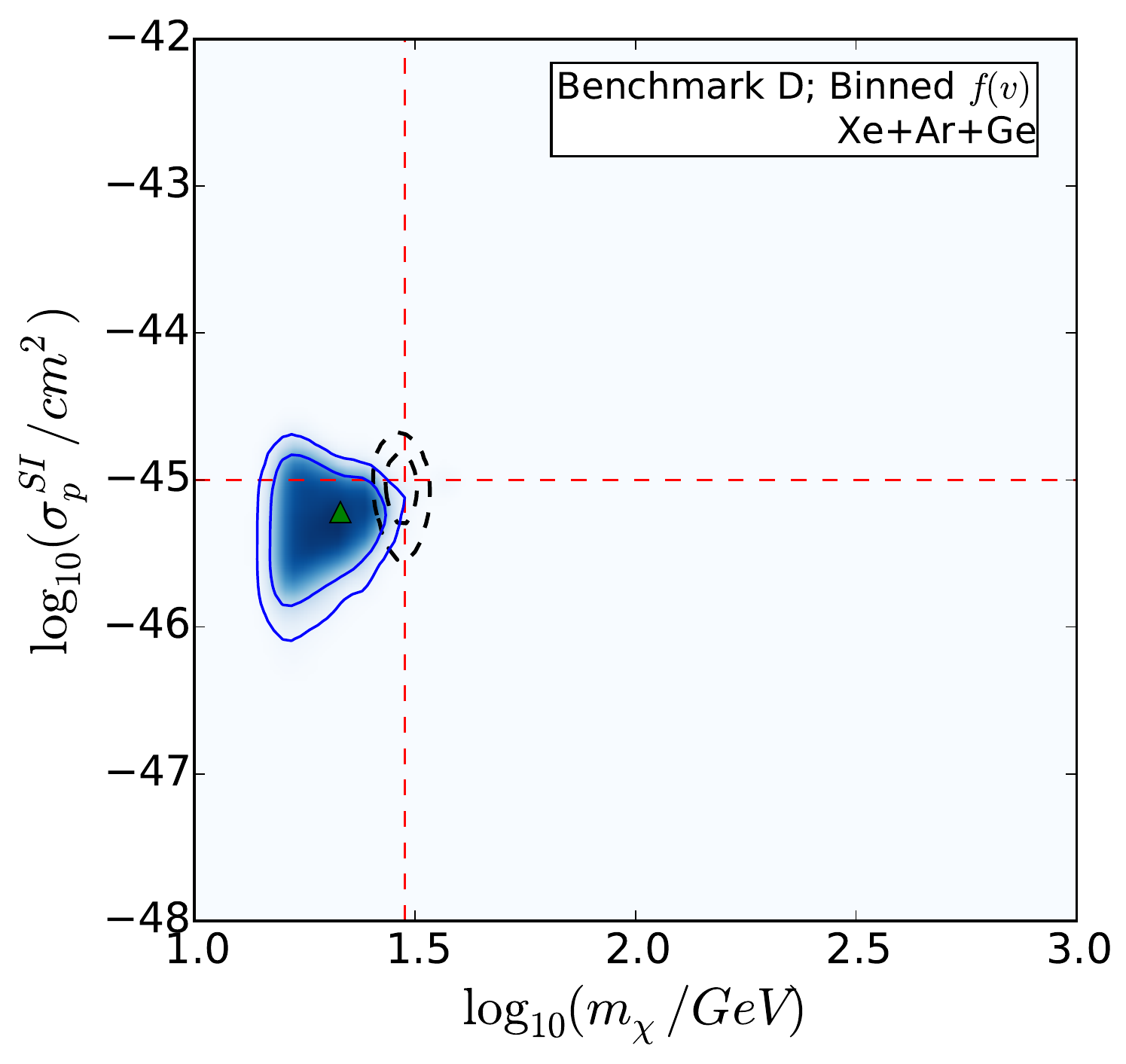}
\includegraphics[width=0.32\textwidth]{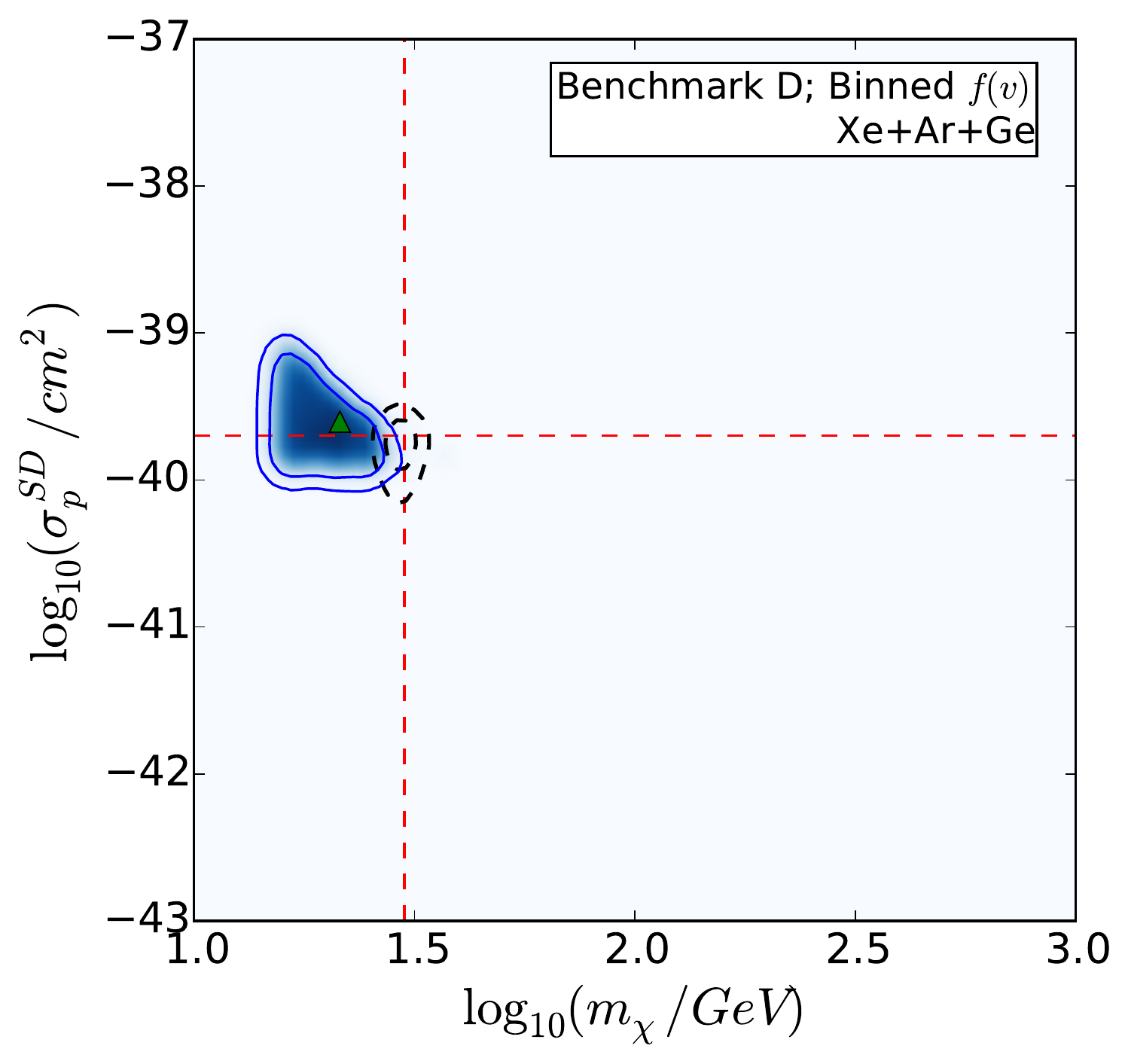}
\includegraphics[width=0.32\textwidth]{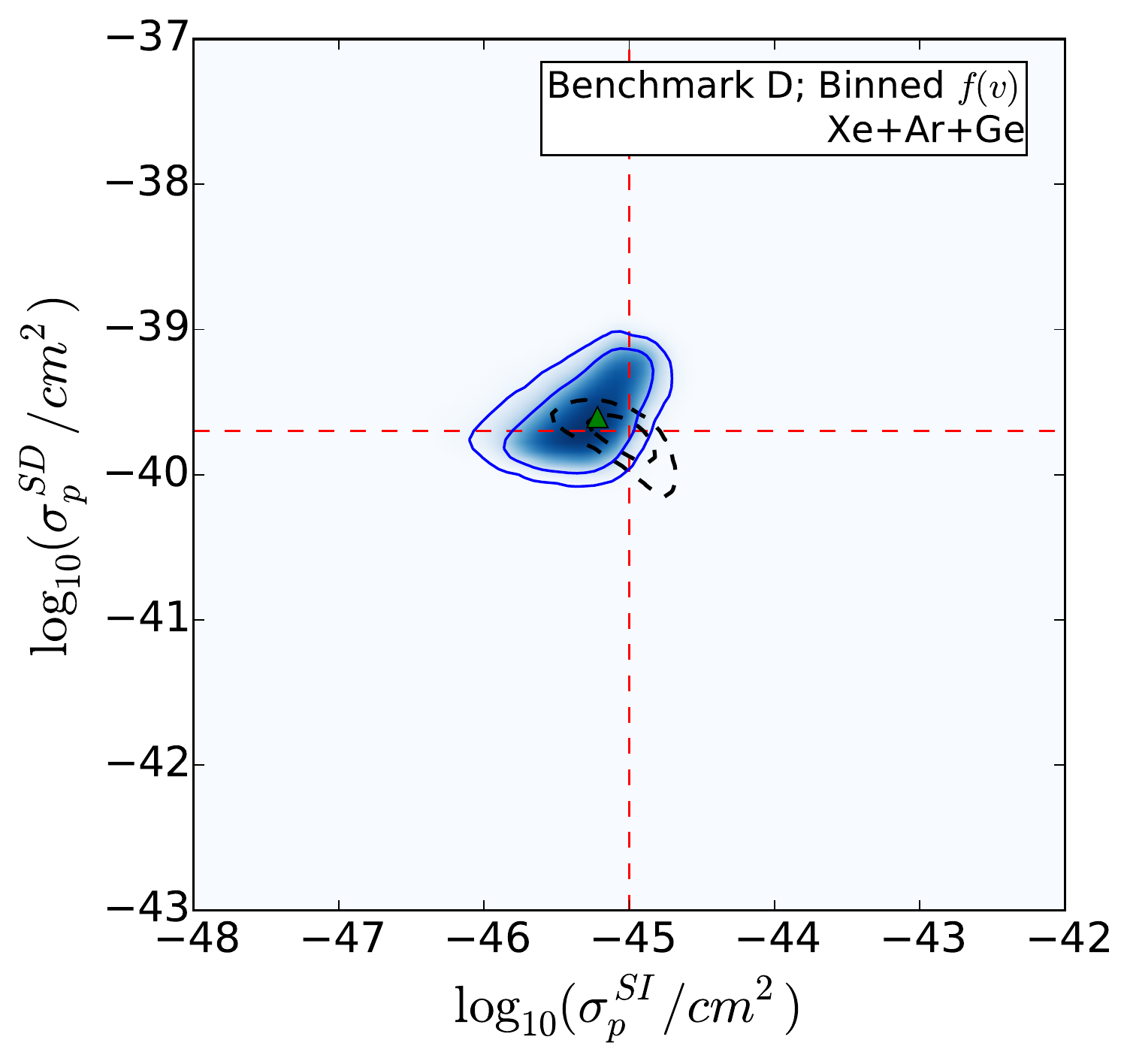}
\includegraphics[width=0.32\textwidth]{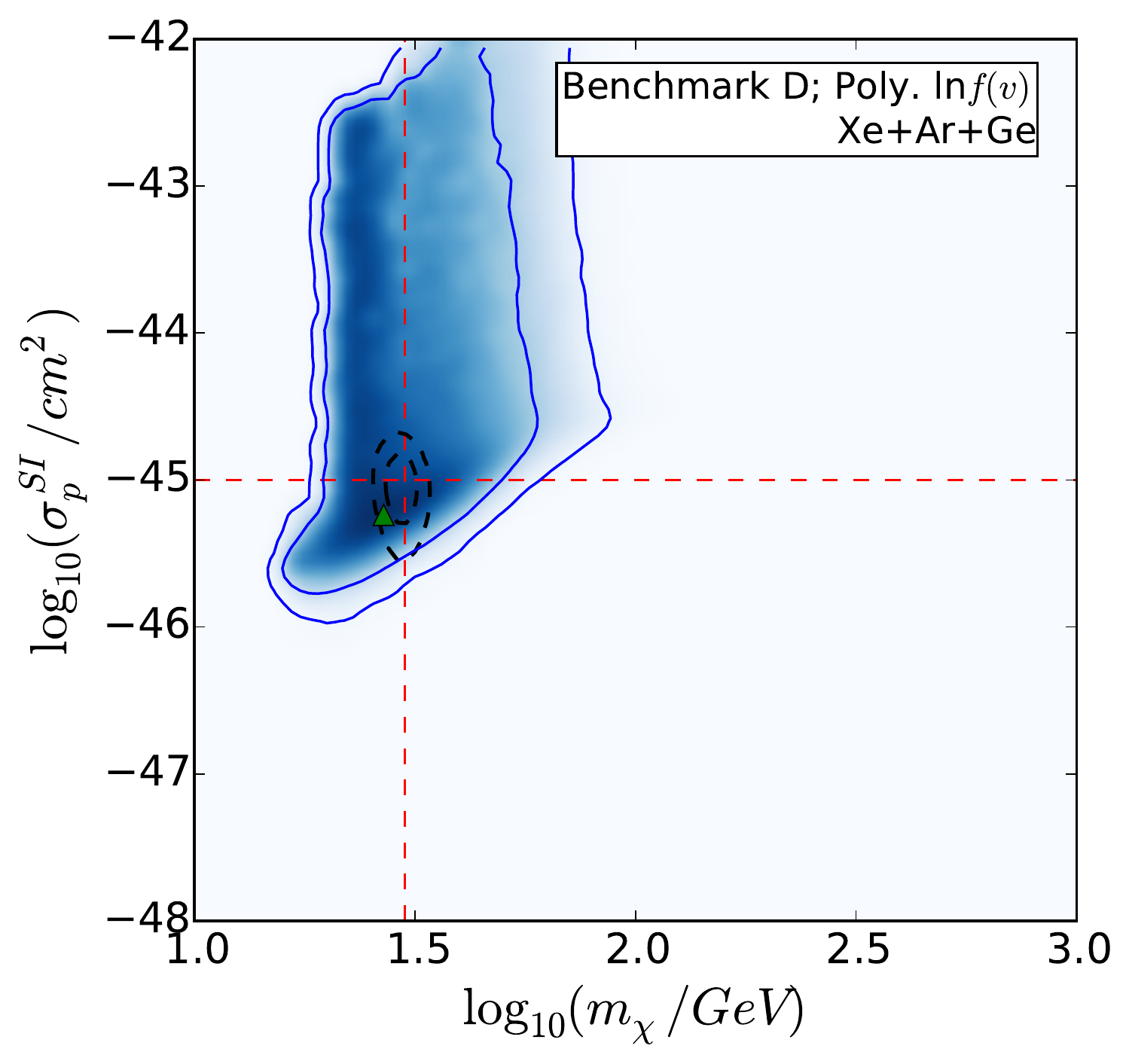}
\includegraphics[width=0.32\textwidth]{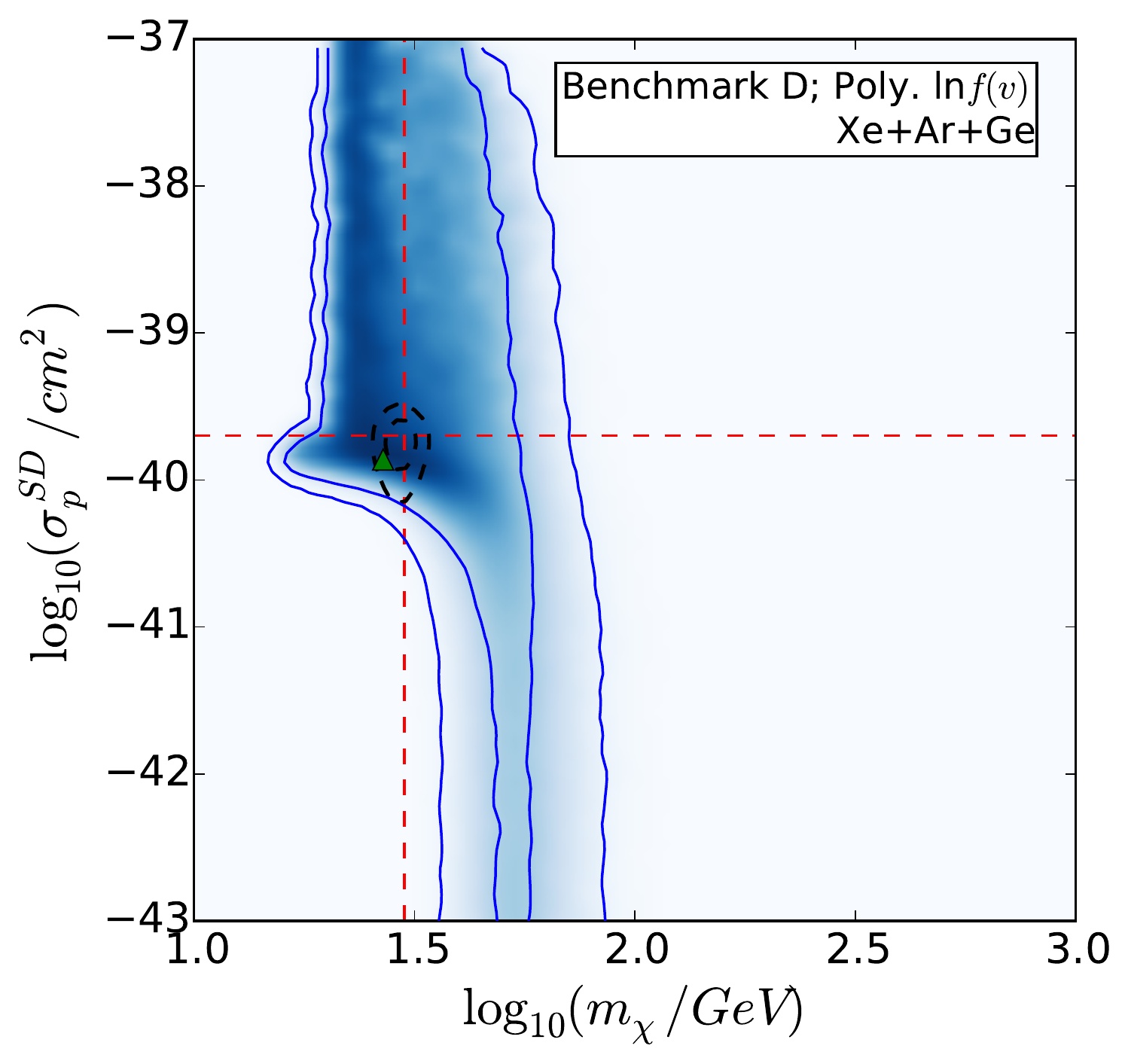}
\includegraphics[width=0.32\textwidth]{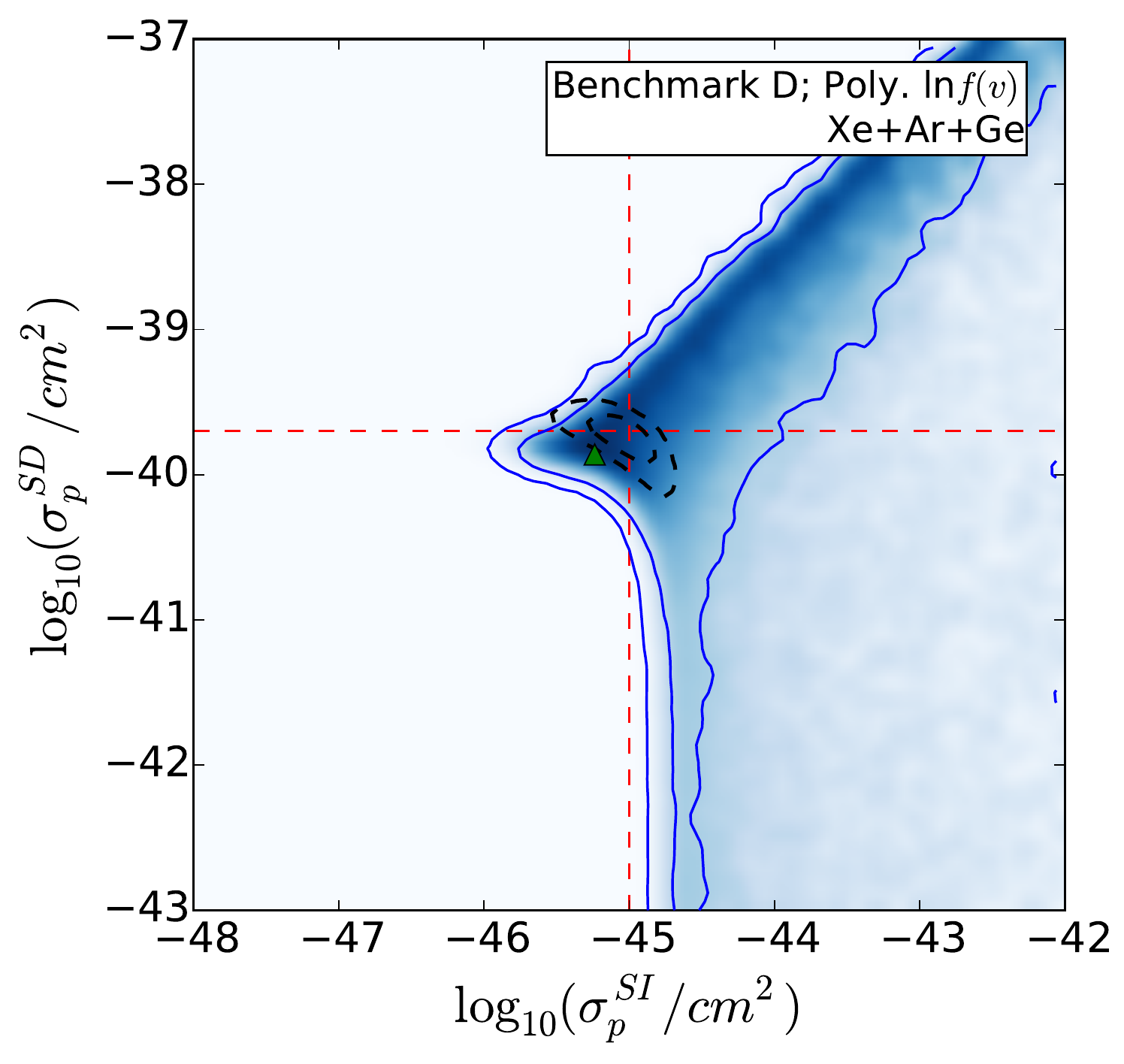}
\caption{Same as Fig. \ref{fig:DDonly_benchmarkA} but for benchmark D.}
\label{fig:DDonly_benchmarkD}
\end{figure*}

\subsection{Benchmark C}
Figure~\ref{fig:DDonly_benchmarkC} shows the results for benchmark C, for 
which the mass is reduced to 30 GeV, with cross sections of 
\sigmapsi= $10^{-45} \textrm{ cm}^2$ and 
\sigmapsd= $2 \times 10^{-40} \textrm{ cm}^2$ and a SHM $f(v)$. As for 
benchmarks A and B, using a fixed speed distribution (black dashed) leads to 
closed contours and tight constraints on the WIMP parameters and, with the 
binned parametrisation (top row), there again appears to be a slight bias 
towards lower WIMP masses, although the contours are not significantly 
widened. Indeed, for a binned $f(v)$, the reconstruction works quite well and 
all three quantities are determined with a good precision (approximately one 
order of magnitude for the cross sections and a factor of 2 for the WIMP mass). 
However, the results of the scan using the polynomial parametrisation (bottom 
row) are dramatically different. The 95\% confidence contours now extend up 
to $m_\chi \approx 100 \textrm{ GeV}$, owing to the wide range of functional 
forms which can be explored by this parametrisation. The degeneracy in the 
cross sections up to large values is even more pronounced than in the case of 
benchmark B. The lower input WIMP mass of benchmark C means that the region 
not covered by direct detection experiments extends up to $v\sim 200 \kms$, 
giving more freedom to the velocity integral to increase at low $v$.

For the polynomial parametrisation, the contours extend down to arbitrarily 
small values of $\sigmapsd$. As in the case of the higher mass benchmarks, 
explaining the data with only SI interactions requires a steeper velocity 
integral. For the low mass benchmarks, the fiducial spectrum is already 
relatively steep, requiring a velocity integral which is even steeper to give 
a good fit to the data at higher values of $m_\chi$. This is possible using 
the rapidly-varying polynomial parametrisation but not using the binned 
parametrisation, allowing the low \sigmapsd region to enter the confidence 
contours only in the former case.

\subsection{Benchmark D}
Finally, in Fig.~\ref{fig:DDonly_benchmarkD} we show the results for benchmark 
D. These are very similar to those for benchmark C. This is because the 
contribution of the dark disk is predominantly below $v\sim 200 \kms$, and 
for a 30 GeV WIMP this is beneath the lowest speed probed by the direct 
detection experiments considered. The main difference with respect to 
benchmark C is the fact that now there is a very clear bias in the WIMP mass 
for the binned speed distribution. The reconstruction prefers low values of 
\mwimp and the input parameter values are now outside the 95\% contours.

In this section, we have presented the results of parameter reconstructions 
for $(m_\chi, \sigmapsi, \sigmapsd)$ using data from multiple direct detection 
experiments only. As expected, when astrophysical uncertainties are neglected, 
the reconstruction of these parameters is very precise (apart from well known 
degeneracies between the WIMP mass and cross section). When $f(v)$ is allowed 
to vary in the fit, the confidence contours for the parameters are 
significantly widened. Using the 10-bin parametrisation, there is a clear 
bias in the WIMP mass for certain choices of benchmark parameters (in 
particular, benchmark D which has a light WIMP and a dark disk). This bias 
arises because bins of a fixed width in $v$ correspond to smaller bins in 
$E_R$ for smaller WIMP masses. This means that reducing the reconstructed 
WIMP mass allows a closer fit to the data, leading to the observed bias (see 
Ref.~\cite{Kavanagh:2012} for a detailed discussion).

The 6-polynomial parametrisation does not exhibit such a bias, but leads to 
even larger parameter uncertainties than for the binned parametrisation. Most 
notably, the confidence contours for the cross sections extend up to 
arbitrarily large values. Nonetheless, we obtain closed intervals for the 
WIMP mass when the input value is light (30 GeV, benchmarks C and D) relative 
to the mass of the detector nuclei, for both of the parameterisations of 
$f(v)$.

\section{Direct detection and IceCube data}
\label{sec:DDwithIC}
\makeatletter{}\begin{figure*}[th]
\centering
\includegraphics[width=0.32\textwidth]{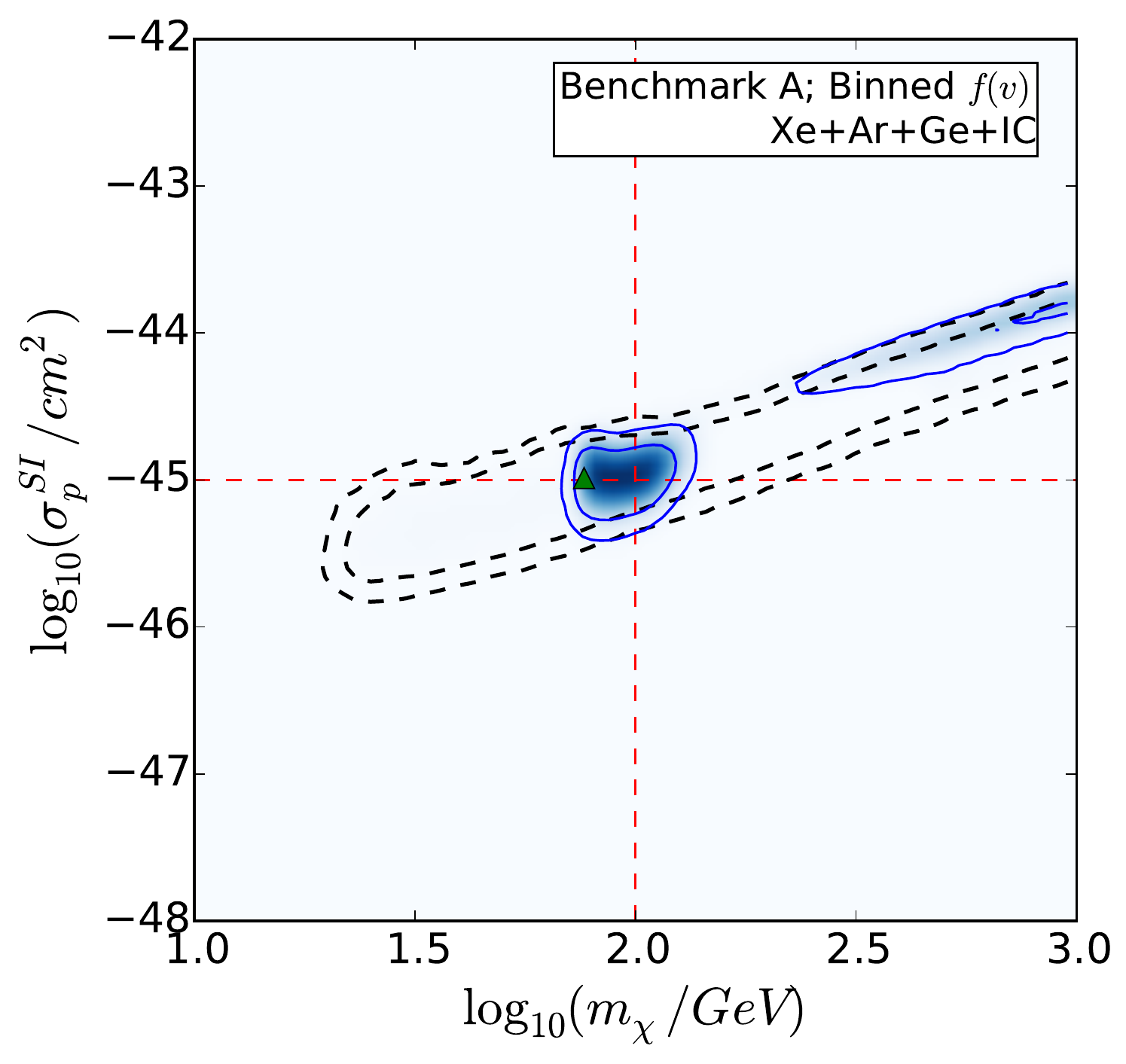}
\includegraphics[width=0.32\textwidth]{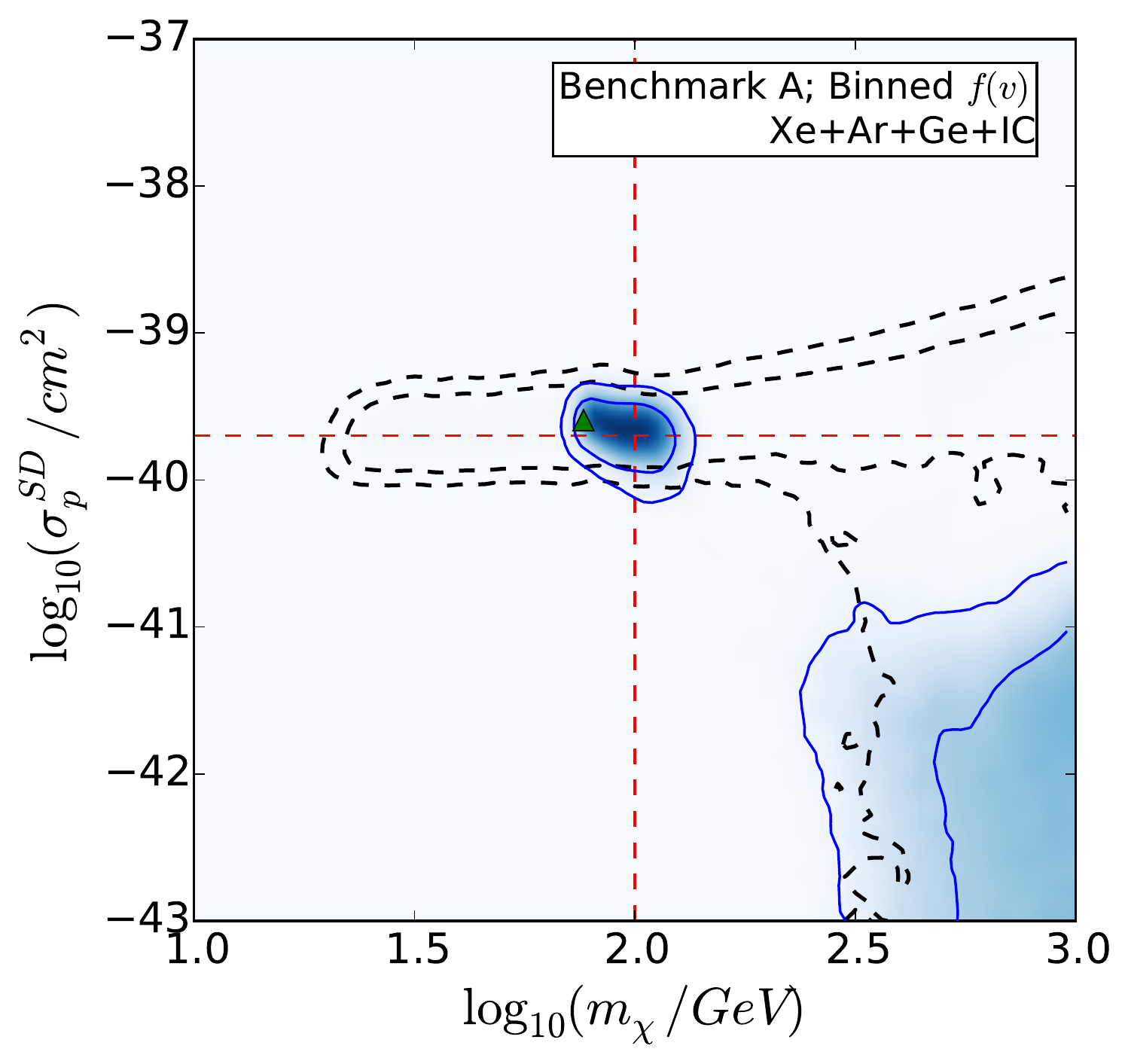}
\includegraphics[width=0.32\textwidth]{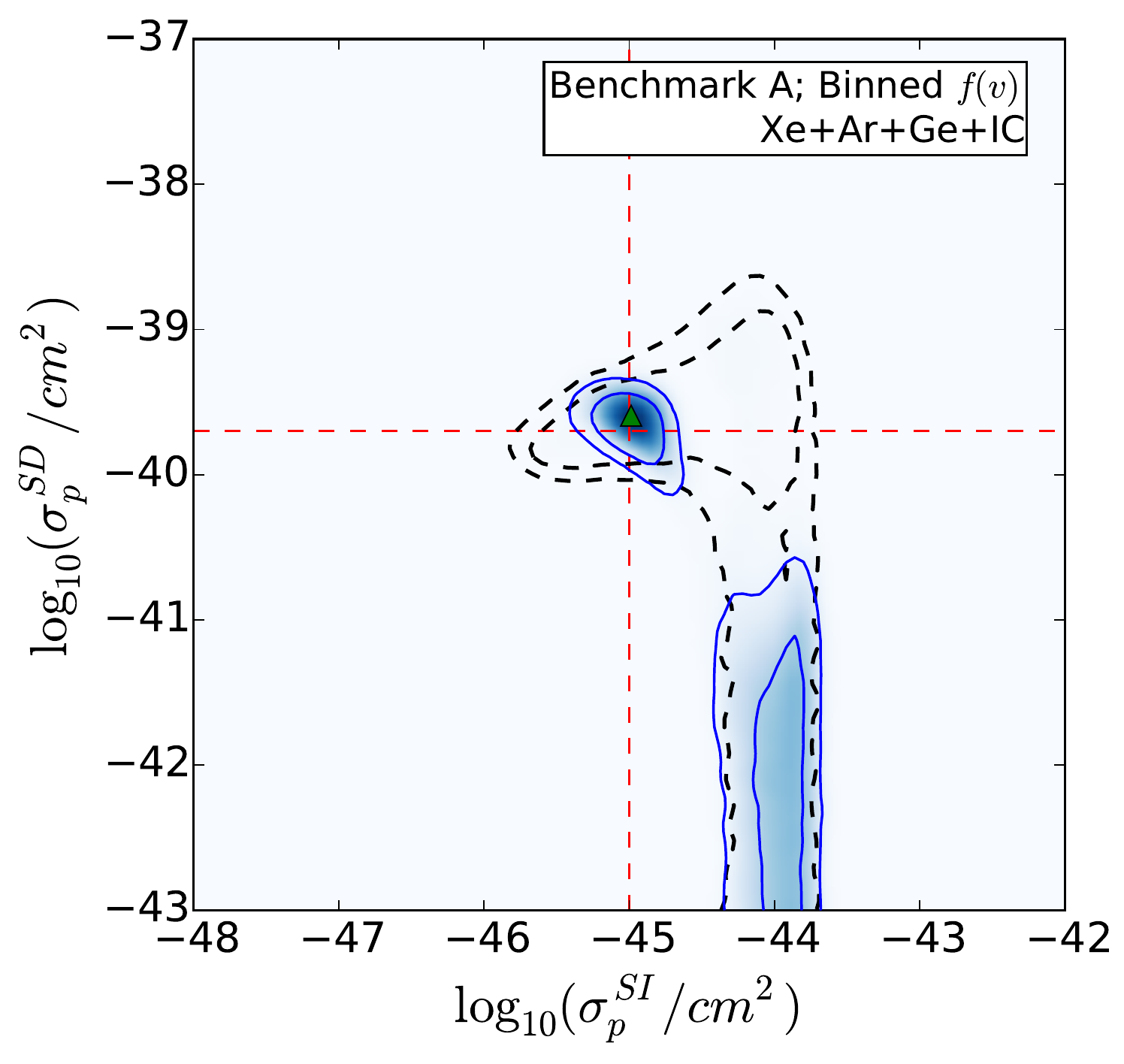}
\includegraphics[width=0.32\textwidth]{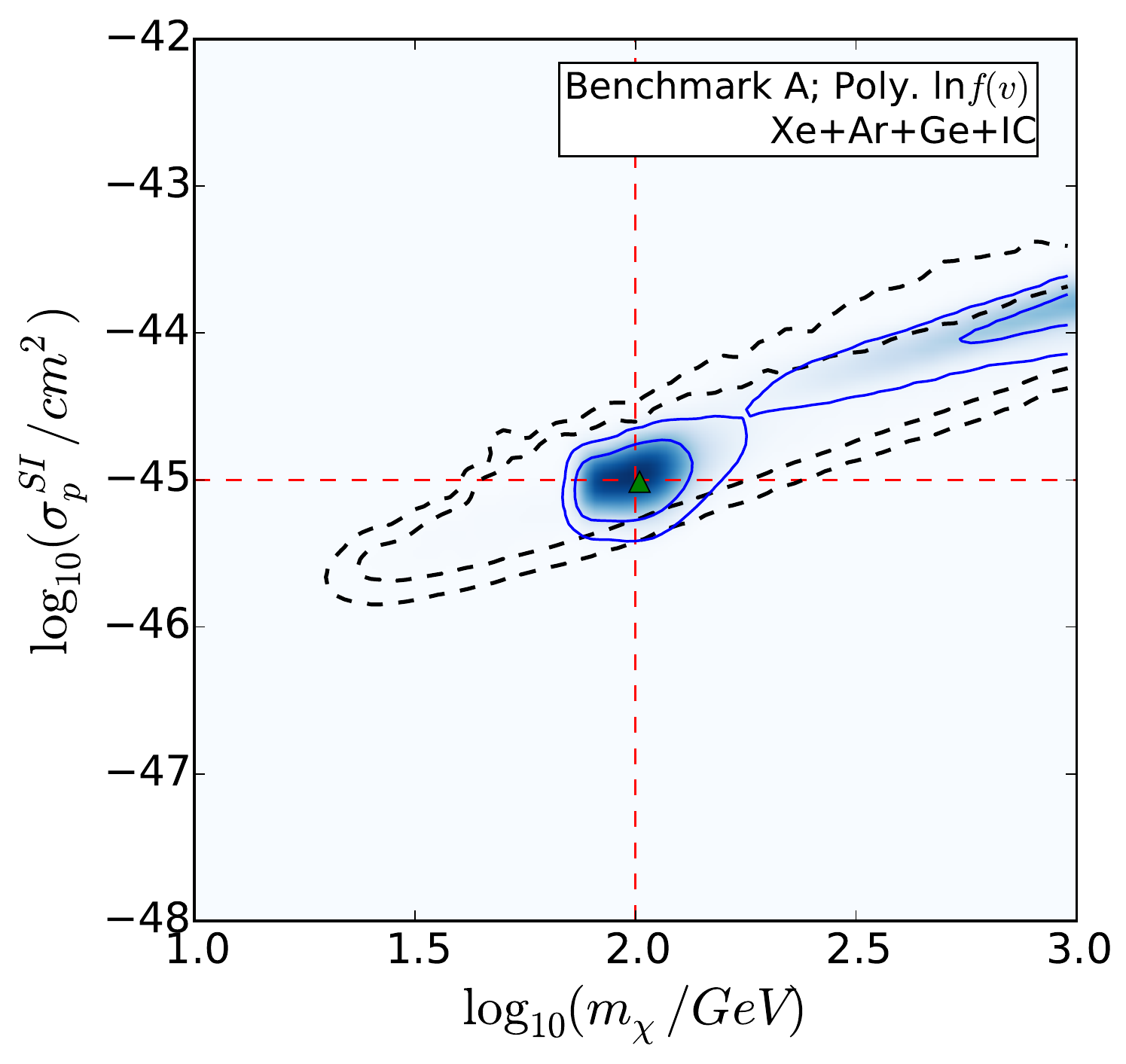}
\includegraphics[width=0.32\textwidth]{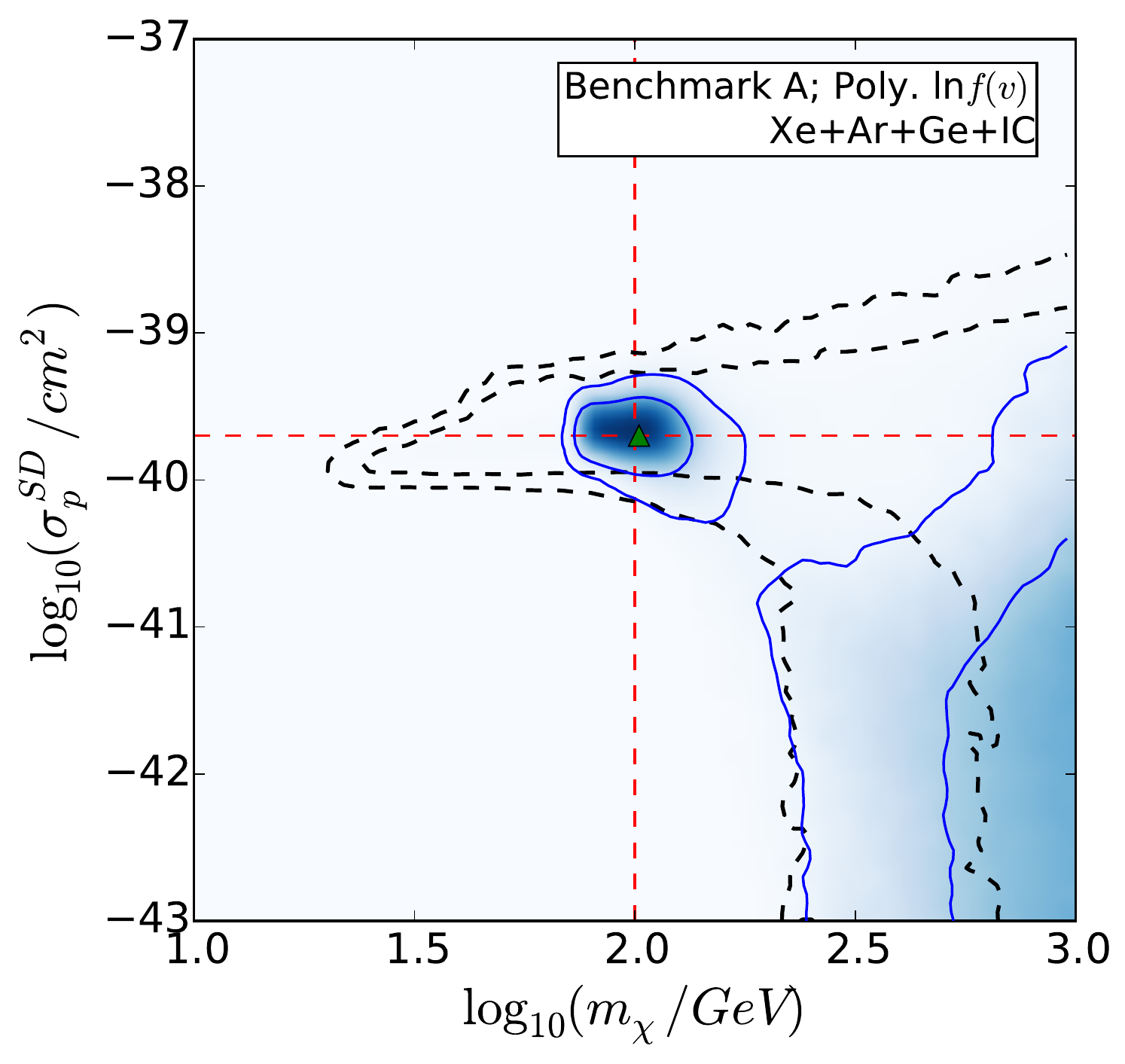}
\includegraphics[width=0.32\textwidth]{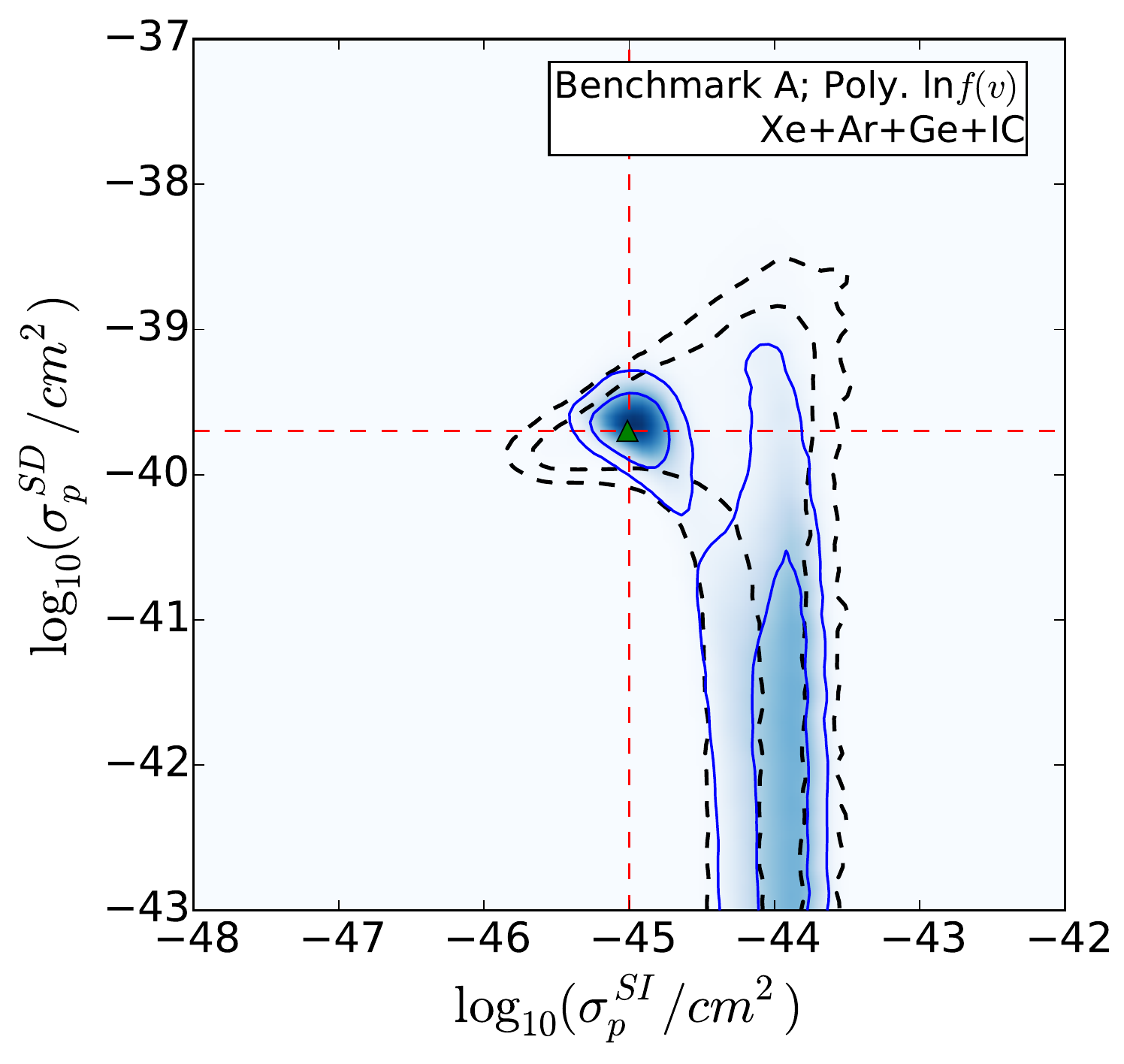}
\caption{Same as Fig.~\ref{fig:DDonly_benchmarkA} for benchmark A but using data from both direct detection and IceCube experiments. In this case, dashed black contours show the 68\% and 95\% confidence contours obtained using direct detection data only (i.e.~they correspond to the blue contours of Fig.~\ref{fig:DDonly_benchmarkA}).}
\label{fig:DDIC_benchmarkA}
\end{figure*}

In this section, we present the results of scans performed using both direct
detection and IceCube data. 

\subsection{Benchmark A}
Figure~\ref{fig:DDIC_benchmarkA} shows the 2-dimensional profile likelihood 
distributions in the case of benchmark A when IceCube data is included, in 
addition to the three direct detection experiments. The dashed black contours 
now correspond to the 68\% and 95\% confidence contours obtained using direct 
detection data only for comparison. The blue contours are with IceCube data 
and are considerably smaller than the dashed black ones.

When astrophysical uncertainties are included, the profile likelihood is 
multimodal. There is a small region of allowed parameter space around the 
input parameter values, and a second region at large masses, large SI and 
negligible SD cross sections. This is true for both parametrisations, with the 
only slight difference being that, for the polynomial parametrisation, the two 
regions are almost connected at the 95\% confidence level. This is because the 
polynomial parametrisation can explore a wider range of shapes for the speed 
distribution, allowing the data to be fit reasonably well with a wider range 
of WIMP masses.

The strong degeneracy in the WIMP mass, which occurs when only direct 
detection data is used, has been substantially reduced with the inclusion of 
IceCube data. Low mass WIMPs are no longer viable as they cannot produce the 
observed number of events above the threshold of IceCube. As discussed in 
Sec.~\ref{sec:DDonly}, at large masses two scenarios were possible with direct 
detection data only: a region at low \sigmapsd, where the observed events were 
explained in terms of SI interactions only, and a mixed SI/SD scenario (top 
right corner of all the panels of Fig.~\ref{fig:DDonly_benchmarkA}), with 
velocity integrals slightly steeper than the input one of the SHM. Including 
the information from IceCube eliminates this second possibility, as it 
produces too many neutrinos in IceCube. The number of neutrinos produced in 
IceCube could be reduced with a $f(v)$ which goes rapidly to zero below 
$\sim 200 \, {\rm km \, s}^{-1}$. An example of such a distribution is shown in 
Fig.~\ref{fig:SpeedExamples}, labeled `iv'. However, the shape of the 
resulting velocity integral cannot be reconciled with the spectrum of direct 
detection events, especially in the xenon experiment.

With small $\sigmapsd$ and large $m_{\chi}$, good fits to the direct detection 
data are obtained by making the velocity integral even steeper and, since 
$\sigmapsd$ is small, the expected number of neutrinos is compatible with 
the number observed in IceCube. Therefore the region of parameter space at 
large WIMP masses and small $\sigmapsd$ is still allowed when IceCube data 
is added.

We can again compare to the work of Arina et al.~ (in particular, the middle row of Fig.~3 of Ref.~\cite{Arina:2013}). Our accurate reconstruction of the WIMP mass matches that found in Ref.~\cite{Arina:2013} when direct detection and IceCube are combined. In contrast, we obtain significantly stronger constraints on the SI cross section. As previously stated, this is due to the fact that the present analysis uses an ensemble of different direct detection experiments. However, we note that here we have fully accounted for general uncertainties in the speed distribution and yet we can still obtain constraints similar to those of Arina et al.

\begin{figure*}[t]
\centering
\includegraphics[width=0.32\textwidth]{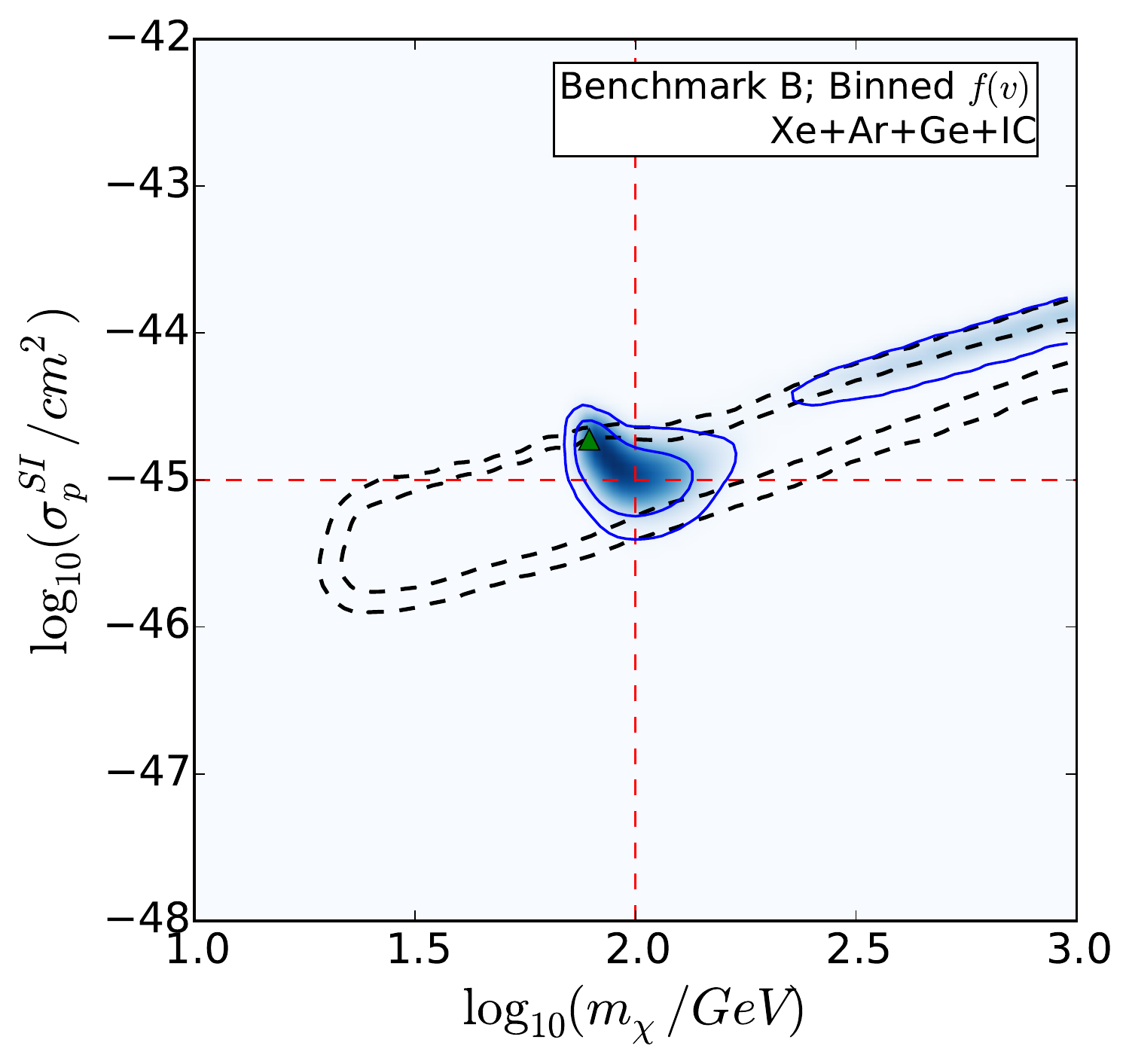}
\includegraphics[width=0.32\textwidth]{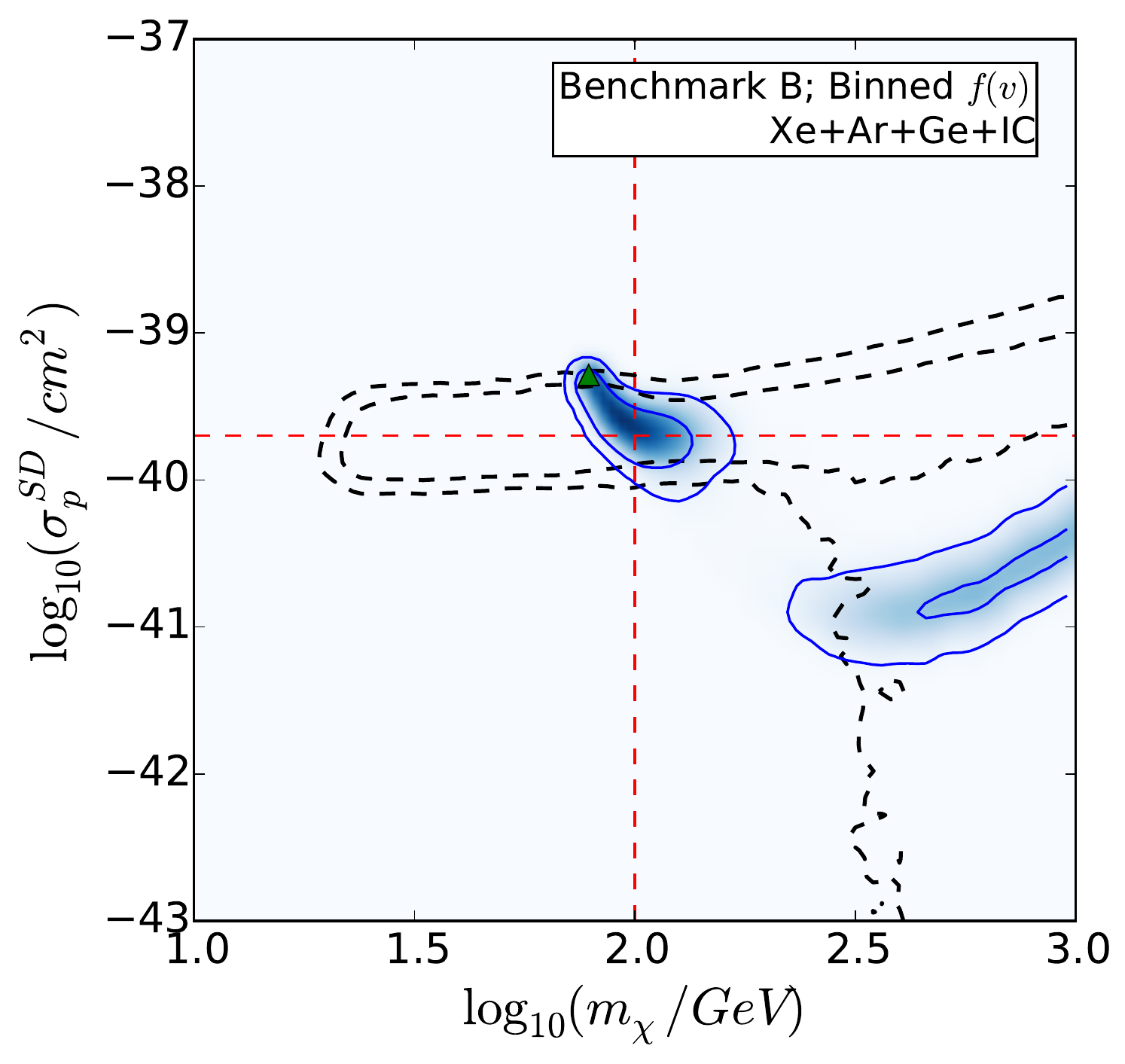}
\includegraphics[width=0.32\textwidth]{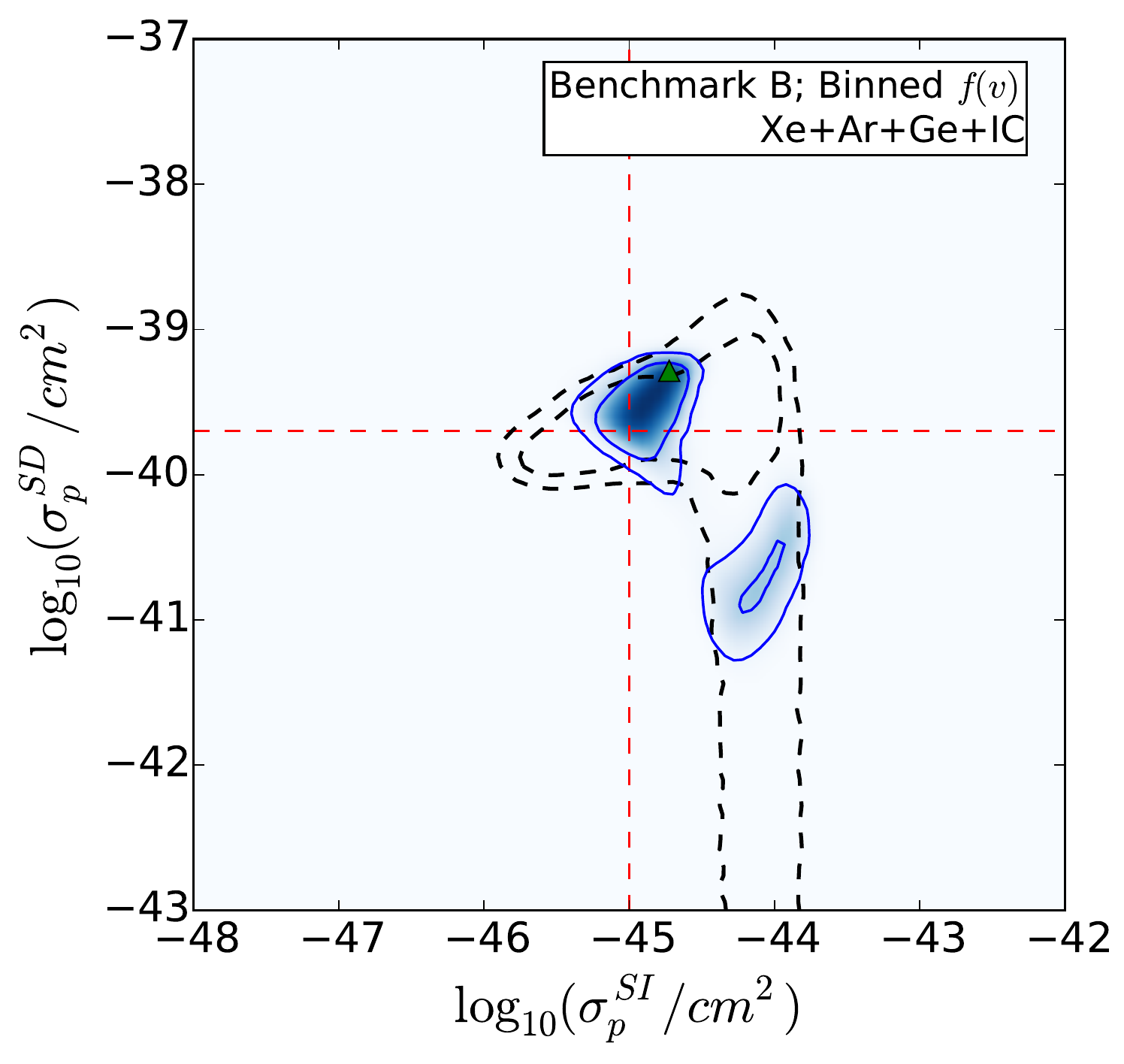}
\includegraphics[width=0.32\textwidth]{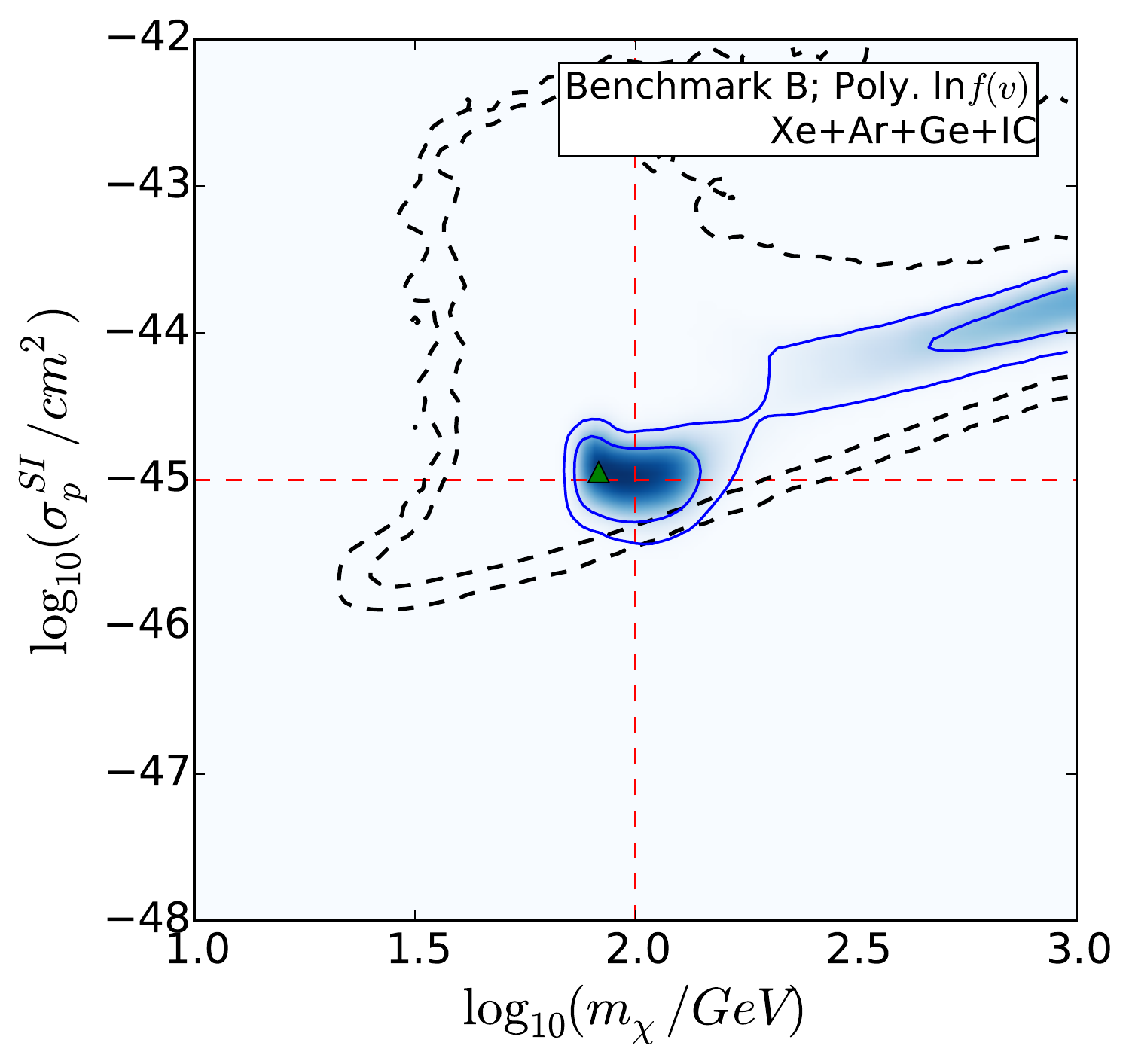}
\includegraphics[width=0.32\textwidth]{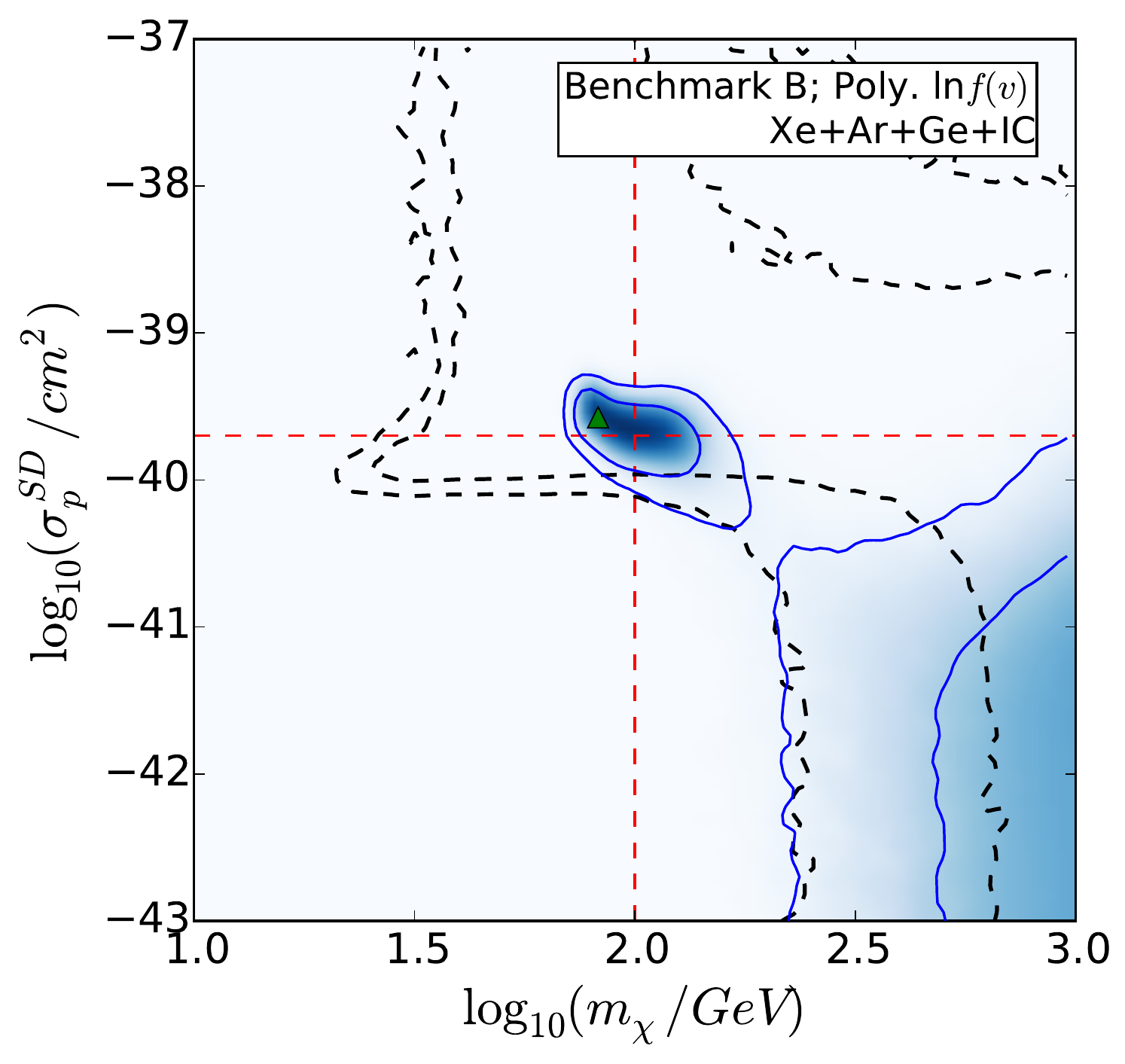}
\includegraphics[width=0.32\textwidth]{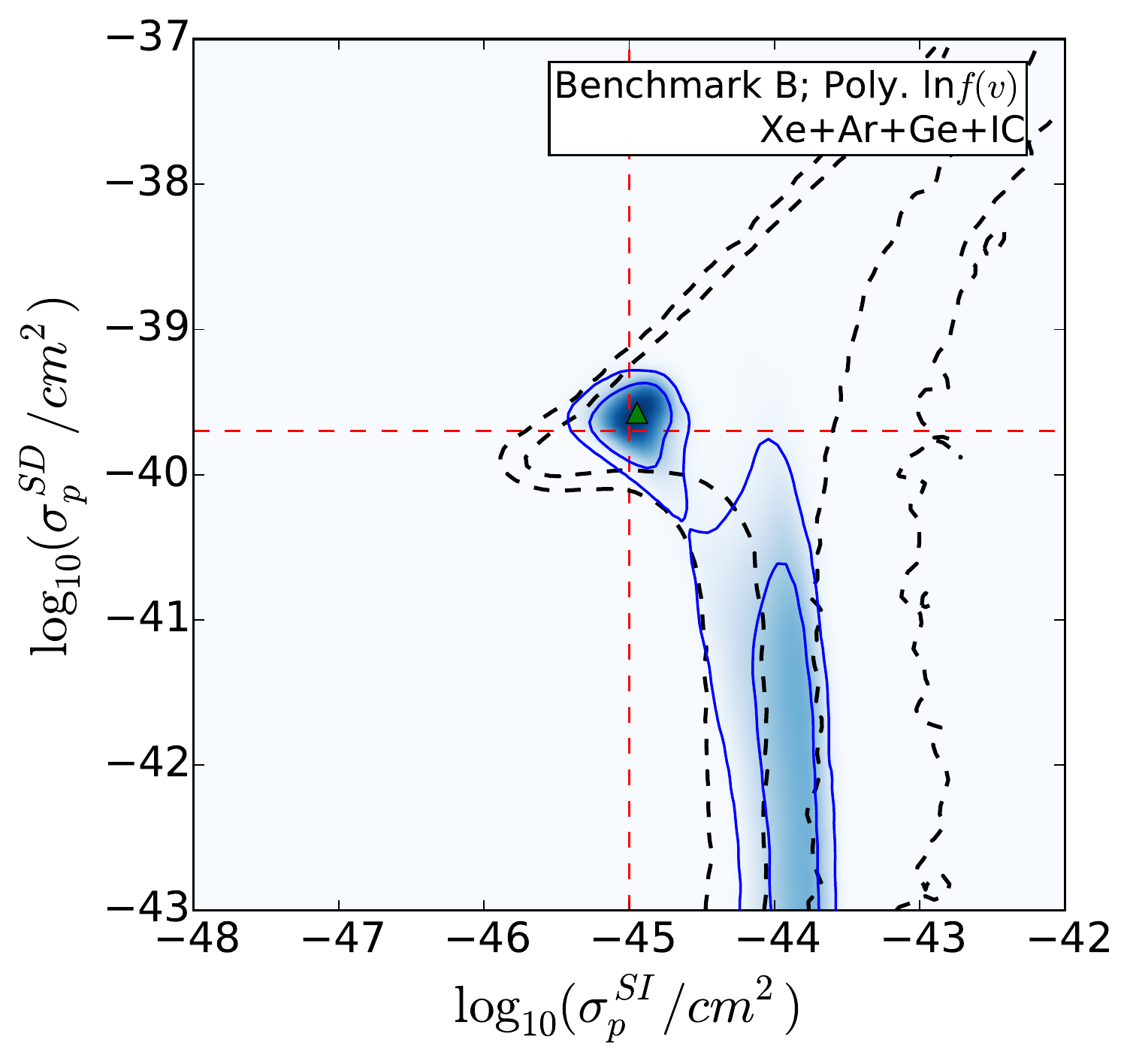}
\caption{Same as Fig.~\ref{fig:DDIC_benchmarkA} but for benchmark B.}
\label{fig:DDIC_benchmarkB}
\end{figure*}

\subsection{Benchmark B}
Figure~\ref{fig:DDIC_benchmarkB} contains the same plots as 
Fig.~\ref{fig:DDIC_benchmarkA}, but for benchmark B, which has an additional 
contribution to the speed distribution from a dark disk. For both 
parametrisations, the results are similar to those for benchmark A. The only 
notable difference is that for the binned distribution (top row), the allowed 
region of parameter space at large masses is now bounded from below in 
$\sigmapsd$. In this region all the events in the direct detection experiments 
are due to SI interactions (and the velocity integral is quite steep) and the
SD cross section is small enough not to overproduce neutrinos in IceCube. 
Decreasing $\sigmapsd$ further has no effect on the direct detection 
experiments, while underproducing the signal in IceCube. With the polynomial 
parameterisation it is possible to compensate for this underproduction of 
neutrinos by increasing the speed distribution below 50 -- 80 
${\rm km \, s}^{-1}$, where it has no effect on the direct detection 
experiments. However this is not possible with the binned parameterisation, 
since the first bin extends up to 100 ${\rm km \, s}^{-1}$ and any change 
would also affect the number of events in xenon. This lower bound on the SD 
cross section for the binned parameterisation was not present for benchmark A 
as that benchmark only predicts 43.3 neutrinos (compared to the 242.9 of 
benchmark B). The smaller number of expected neutrinos means that benchmark A 
is less sensitive to changes in the number of neutrinos, and regions of 
parameter space with few neutrinos are still allowed.

Comparing the results with and without IceCube data (solid blue and dashed 
black contours respectively) using the polynomial parametrisation for 
benchmark B, we see that the regions of parameter space which extended up to 
large values of the cross sections are eliminated when IceCube data is 
included. This is most clear in the bottom right panel of 
Fig.~\ref{fig:DDIC_benchmarkB}, in which the contours no longer extend into 
the upper right hand corner. With direct detection data only, this region 
was allowed due to the possibility of having steeply rising velocity 
integrals for speeds which direct detection experiments could not probe. 
However the IceCube event rate is sensitive to these low speeds and such 
distributions would produce too many neutrino events.

\begin{figure*}[htp!]
\centering
\includegraphics[width=0.31\textwidth]{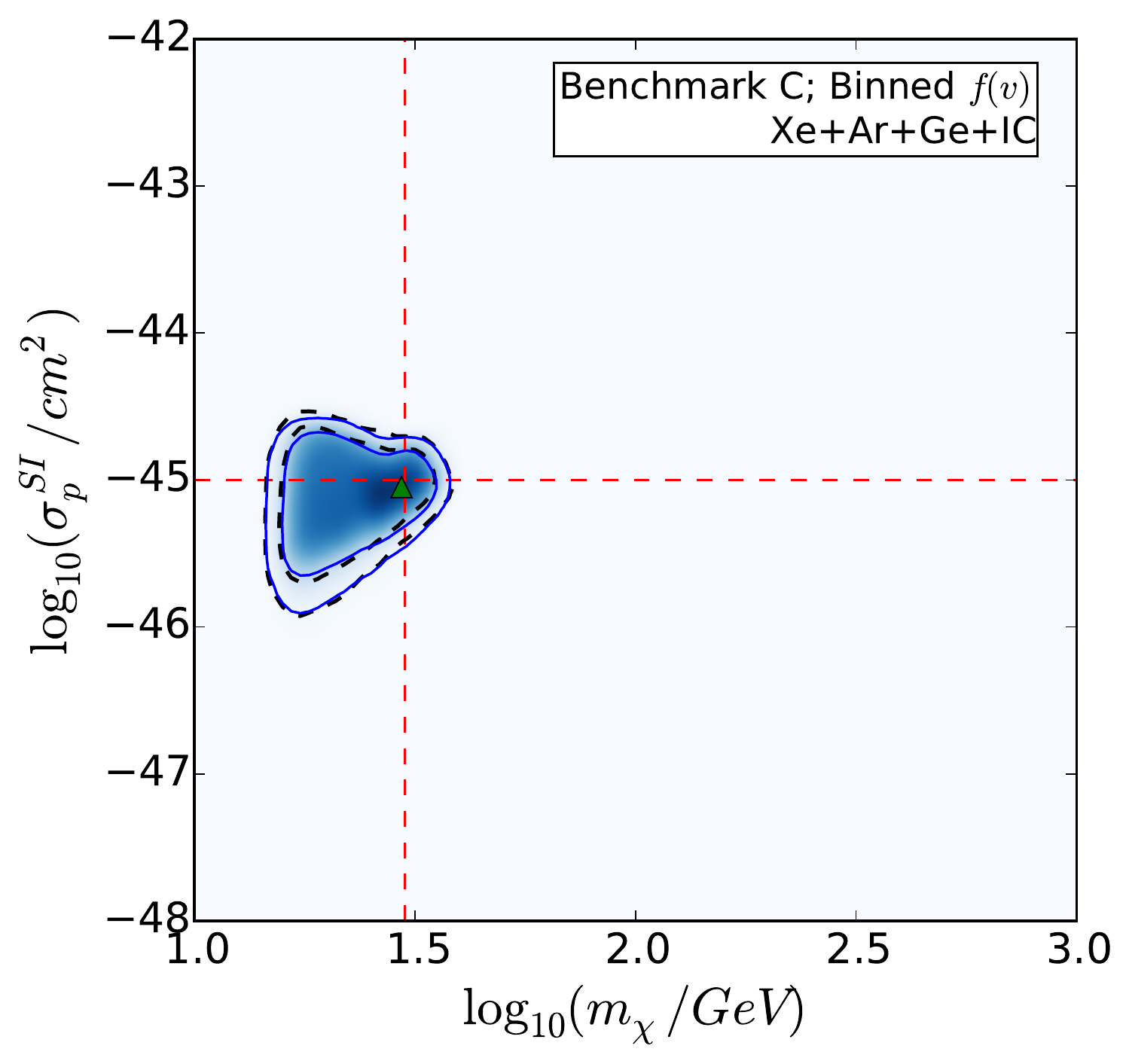}
\includegraphics[width=0.31\textwidth]{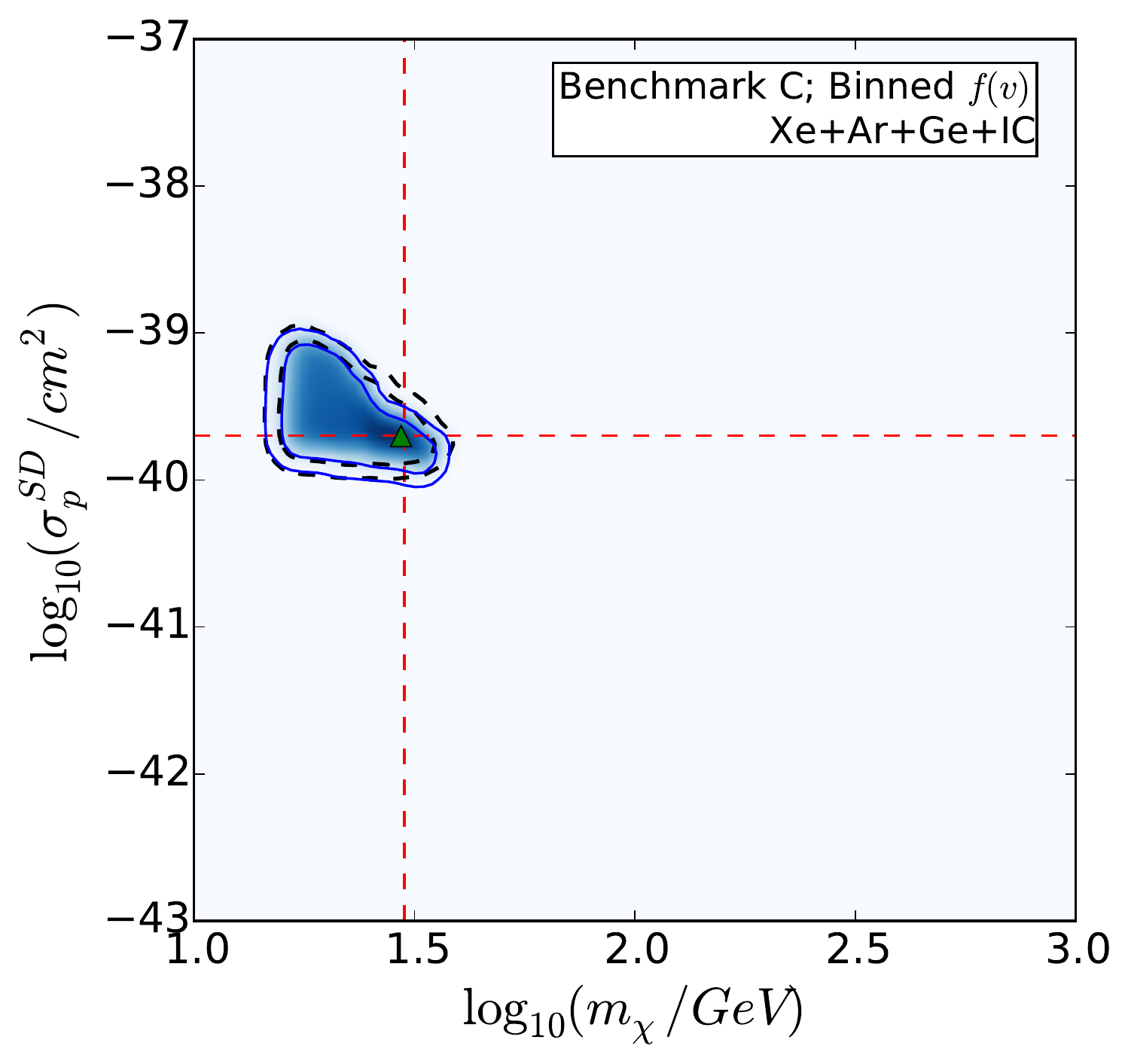}
\includegraphics[width=0.31\textwidth]{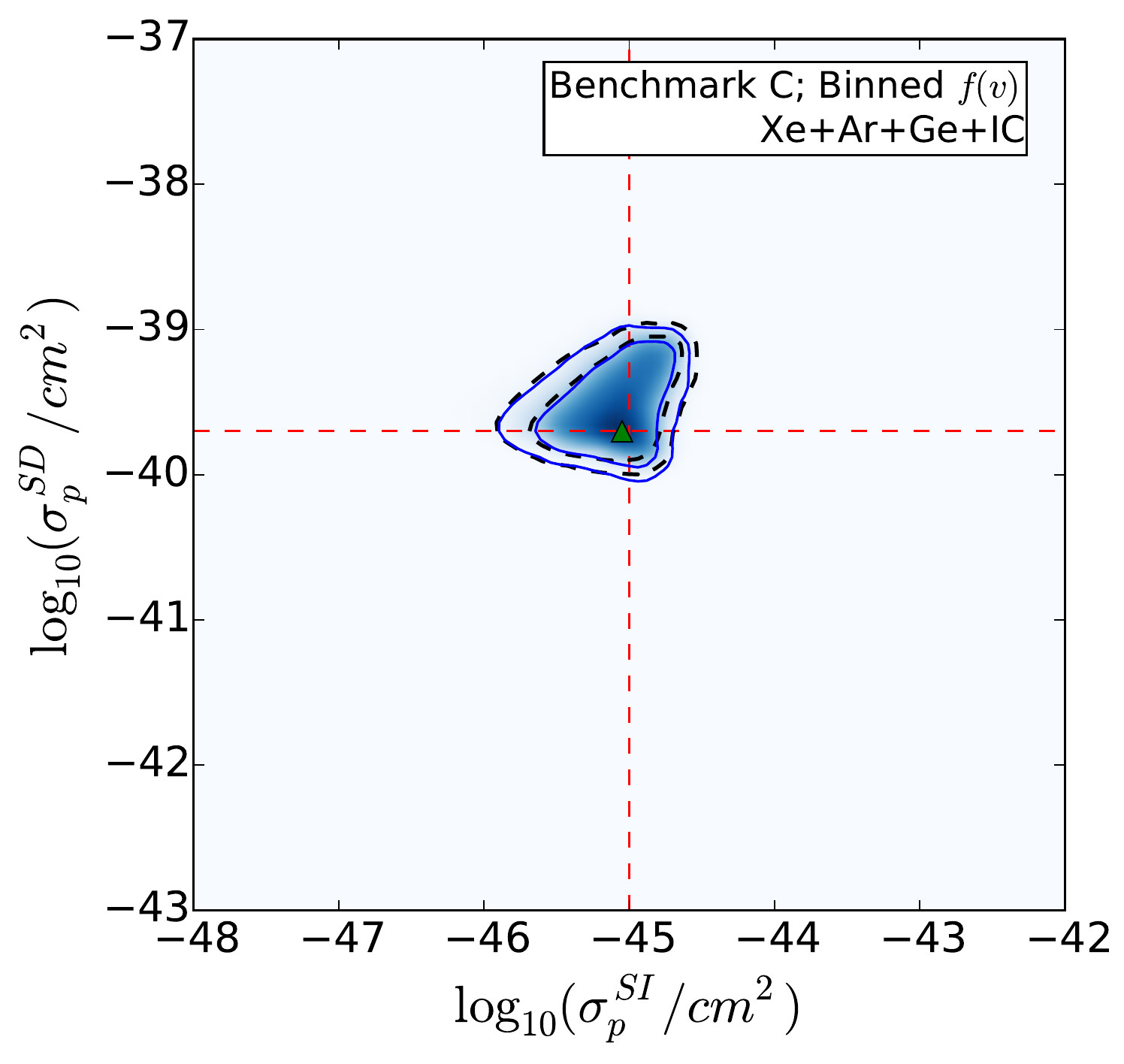}
\includegraphics[width=0.31\textwidth]{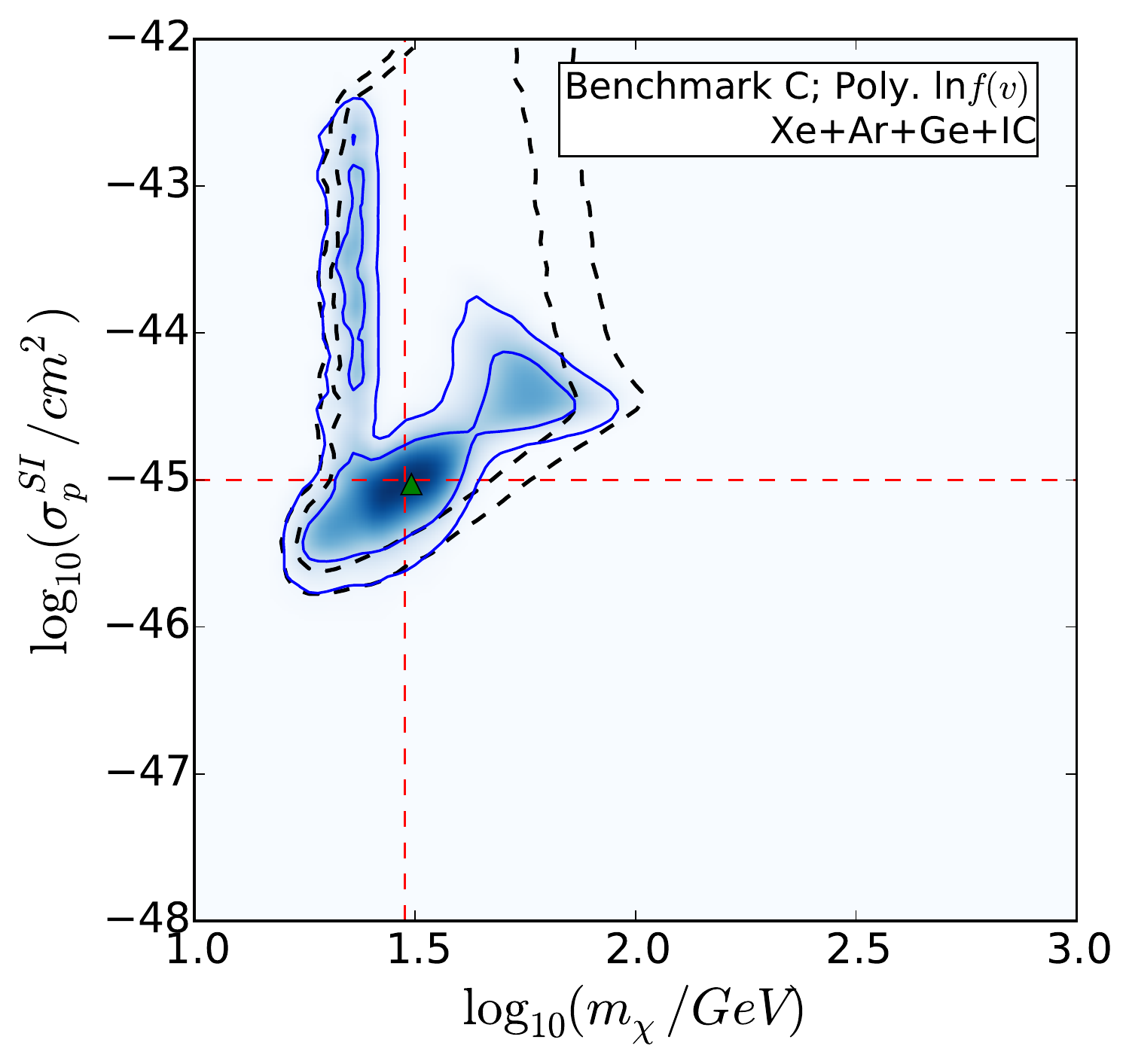}
\includegraphics[width=0.31\textwidth]{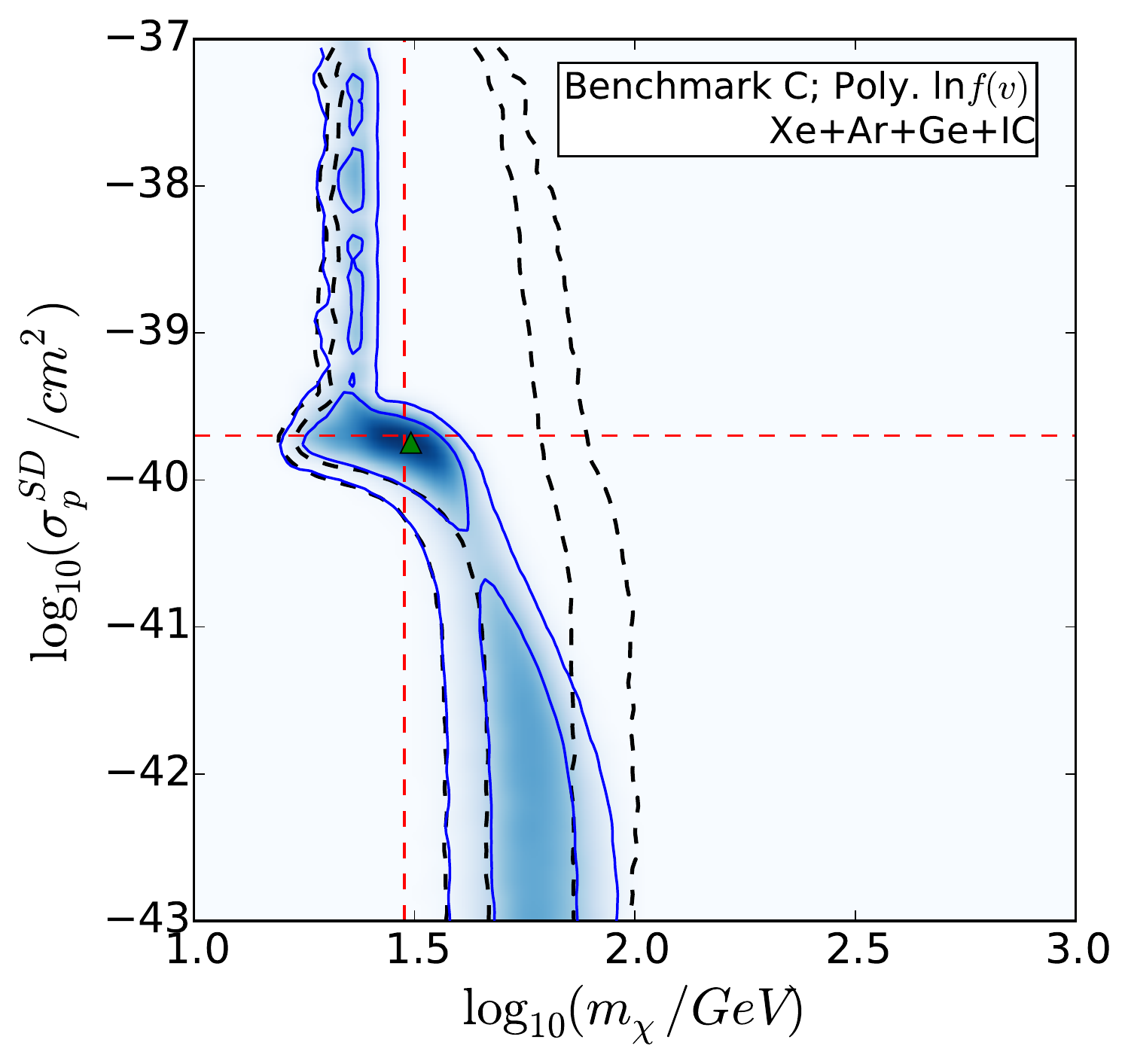}
\includegraphics[width=0.31\textwidth]{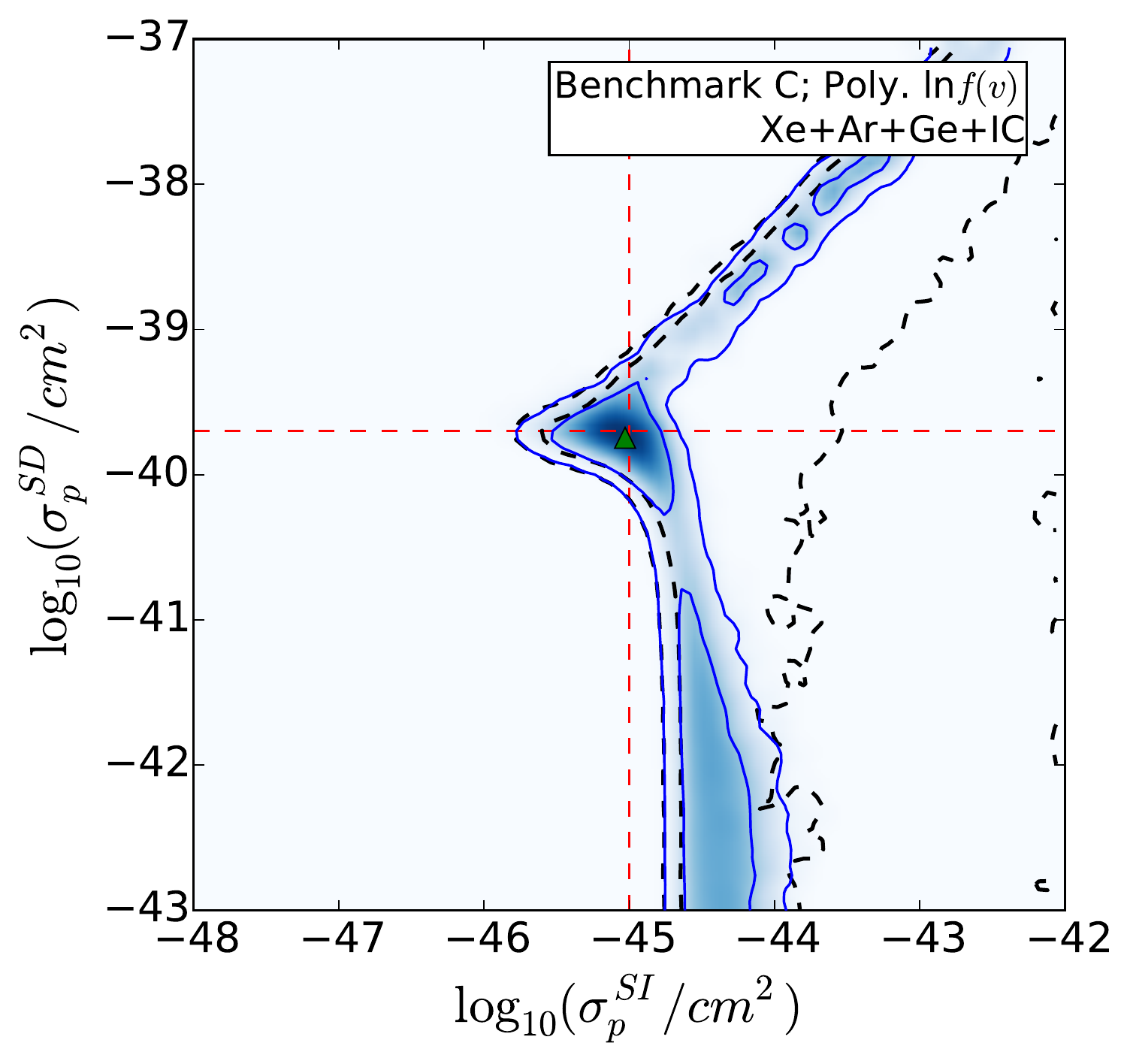}
\caption{Same as Fig.~\ref{fig:DDIC_benchmarkA} but for benchmark C.}
\label{fig:DDIC_benchmarkC}
\end{figure*}

\begin{figure*}[hbp!]
\centering
\includegraphics[width=0.31\textwidth]{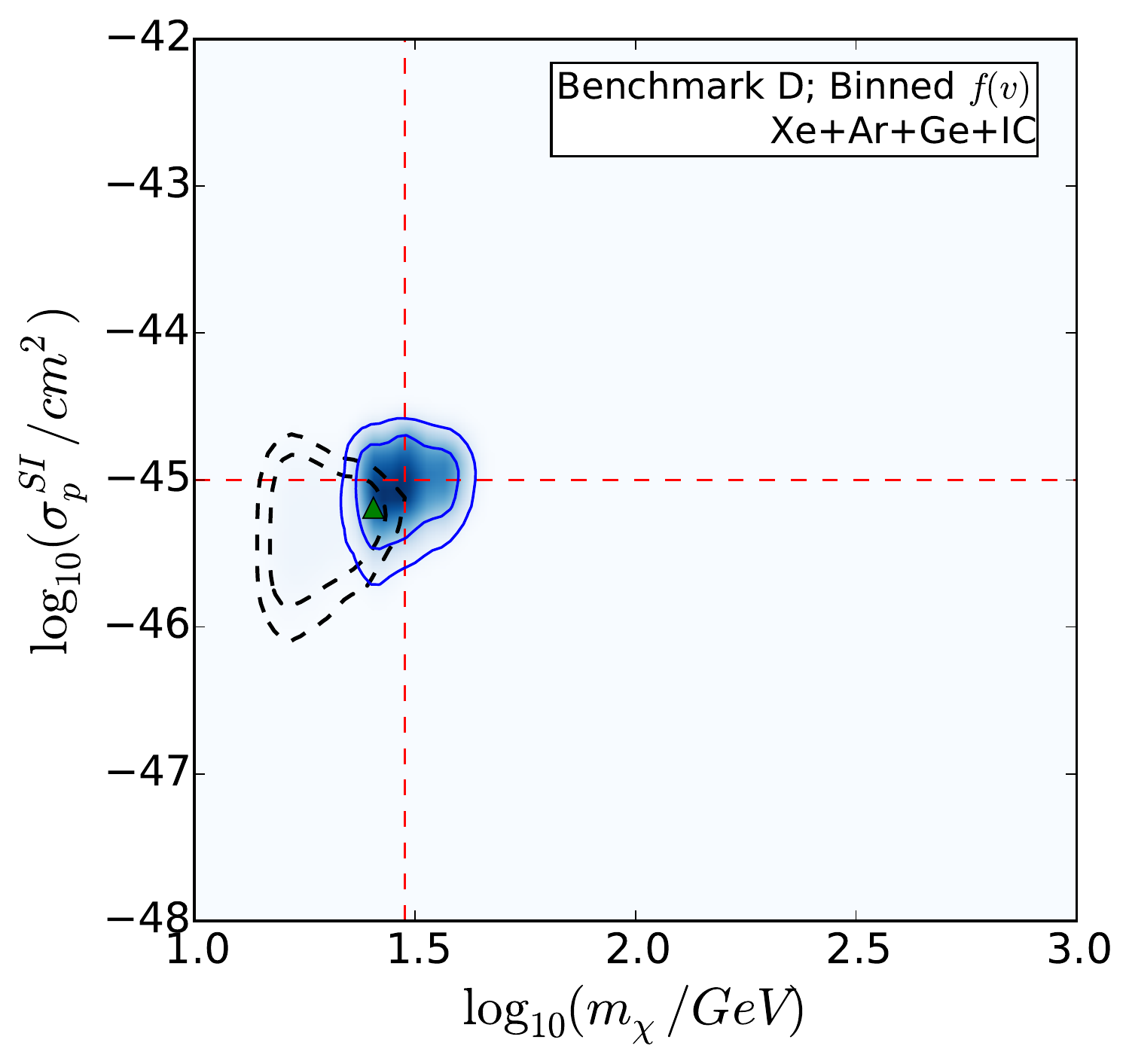}
\includegraphics[width=0.31\textwidth]{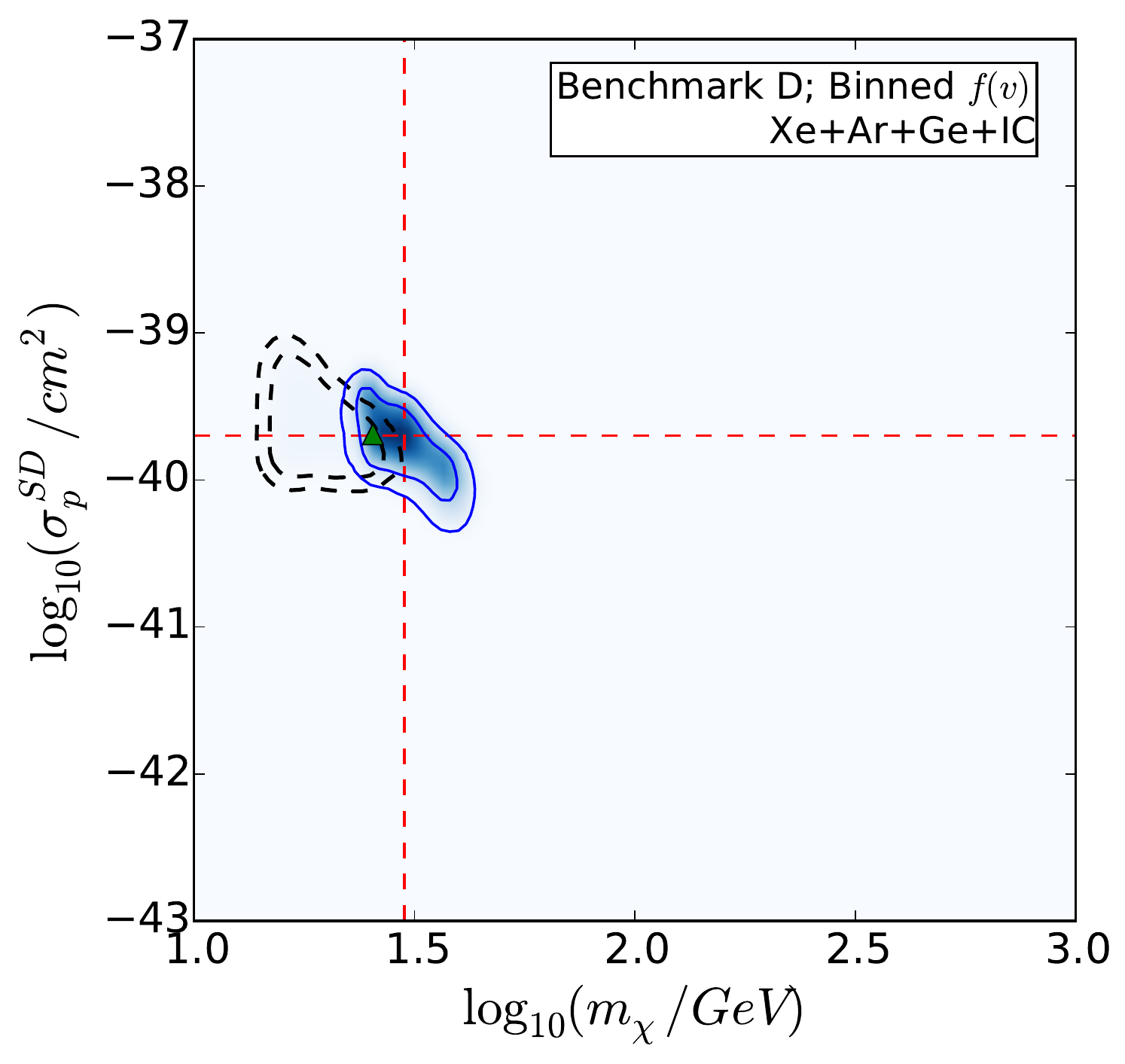}
\includegraphics[width=0.31\textwidth]{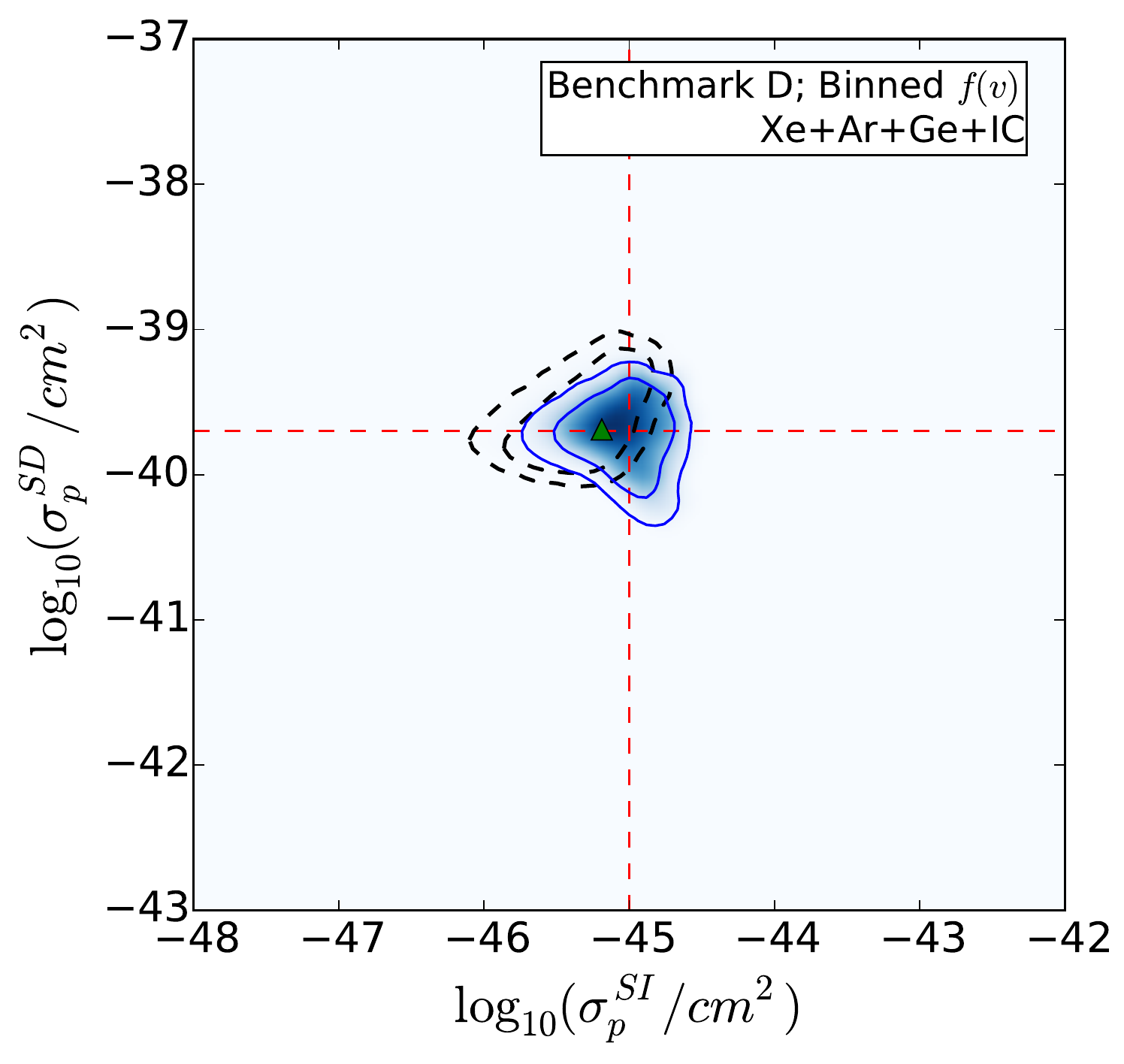}
\includegraphics[width=0.31\textwidth]{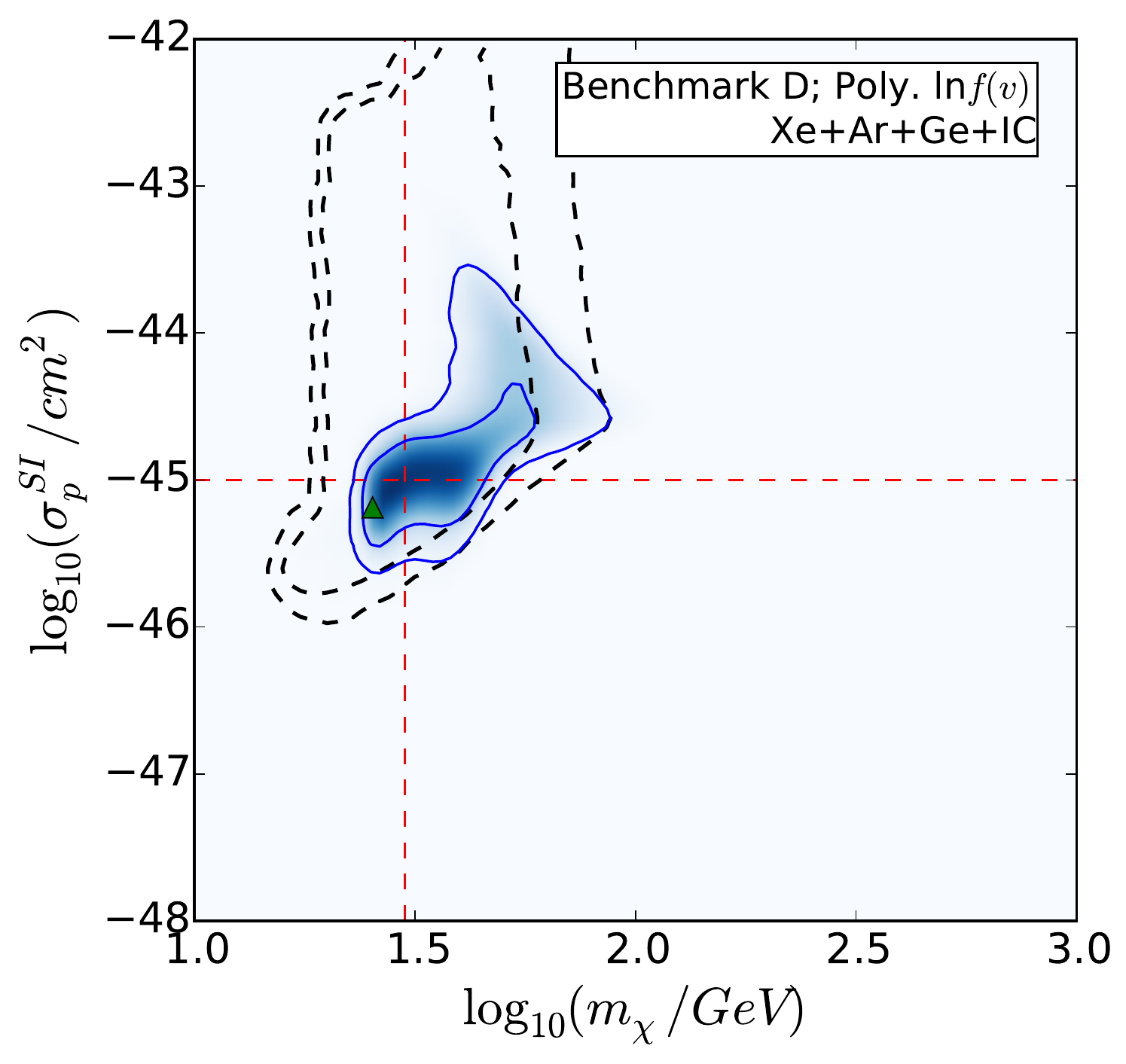}
\includegraphics[width=0.31\textwidth]{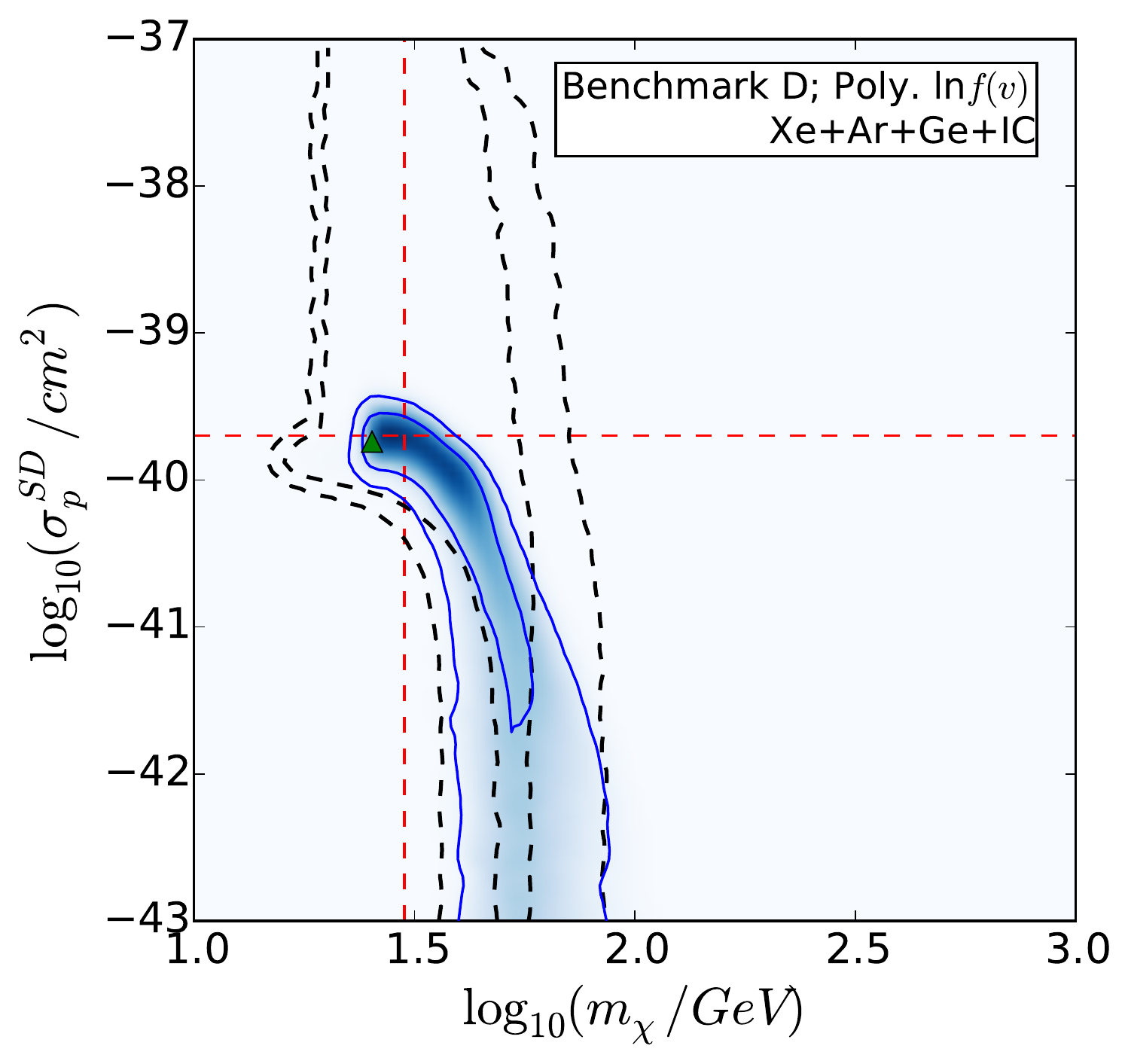}
\includegraphics[width=0.31\textwidth]{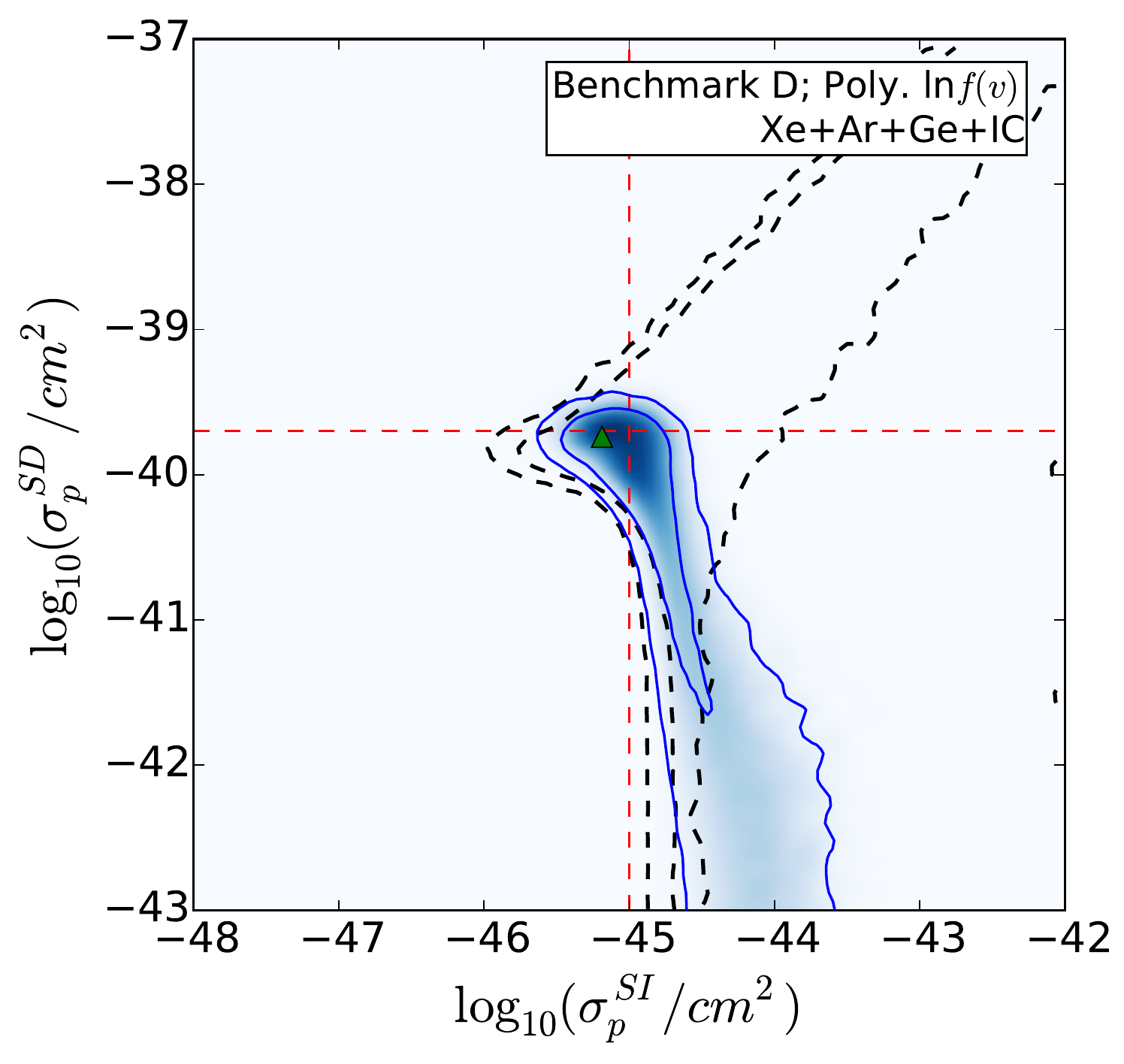}
\caption{Same as Fig.~\ref{fig:DDIC_benchmarkA} but for benchmark D.}
\label{fig:DDIC_benchmarkD}
\end{figure*}

\subsection{Benchmark C}
Figure~\ref{fig:DDIC_benchmarkC} shows the same plots as 
Fig.~\ref{fig:DDIC_benchmarkA}, but for benchmark C, which has a smaller input 
WIMP mass. Using the binned parametrisation (top row), the confidence 
contours are not significantly changed from the case with direct detection 
only. This is because the number of signal events in IceCube is just 13, 
which is consistent with the observed background at just over $1\sigma$. As 
previously discussed, the binned distribution does not probe distribution 
functions which rise rapidly at low $v$. Because these are the only 
distribution functions which would be excluded by the small number of signal 
events, the addition of the IceCube data therefore has little impact on 
parameter reconstruction.

Using the polynomial parametrisation, the region at large \mwimp where both 
SI and SD interactions are significant is now excluded. Similarly, a part of 
the parameter space at high cross section is now also excluded. As for the 
higher mass benchmarks, distributions which rise rapidly at low $v$ will 
overproduce events in IceCube and are therefore excluded. However, in contrast 
to the higher mass benchmarks, there remains a region which extends up to 
large cross sections at the 95\% level. This region only occurs for WIMP 
masses below the sensitivity threshold of IceCube, $m_\chi =  20 \textrm{ GeV}$. 
Such low-mass WIMPs produce no signal events in IceCube and therefore any 
form of $f(v)$ can fit the IceCube data set (which is consistent with 
background at the $1\sigma$ level).

\subsection{Benchmark D}
Figure~\ref{fig:DDIC_benchmarkD} show the same plots as 
Fig.~\ref{fig:DDIC_benchmarkC}, but for benchmark D, which has a contribution 
from a dark disk. This results in roughly three times more IceCube events than 
in benchmark C, which significantly improves parameter reconstruction. For 
the binned parametrisation, there is a noticeable shift in the contours to 
higher masses, so that they are more centered on the input mass value. In 
addition, the best-fit point lies closer to the input parameter values than 
in the case with only direct detection. WIMP masses below 20 GeV are excluded 
by IceCube data as they produce no signal events. Additionally, going to lower 
WIMP masses reduces the size of the bins in $E_R$ (which was the source of 
the previous bias), but requires a flatter form of $f(v)$ to fit the direct 
detection data. This flatter distribution results in a smaller solar capture 
rate and is therefore excluded as it produces too few IceCube events. The 
result is that the bias towards lower WIMP masses has been eliminated by the 
addition of IceCube data and the benchmark values now lie within the 68\% 
contours.

Using the polynomial parametrisation, the high cross section regions of the 
parameter space are now entirely excluded, as the signal has a greater 
statistical significance than for benchmark C and cannot be explained by 
background alone. The shape of $f(v)$ is now reconstructed so as to neither 
overproduce or underproduce IceCube events. The 95\% contours still extend 
down to small values of $\sigmapsd$. This is, as before, because the 
polynomial parametrisation encompasses steep velocity integrals, meaning that 
both direct detection and IceCube data can be explained with relatively small 
values $\sigmapsd$. This is not the case for the binned parameterisation. 

In this Section we have found that for model-independent parametrisations of 
the WIMP speed distribution, using IceCube data in addition to direct 
detection data significantly reduces the size of the allowed particle physics 
parameter space. This extends the previous results of Ref.~\cite{Arina:2013} 
which used a fixed functional form for the speed distribution.

We find that with the addition of IceCube data, the bias in the reconstruction 
of the WIMP mass for the binned parametrisation is substantially reduced. 
The best-fit parameter values now lie close to the input values for all four 
benchmarks. A residual degeneracy between the SI and SD cross sections 
remains and, for some benchmarks, the signals can be explained in terms of 
SI interactions only. However, the large cross section degeneracy which 
arises for the polynomial parametrisation when using only direct detection 
data has been eliminated.

\section{Reconstructing the speed distribution}
\label{sec:SpeedDist}
\makeatletter{}
In this section we present the reconstruction of the speed distribution
using the polynomial and binned parametrisations.

\subsection{Polynomial parametrisation}
\begin{figure*}[tp!]
\centering
\includegraphics[clip,trim=0.5cm 0.5cm 0.5cm 0.5cm,width=0.4\textwidth]{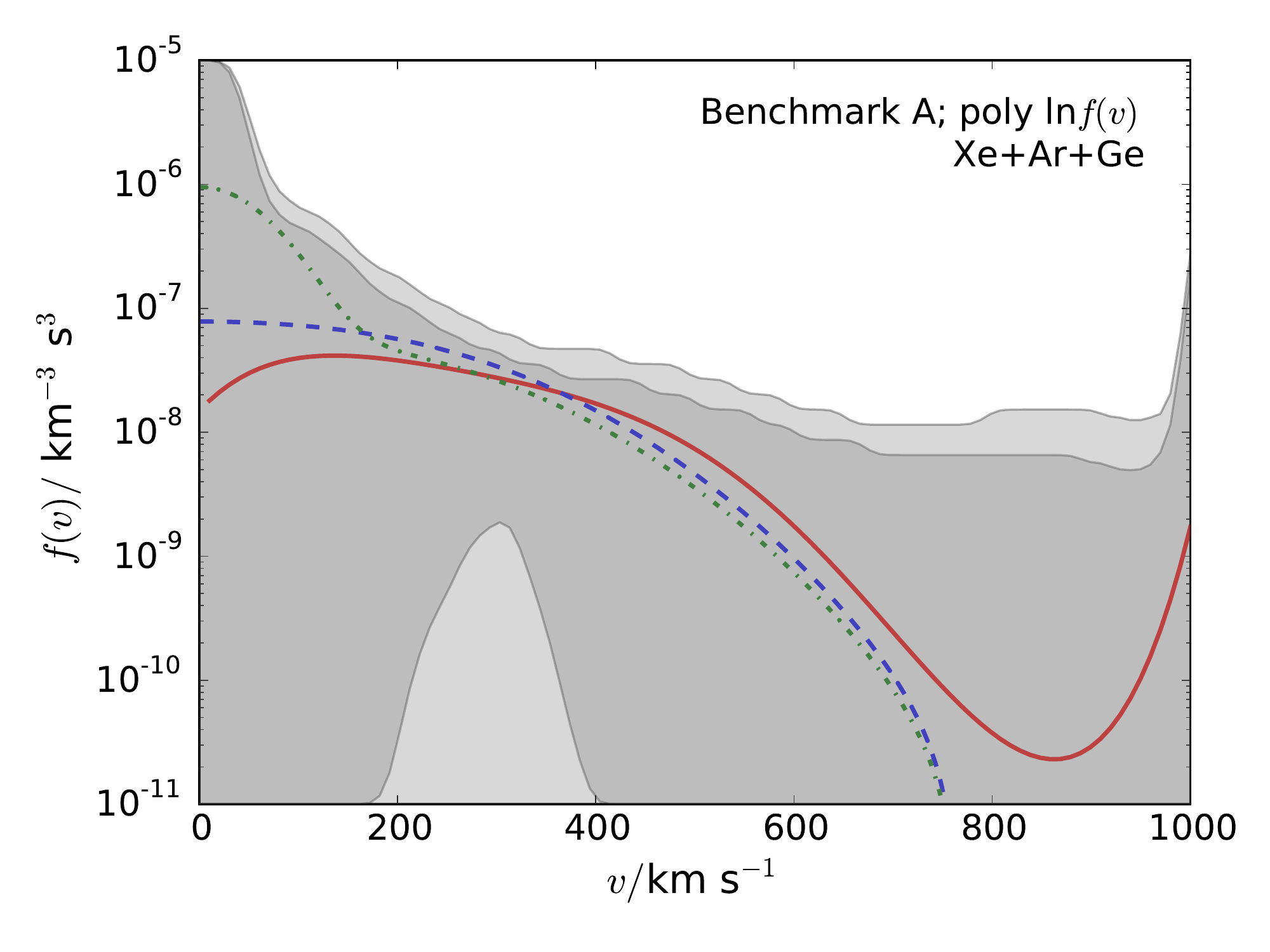}
\includegraphics[clip,trim=0.5cm 0.5cm 0.5cm 0.5cm,width=0.4\textwidth]{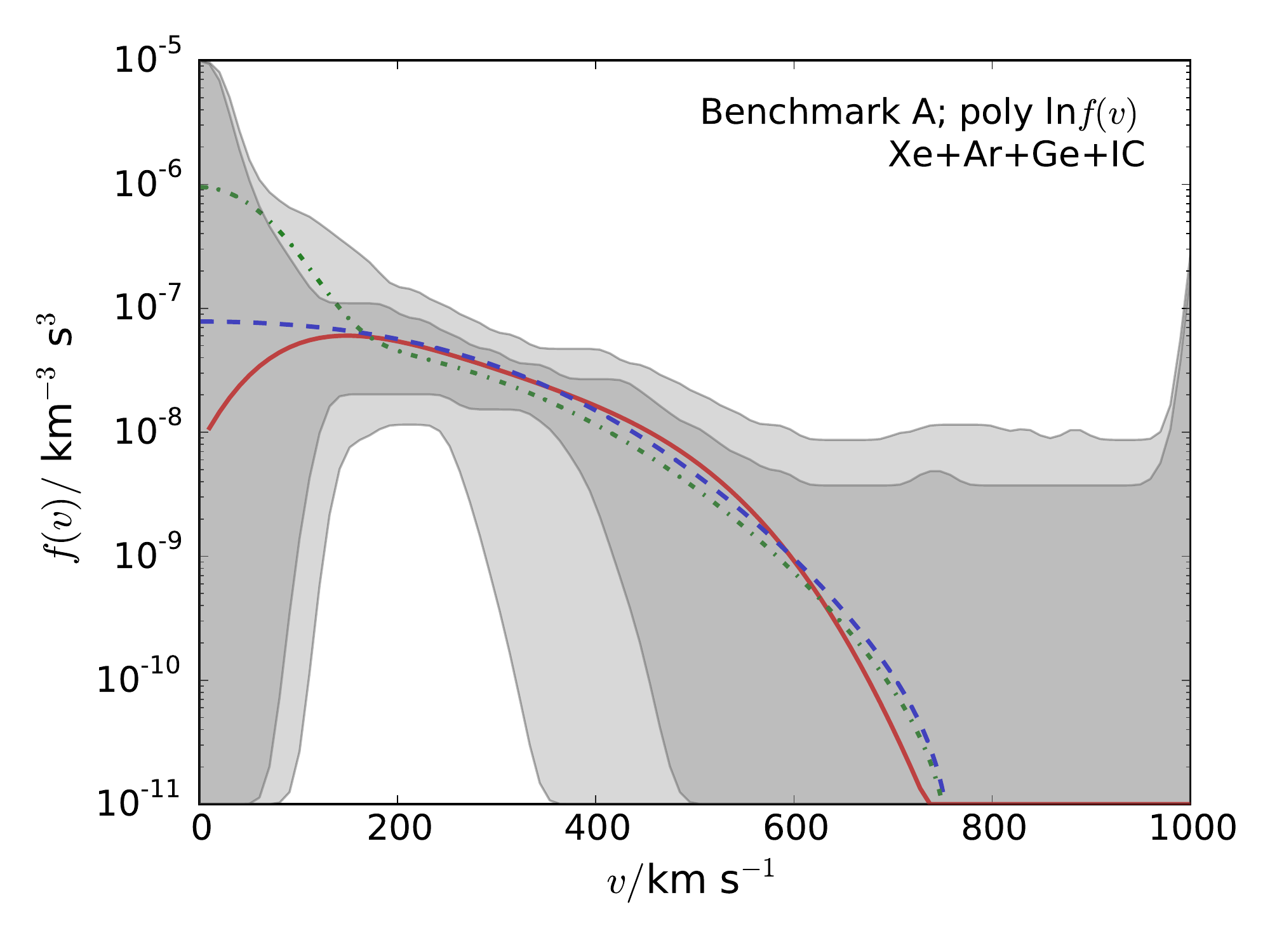}

\includegraphics[clip,trim=0.5cm 0.5cm 0.5cm 0.5cm,width=0.4\textwidth]{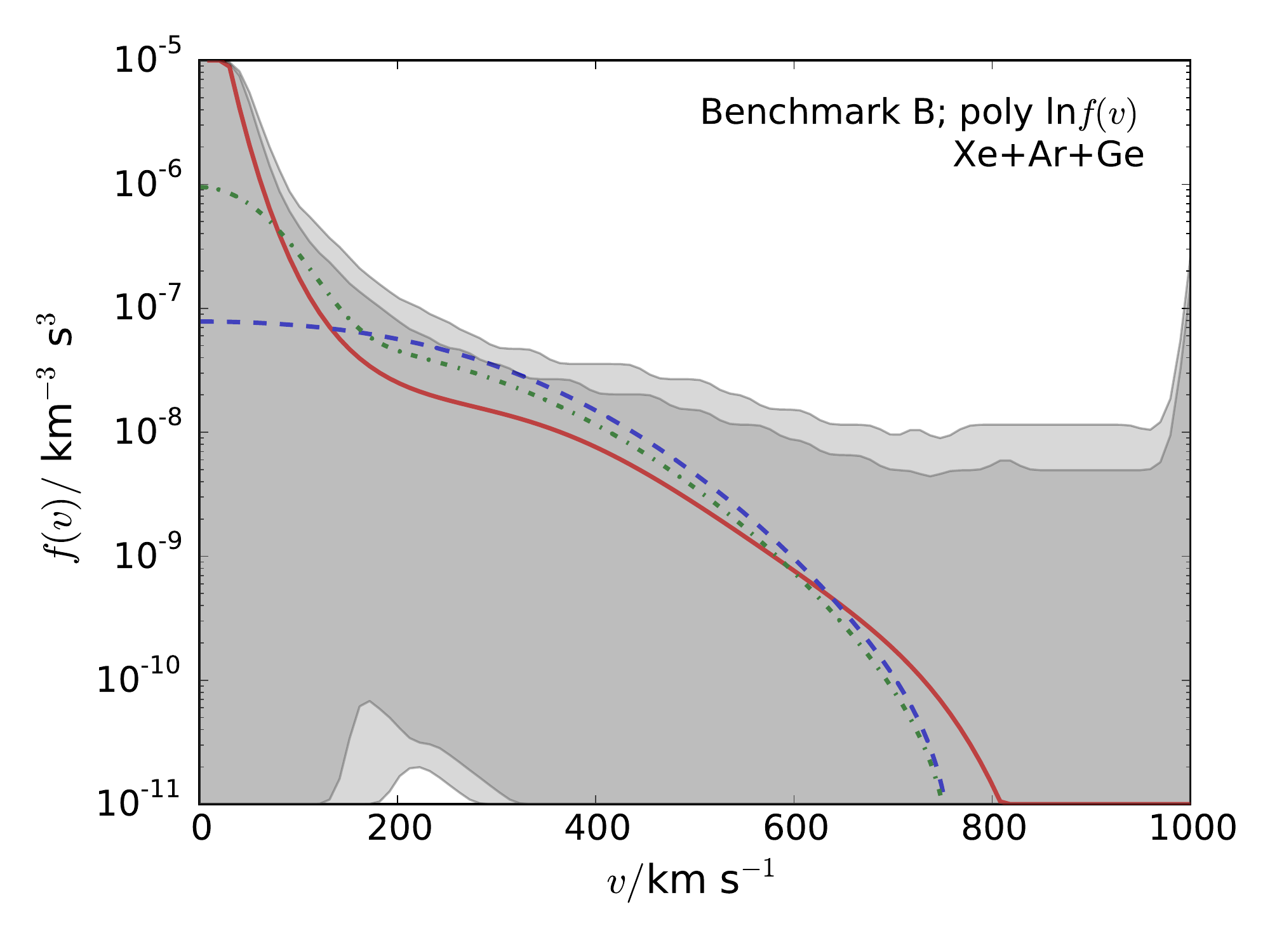}
\includegraphics[clip,trim=0.5cm 0.5cm 0.5cm 0.5cm,width=0.4\textwidth]{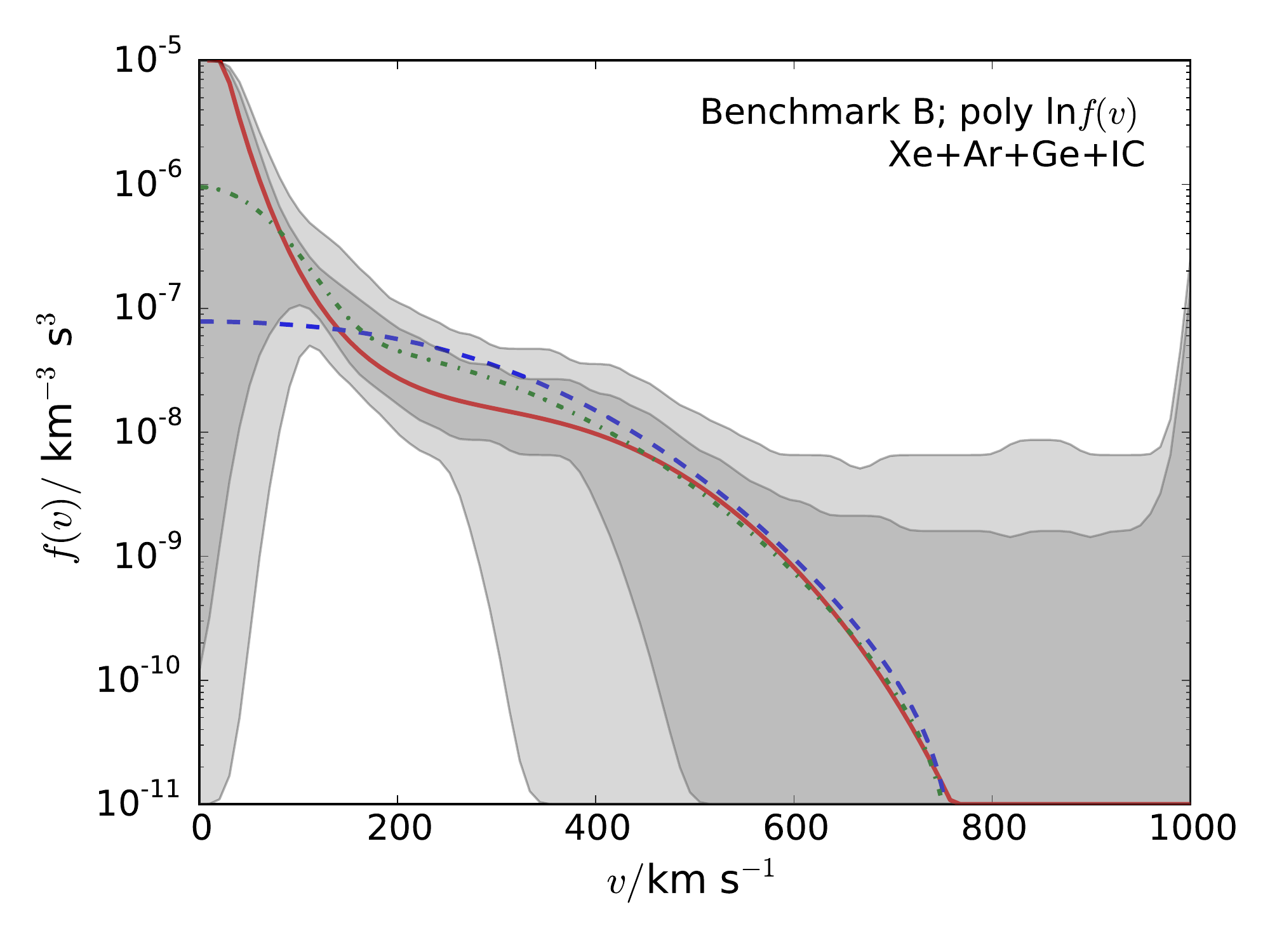}

\includegraphics[clip,trim=0.5cm 0.5cm 0.5cm 0.5cm,width=0.4\textwidth]{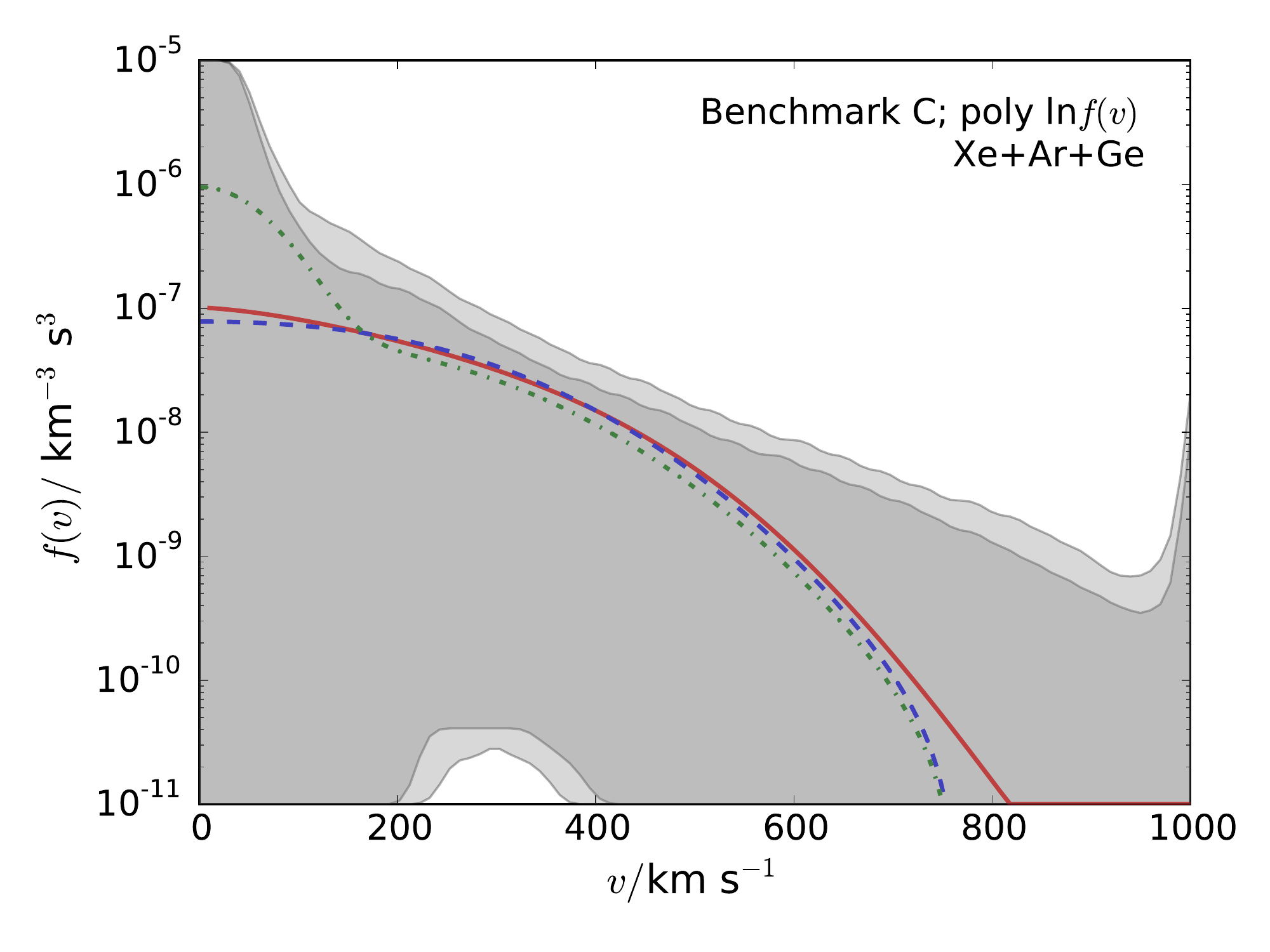}
\includegraphics[clip,trim=0.5cm 0.5cm 0.5cm 0.5cm,width=0.4\textwidth]{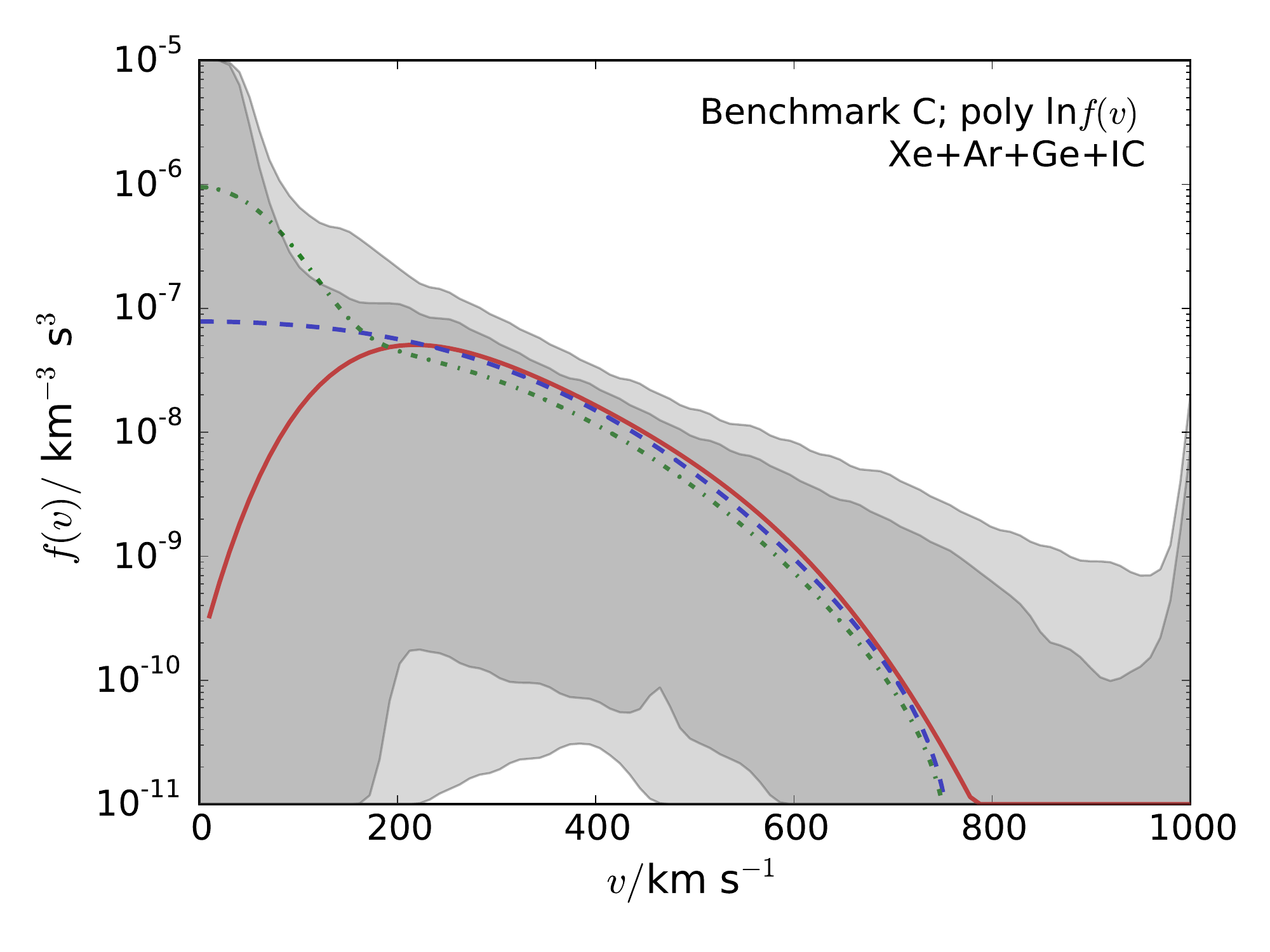}

\includegraphics[clip,trim=0.5cm 0.5cm 0.5cm 0.5cm,width=0.4\textwidth]{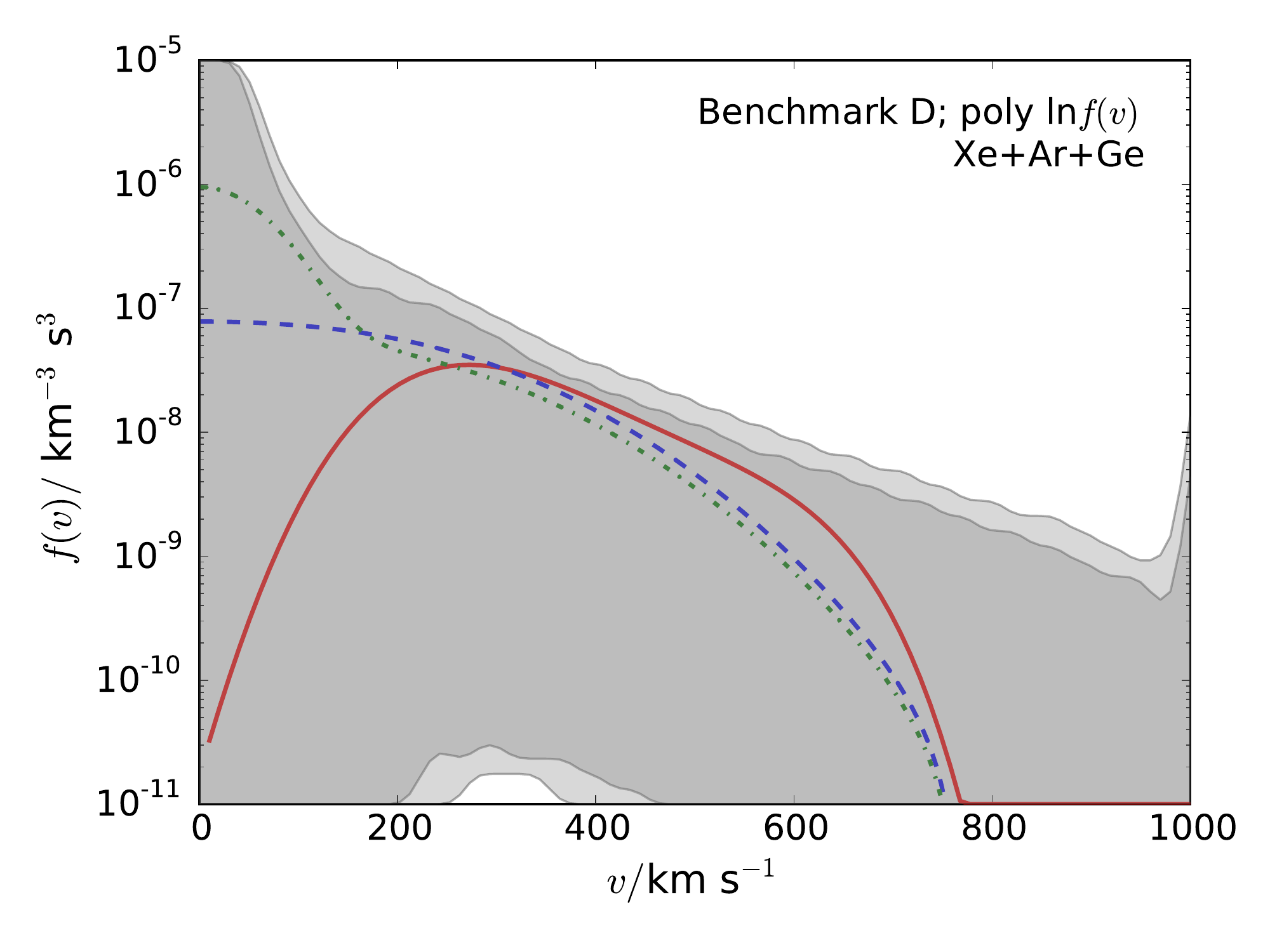}
\includegraphics[clip,trim=0.5cm 0.5cm 0.5cm 0.5cm,width=0.4\textwidth]{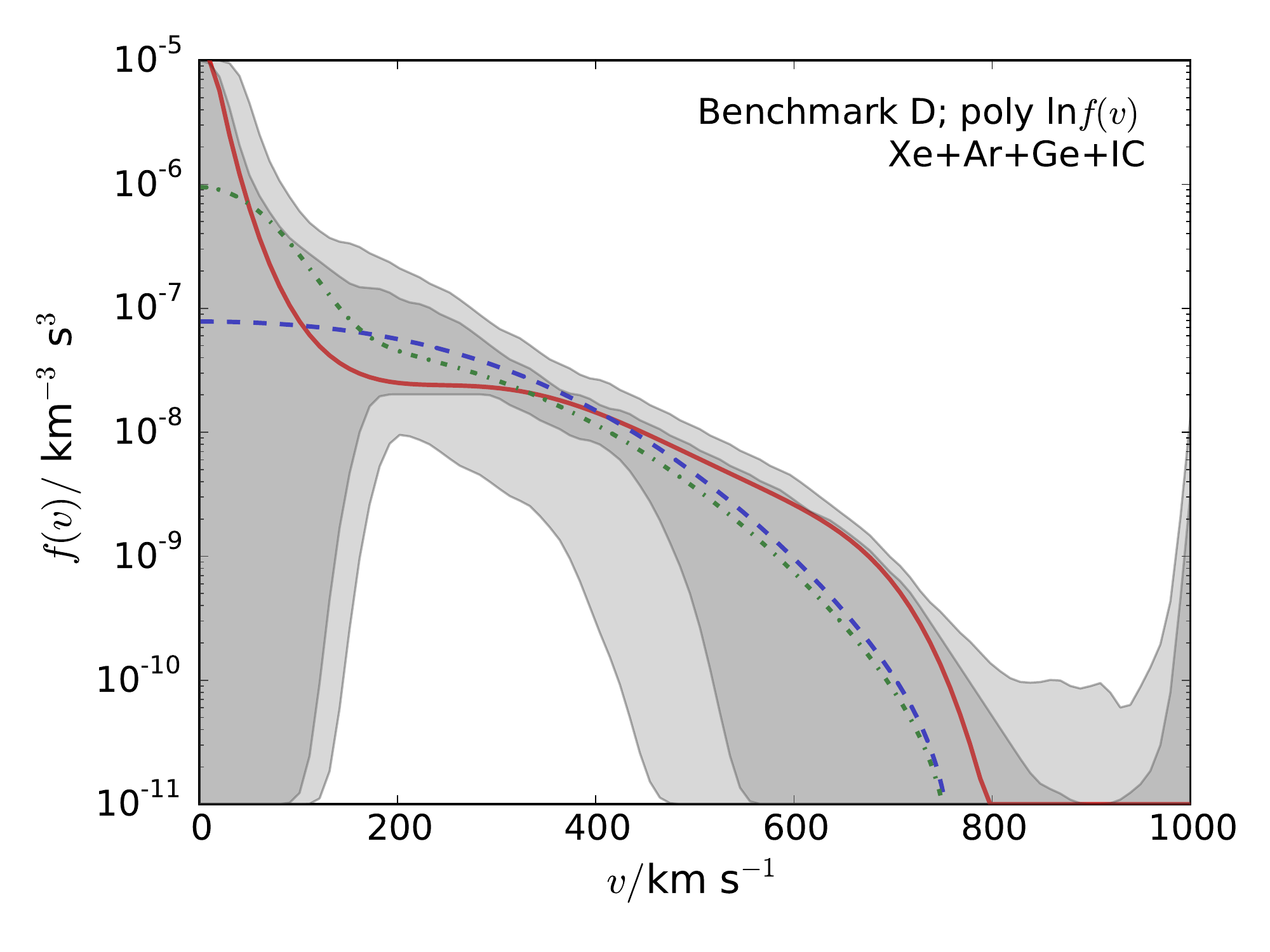}
\caption{Reconstructions of the speed distribution using the polynomial parameterisation with 6 basis functions, using direct detection data only (left column) and using direct detection and IceCube data (right column).  The four rows correspond to benchmarks A-D (from top to bottom). Benchmarks A and C (B and D) have the SHM (SHM+DD) as the input speed distribution. The dashed blue lines indicate the input SHM distribution while the dot-dashed green lines indicate the SHM+DD distribution. In each panel the solid red line shows the best fit, while the grey shaded bands show the 68\% and 95\% confidence intervals.}
\label{fig:f_poly}
\end{figure*}

In Fig.~\ref{fig:f_poly}, we show the results for the polynomial 
parametrisation. The solid red lines indicate the best-fit functions, while 
the grey bands show the 68\% and 95\% confidence intervals. We note that the 
bands are obtained for each value of $v$ by profiling over all other values, 
as well as over the particle physics parameters. For reference, the SHM 
(dashed blue) and SHM+DD (dot-dashed green) distribution functions used in the 
benchmarks are also shown. For the plots in the left column only direct 
detection data are used in the likelihood, while for the right column IceCube 
data is also used.

With only direct detection data, the uncertainties on the speed distribution 
are large. However, some features are apparent. For benchmark A, there is a 
light-grey-shaded region in the range $v \sim 200-400 \, {\rm km \, s}^{-1}$ 
which does not lie within the 68\% band. On the left side of this region, 
there is a contribution from flat speed distributions, such as the one 
labeled `i' in Fig.~\ref{fig:SpeedExamples}, which provide a good fit to the 
data for light WIMPs. On the right side of this region, there is a 
contribution from steeper speed distributions, such as the one labeled `ii' in 
Fig.~\ref{fig:SpeedExamples}, which provides a good fit for heavier WIMPs. 
Values of the speed distribution inside this light-grey region provide a 
poorer fit to the data as they underproduce low-energy events in xenon and/or 
high-energy events in argon.

This feature is not present in benchmark B because speed distributions which 
have low values above $v \sim 100 \, {\rm km \, s}^{-1}$ and rise rapidly 
below this value are also allowed. These steeply-rising distributions are 
those corresponding to the regions of parameter space at very large \sigmapsi 
and \sigmapsd in Fig. \ref{fig:DDonly_benchmarkB}. 

Similar considerations, with the speed distribution decreasing below the 
speeds probed by xenon and above the speeds probed by argon, also explain 
the behaviour of the reconstruction of $f(v)$ for benchmarks C and D. However, 
for these benchmarks, the two regimes are closer to each other (since values 
of \mwimp larger than 100 GeV are not allowed) and the different families of 
$f(v)$ overlap more.

When IceCube data is included (right panels), the constraints on $f(v)$ are 
significantly improved over some ranges of speed. The speed distributions 
which decrease below 200 ${\rm km \, s}^{-1}$ (as in the left panels) are 
eliminated, since the low \mwimp values they correspond to are ruled out, as 
they would not produce neutrino events in IceCube. The IceCube data also 
disfavours the distributions which rise steeply below 
$v \sim 400 \, {\rm km \, s}^{-1}$ which correspond to large values of both 
\sigmapsi and \sigmapsd  (especially for benchmarks B, C and D). The net 
effect is that there is a range of speeds (around 200 ${\rm km \, s}^{-1}$  
for benchmarks A and B and around 300 ${\rm km \, s}^{-1}$ for benchmark D)  
where $f(v)$ can be reconstructed within a factor of $\sim 4$. These speeds 
are just above the direct detection thresholds, where the most information 
about the shape of the recoil energy spectrum is available. At high speeds, 
$v > 600 \, {\rm km \, s}^{-1}$, the uncertainties remain large, as the direct 
detection experiments have no sensitivity to the shape of $f(v)$ above 
$E_\textrm{max}$. Finally, we note that the small number of IceCube signal 
events in benchmark C results in only a slight improvement in the bounds on 
$f(v)$. 

Within the range of speeds where $f(v)$ is reconstructed well, the best-fit 
shape for benchmarks B and D (SHM+DD) is clearly steeper than for benchmarks 
A and C (SHM); the reconstruction has correctly recovered the rapidly rising 
dark disk contribution at low speeds. 

We would now like to determine whether or not it is possible to exclude any 
particular form for the speed distribution using the mock data. The bands in 
Fig.~\ref{fig:f_poly} are calculated from the 1-dimensional profile likelihood 
separately at each value of $v$. However, the uncertainties at different 
values of $v$ are strongly correlated, due to the normalisation of $f(v)$. 
This means that not all shapes falling within the 68\% band are consistent 
with the data at the 68\% level. However, if a speed distribution falls 
outside the 68\% band at some value of $v$, it can be rejected with at least 
68\% confidence.

A more powerful approach is to determine whether the data prefer a particular 
distribution over another. We focus on benchmark D, which has an input speed 
distribution with a dark disk and a low \mwimp, which allows a more accurate 
reconstruction of the speed distribution as discussed above.  We compare the 
relative log-likelihoods of the best-fit point using the polynomial 
parametrisation with the best-fit assuming a fixed SHM distribution. In order 
to meaningfully compare the log-likelihoods of the two best-fit points, the 
two scans must be performed on parameter spaces that are nested one inside 
the other. Therefore, we must determine the combination of Chebyshev 
polynomials that provides a good fit to the SHM~\footnote{Decomposing 
$\ln f(v)$ for the SHM into six Chebyshev polynomials provides a fit which 
is accurate to better than 0.1\%.}. We then fix the polynomial coefficients 
to the values obtained from this fit and perform a parameter scan over the 
remaining particle physics parameters. The best fit for the full $\ln f(v)$ 
parametrisation has a slightly higher log-likelihood value than the best 
fit when the coefficients are fixed to the SHM values. The relative 
log-likelihood between the two models corresponds to a value of 
$\Delta \chi^2 = 2.48$. This is perhaps not surprising, as the significance of the IceCube signal is not very large ($<3\sigma$), and it is this data which distinguishes the two distributions. For 5 degrees of freedom (the 5 free polynomial 
coefficients in the full parametrisation), the significance of this result is below the $1\sigma$ level and we therefore cannot reject the SHM speed distribution.  However, this method allows us to make robust statements about the different speed distributions and may, with greater statistics, allow us to distinguish between them.

\begin{figure*}[tp!]
\centering
\includegraphics[clip,trim=0.5cm 0.5cm 0.5cm 0.5cm,width=0.4\textwidth]{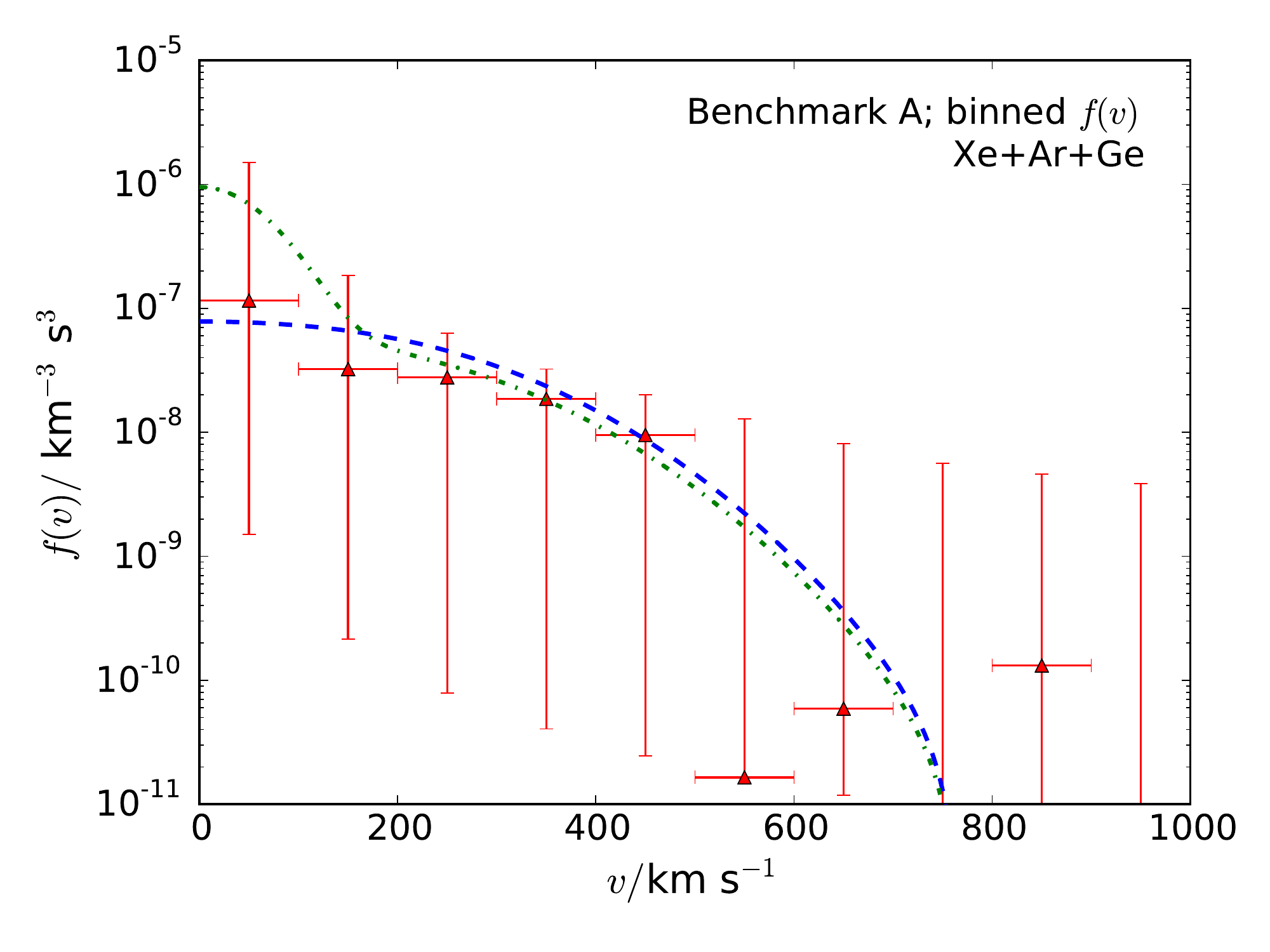}
\includegraphics[clip,trim=0.5cm 0.5cm 0.5cm 0.5cm,width=0.4\textwidth]{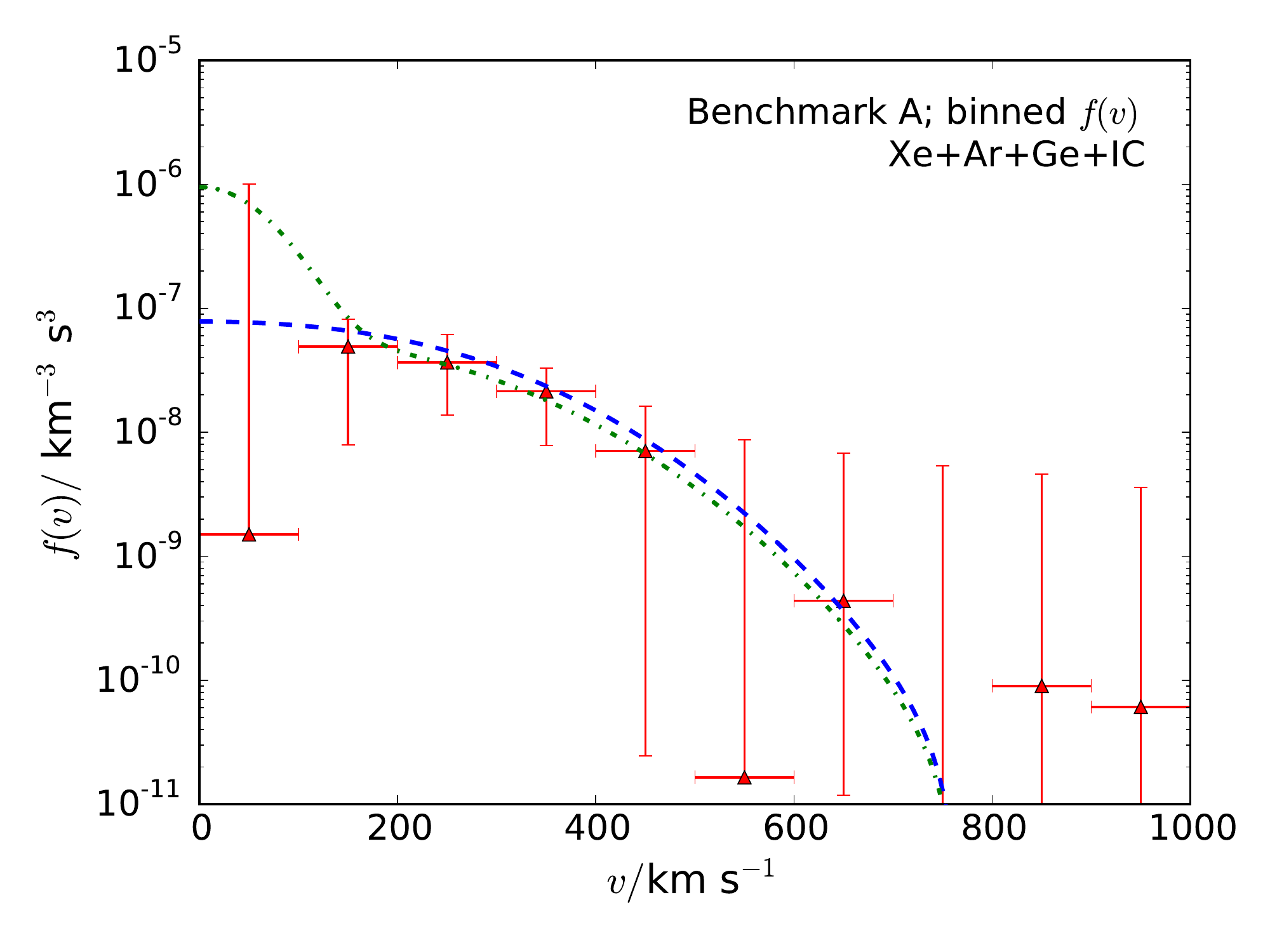}

\includegraphics[clip,trim=0.5cm 0.5cm 0.5cm 0.5cm,width=0.4\textwidth]{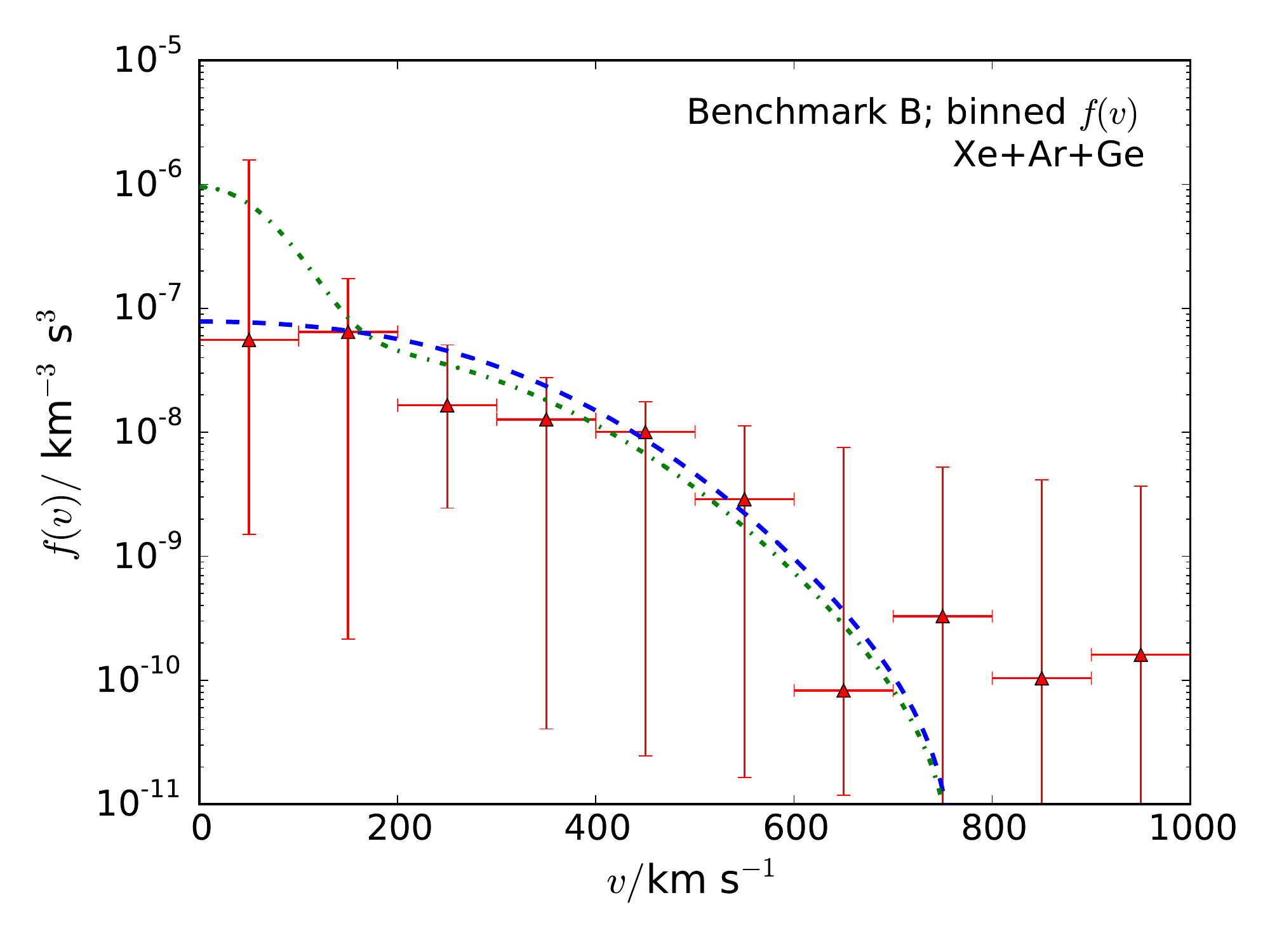}
\includegraphics[clip,trim=0.5cm 0.5cm 0.5cm 0.5cm,width=0.4\textwidth]{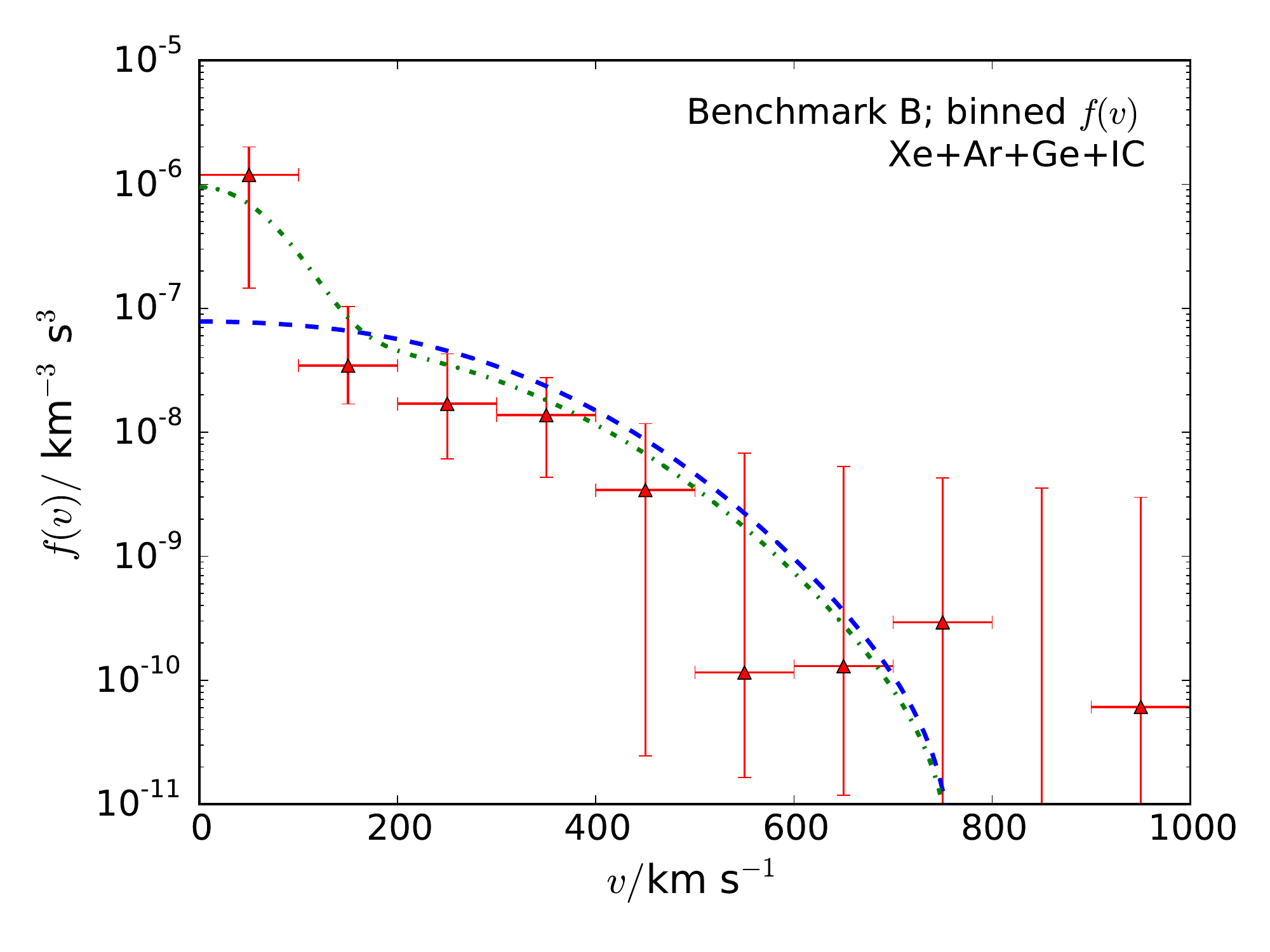}

\includegraphics[clip,trim=0.5cm 0.5cm 0.5cm 0.5cm,width=0.4\textwidth]{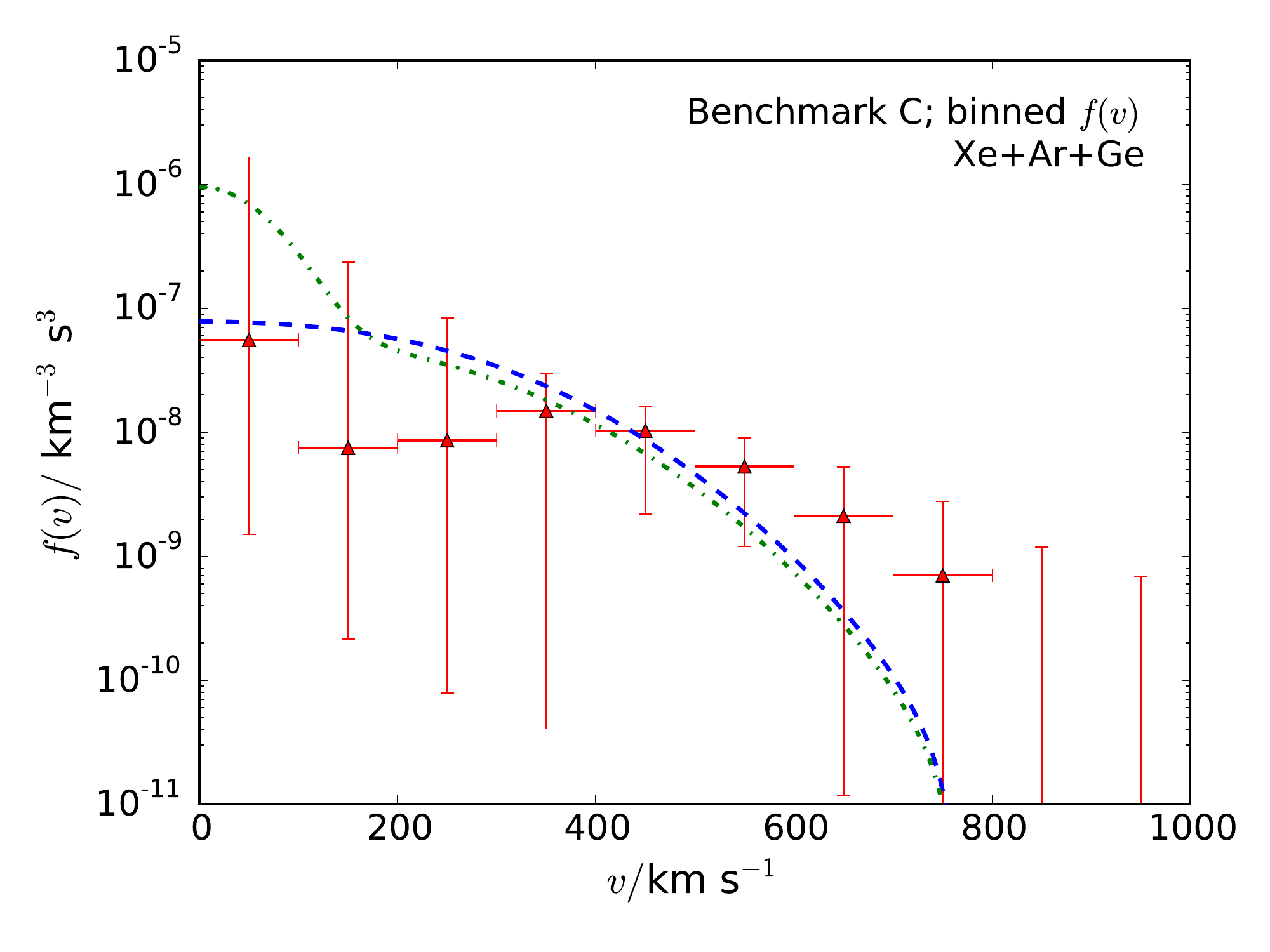}
\includegraphics[clip,trim=0.5cm 0.5cm 0.5cm 0.5cm,width=0.4\textwidth]{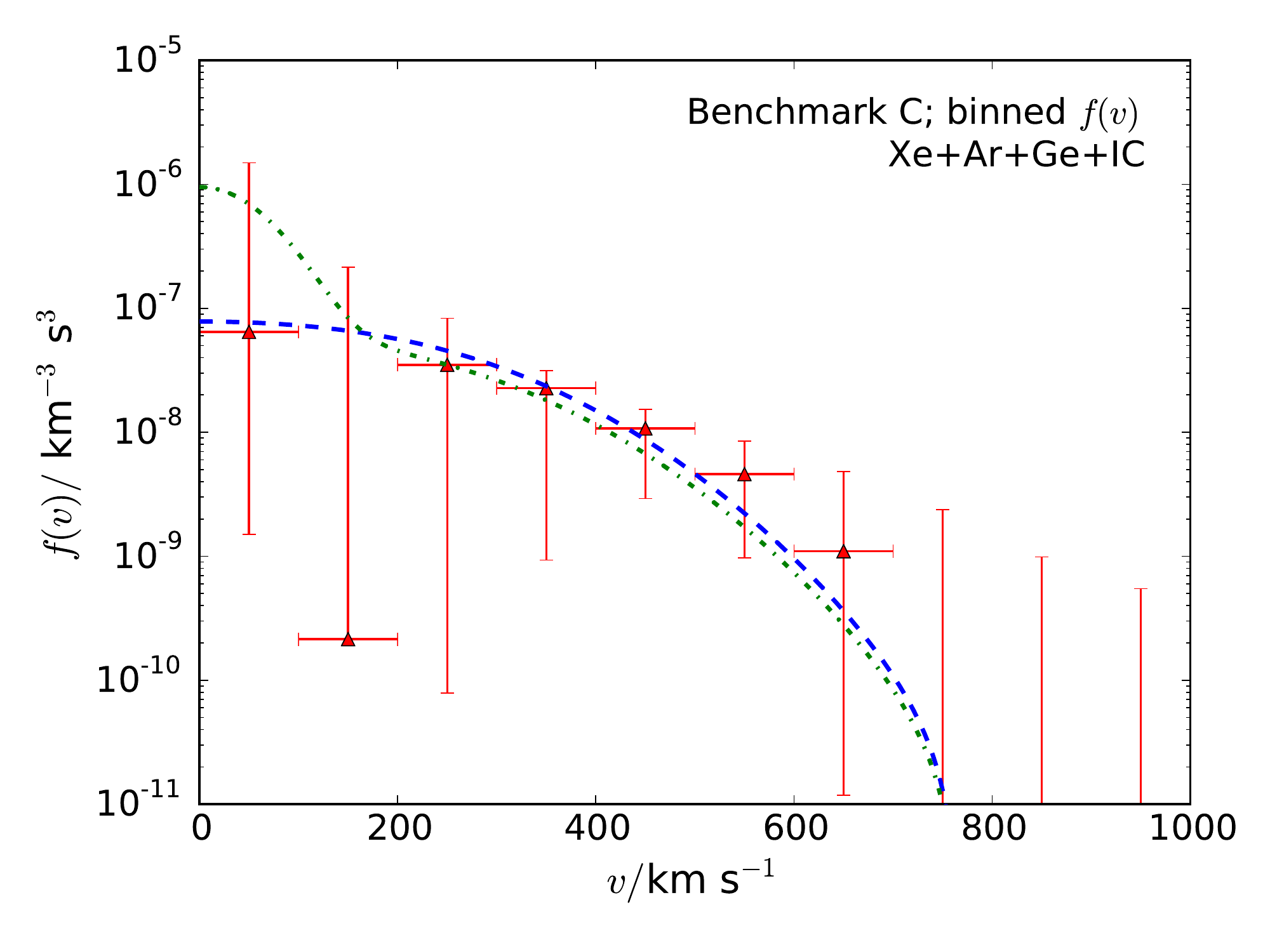}

\includegraphics[clip,trim=0.5cm 0.5cm 0.5cm 0.5cm,width=0.4\textwidth]{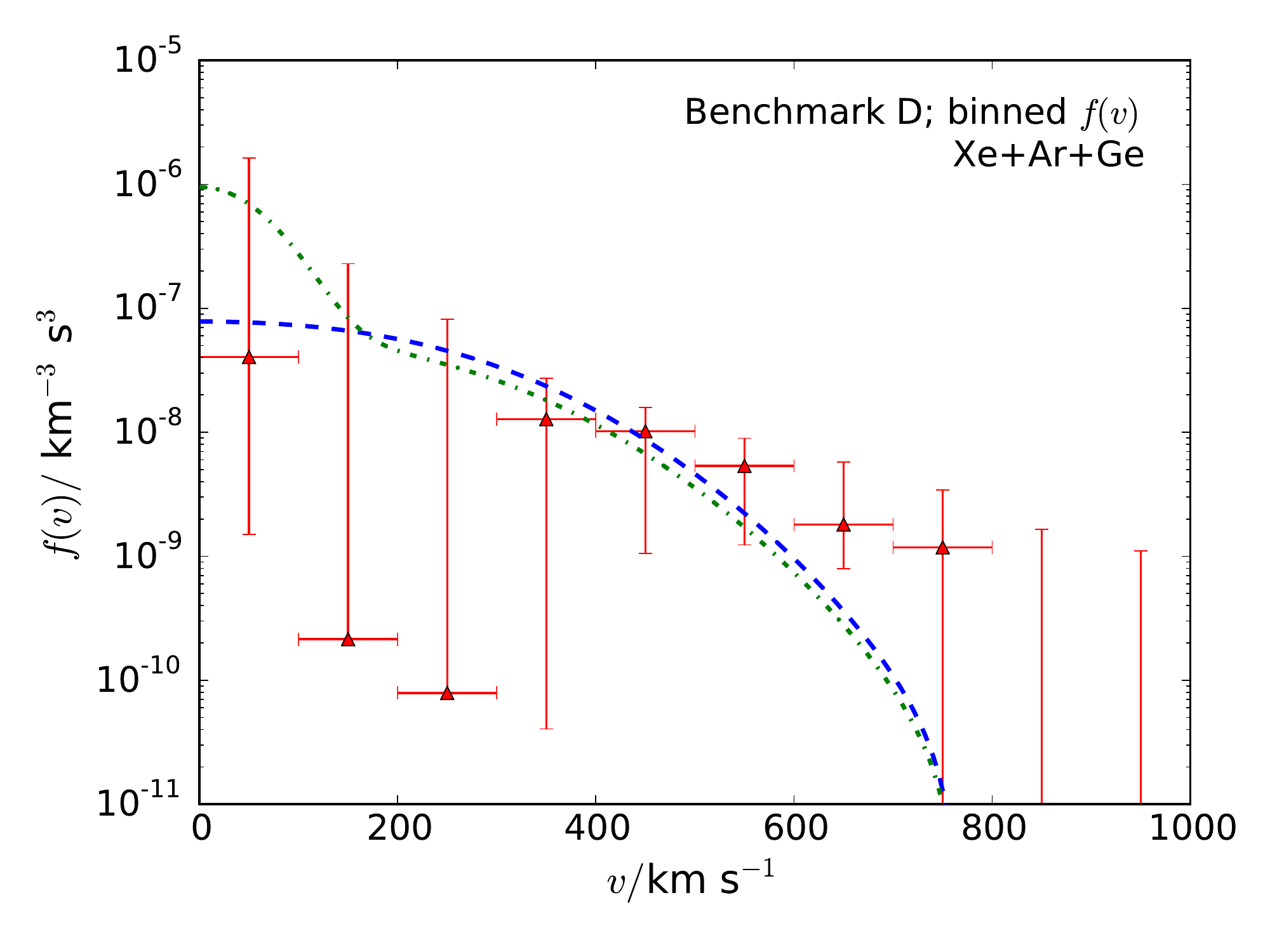}
\includegraphics[clip,trim=0.5cm 0.5cm 0.5cm 0.5cm,width=0.4\textwidth]{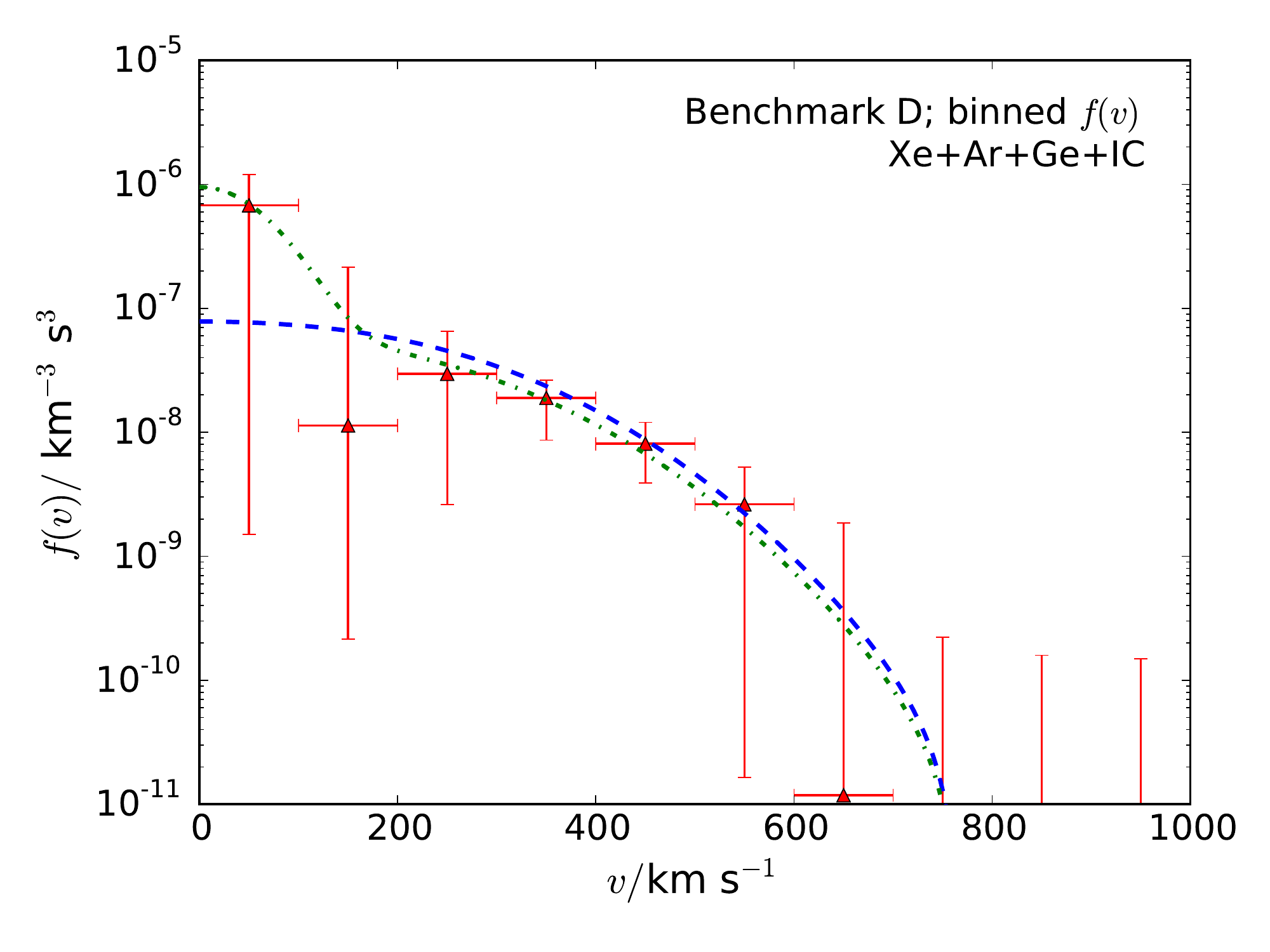}
\caption{Same as Fig. \ref{fig:f_poly} but for the binned parametrisation. The red triangles correspond to the best-fit point, while the red bars indicate the 68\% confidence limits.}
\label{fig:f_binned}
\end{figure*}

\subsection{Binned parametrisation}
In Fig. \ref{fig:f_binned}, we show the speed distributions reconstructed 
using the binned parametrisation. The red triangles show the best-fit bin 
heights, while the error bars indicate the 68\% confidence limits. As for the 
polynomial parametrisation, the error bar on each bin is obtained by profiling 
over all other parameters in the scan.

With only direct detection data, the uncertainties on the bin heights are 
large, with some of the 68\% limits extending down to below 
$10^{-10} \textrm{ km}^{-3} \textrm{ s}^3$. In certain bins, however, the 
constraints are much stronger, with the bin heights constrained within a 
factor of roughly 5. For benchmarks C and D, $f(v)$ is reconstructed with 
better precision for large speeds (500 -- 700 \kms) than for the polynomial 
parameterisation. This is because very large \sigmapsi and \sigmapsd are not 
allowed with the binned parameterisation (see 
Figs. \ref{fig:DDonly_benchmarkC} and \ref{fig:DDonly_benchmarkD} and the 
discussion in Sec. \ref{sec:DDonly}). There is, however, mild tension with 
the input speed distribution for benchmark D (bottom left panel). In the 
range 500 -- 700 \kms, the reconstruction appears to overestimate the bin 
heights, resulting in a flatter shape for $f(v)$ than the input form. In this 
case, the reconstructed WIMP mass was lower than the input value. As 
discussed in Ref.~\cite{Kavanagh:2012}, this is because reducing the WIMP 
mass reduces the size of the bins when converting to energy space, which 
leads to a better fit to the data. When the WIMP mass is decreased, the 
velocity integral needs to become less steep in order to counterbalance the 
steepening of the spectrum. The increased WIMP population in the range 500 -- 
700 \kms is then balanced by a depleted $f(v)$ at lower speeds in order to 
maintain the overall normalisation to unity. This results in very low bin 
heights in the range 100 -- 200 \kms.

As in the case of the polynomial parameterisation (Fig. \ref{fig:f_poly}), 
without IceCube data the reconstructed speed distributions for the pairs of 
benchmarks with and without a dark disk (A \& B and C \& D) are almost 
indistinguishable, as the direct detection experiments have little 
sensitivity at low speeds where they differ. When IceCube data is added, 
benchmarks B and D, which have a dark disk, show a clear spike in the lowest 
bin, which is not present for benchmarks A and C. We also note that the 
best-fit bin heights now trace the input speed distributions closely. As 
for the polynomial parameterisation, the uncertainties are smallest for 
speeds close to the direct detection thresholds (0 -- 300 \kms for benchmarks 
A and B and 200 -- 500 \kms for the lighter benchmarks C and D).

We note that a likelihood comparison between the binned distribution and 
some fixed $f(v)$ such as the SHM (as was performed for the polynomial 
parametrisation) may not be appropriate. This is because the 10-bin 
distribution does not provide as close an approximation to the shape of the 
SHM. Thus, such a likelihood comparison would not necessarily be meaningful.

\section{Discussion}
\label{sec:discussion}
\makeatletter{}
In Sec.~\ref{sec:DDonly}, we examined the reconstruction of the WIMP mass and 
cross sections with binned and polynomial parametrisations of the speed 
distribution, using mock data from direct detection experiments only. As 
found in Refs.~\cite{Peter:2011,Kavanagh:2012,Kavanagh:2013a,Kavanagh:2013b}, 
with the binned parametrisation there can be a bias in the WIMP mass. We also 
saw that there is a strong degeneracy between the SI and SD cross sections 
and the shape of the speed distribution when using only direct detection data. 
In particular, large cross sections can be accommodated by increasing the 
fraction of the WIMP population which lies below the direct detection energy 
thresholds. Even though this degeneracy was only apparent when using the 
polynomial $\ln f(v)$ parametrisation, it will affect all methods which make 
no astrophysical assumptions. The binned parametrisation of the speed 
distribution (top rows of Figs.~\ref{fig:DDonly_benchmarkA}, 
\ref{fig:DDonly_benchmarkB}, \ref{fig:DDonly_benchmarkC} and 
\ref{fig:DDonly_benchmarkD}) appears to lead to closed contours for the 
particle physics parameters. However, in the left column of 
Fig.~\ref{fig:f_binned}, we demonstrate that this parametrisation is 
insensitive to the presence of a dark disk at low speeds, and it is this lack 
of sensitivity that leads to the spurious upper limits on the cross sections. 

With the inclusion of IceCube data in Sec.~\ref{sec:DDwithIC}, the situation 
is significantly improved. The degeneracy to large cross sections is 
eliminated for all four of the benchmarks that we consider. The sensitivity of 
Solar capture to the low-speed WIMP population allows us to exclude the region 
of parameter space with large WIMP mass and large SI and SD cross sections, 
as it overproduces neutrino events at IceCube. The low mass region is also 
much more tightly constrained as if the mass is too small too few neutrinos 
are seen in IceCube. As seen in the top rows of 
Figs.~\ref{fig:DDIC_benchmarkA}, \ref{fig:DDIC_benchmarkB}, and 
\ref{fig:DDIC_benchmarkD}, the inclusion of IceCube data removes the bias in 
the reconstruction of the WIMP mass which occurs for the binned 
parameterisation with direct detection data only~\footnote{In benchmarks A and 
B, two distinct regions of parameter space are allowed, but the one around 
the input parameter values is significantly preferred over the other.}.

\begin{figure}[!t]
  \centering
  \includegraphics[width=0.45\textwidth]{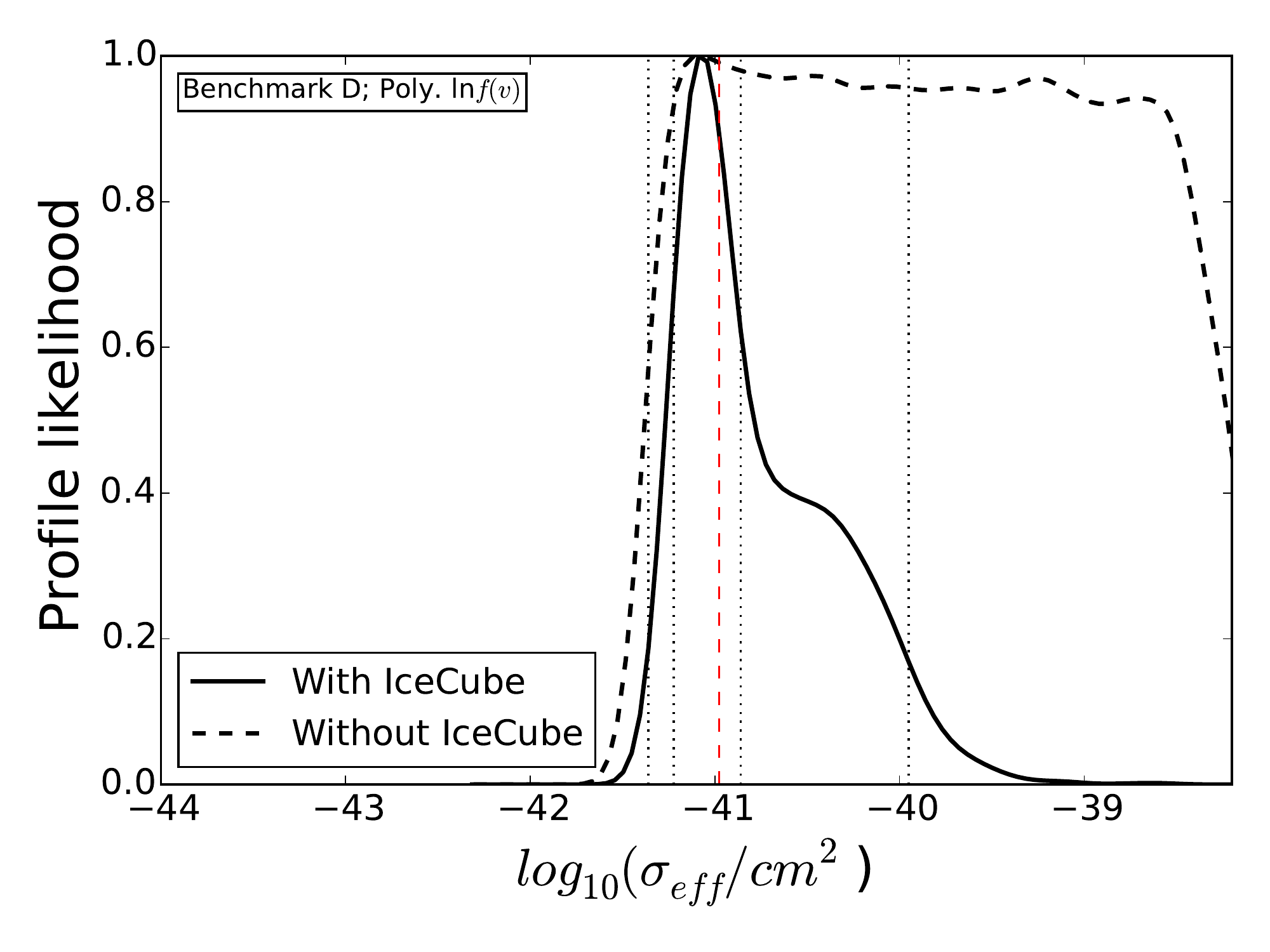}
\caption{Profile likelihood for the effective cross section of germanium, $\sigma_\textrm{eff}$, (defined in Eq.~(\ref{eq:NT:sigeff})) for benchmark D, obtained using the polynomial parametrisation with and without IceCube data (solid and dashed black lines respectively). The vertical red dashed line corresponds to the input value for this benchmark while the vertical dotted black lines correspond to the 68\% and 95\% confidence intervals for the case \textit{with} IceCube data.}
\label{fig:NT:sigeff}
\end{figure}

Some degeneracy still remains. In particular, it is possible to reduce the 
SD cross section significantly and compensate by increasing the SI 
contribution if, at the same time, the velocity integral is also made steeper 
at low speeds. It is possible to define an effective cross section, 
$\sigma_\textrm{eff}$, which incorporates both cross sections and controls the 
overall event rate. Due to the different response of each detector to SI and 
SD couplings, each experiment (including IceCube) will have a different 
$\sigma_\textrm{eff}$. Here, for simplicity, we focus on the case of a 
germanium detector for which
\begin{equation}
\label{eq:NT:sigeff}
\sigma_\textrm{eff} = \sum_{i} f_i A_i^2 \sigmapsi + f_{73} \frac{16\pi}{3} \frac{\sigmapsd}{2J+1} S_{00}(0)\,,
\end{equation}
where $f_i$ and $A_{i}$ are the mass fraction and mass number of isotope $i$. 
Figure~\ref{fig:NT:sigeff} shows the profile likelihood for 
$\sigma_\textrm{eff}$ with and without IceCube data (as solid and dashed lines 
respectively) for benchmark D, obtained using the polynomial parametrisation. 
Without IceCube data there is only a lower limit on $\sigma_\textrm{eff}$, 
below which there are too few events produced in the detector. For larger 
values of $\sigma_\textrm{eff}$, the profile likelihood is almost completely 
flat with an uncertainty of roughly three orders of magnitude. Including 
IceCube data, the profile likelihood becomes sharply peaked, with the value of 
$\sigma_\textrm{eff}$ constrained to within a factor of four at the 68\% level. 
Clearly, the inclusion of IceCube data means that we can now reconstruct 
the value of the effective cross section, rather than only placing a lower 
limit.

The residual degeneracy between the SI and SD cross sections could be broken 
by the inclusion of an additional direct detection experiment using a nuclear 
element that is sensitive mainly to SD interactions \cite{Cerdeno:2013gqa,
Cerdeno:2014uga}. This approach is currently taken by the COUPP and ROSEBUD 
collaborations \cite{Behnke:2008zza,Coron:2011zz} and is also proposed for 
the EURECA experiment \cite{Kraus:2011zz}.

As in Ref.~\cite{Arina:2013}, the conclusions we draw apply even in the 
absence of a significant signal at IceCube, provided the WIMP is not too 
light. In benchmark C, the number of signal events is just 13, which is 
consistent with the observed background at just over $1\sigma$. Even with a 
signal of such low significance, we can still break the degeneracy between 
the cross section and $f(v)$, as explained above. However if the WIMP mass is 
smaller than the IceCube detection threshold, no neutrino events will be 
produced, regardless of the scattering cross sections and speed distribution. 
There is therefore no improvement in the reconstruction of the WIMP mass and 
the cross section degeneracy remains.

In Sec.~\ref{sec:DDonly} we demonstrated that the binned speed parametrisation 
is not suitable when only direct detection data is used, as it may lead to a 
bias in the WIMP mass for certain benchmark parameters (see also 
Refs. \cite{Peter:2011,Kavanagh:2012}). With the polynomial parametrisation, 
no such bias occurs. However, this parametrisation typically results in much 
larger parameter uncertainties than the binned one, including a significant 
degeneracy to large values of the cross sections. This is due to the fact that 
the polynomial parametrisation can explore a wider range of forms for $f(v)$, 
including distributions which are rapidly varying.

When IceCube data is included, the bias in the WIMP mass for the binned 
parametrisation is significantly reduced, as low WIMP masses underproduce
events in IceCube. A binned parametrisation also leads to narrow parameter 
uncertainties (with closed contours in benchmarks B, C and D, see 
Figs.~\ref{fig:DDIC_benchmarkB}, \ref{fig:DDIC_benchmarkC} and 
\ref{fig:DDIC_benchmarkD}) and tighter constraints on $f(v)$ than with the 
polynomial parametrisation. This is because the reconstruction using a 
polynomial decomposition encompasses qualitatively different, and larger, 
regions of the parameter space than that using the binned $f(v)$. For example, 
small values of $\sigmapsd$ are allowed at the 68\% level in benchmarks B, C 
and D, only when the polynomial parametrisation is used. This is due to the 
fact that this parametrisation can encompass steep or rapidly varying 
distributions which, in these regions of the parameter space, are required to 
produce a good fit to the data.

It is not clear which speed parametrisation is optimal when direct detection 
and neutrino telescope data are combined. In this case, where the combined 
data are sensitive to the full range of WIMP speeds, the flexibility of the 
polynomial parametrisation may not be a benefit. In particular the rapidly 
varying shapes it probes may not be physically well motivated. For example, 
it is not clear how a $f(v)$ that is rapidly varying or rising at low speeds 
could arise in an equilibrium model of the Milky Way. If we are confident 
that the speed distribution does not contain sharp features, then the binned 
method, which is not sensitive to such features and produces tighter 
constraints on the particle physics parameters, is most suitable. However 
if we want to allow for a more general shape, with the possibility of sharp 
features, such as high density streams, the polynomial parametrisation is 
more appropriate. A pragmatic approach would be to use both parametrisations.

We note that we have made several assumptions in this work. We have neglected 
uncertainties in the SD form factors, which may lead to wider uncertainties 
on the particle physics parameters. Using the parametrisation in 
Ref. \cite{Cerdeno:2012} would allow us to take this into account, and also 
compare the relative importance of nuclear and astrophysical uncertainties. 
Further simplifications include the assumptions of equilibrium between the 
capture and annihilation rates in the Sun, and the approximation that 
annihilations occur into a single channel. These uncertainties could be 
relaxed and incorporated as free parameters in the fit. In this paper, we 
focused on an idealized scenario which neglects these uncertainties in order 
to highlight the improvement in the determination of the WIMP particle 
physics and astrophysics parameters that can be achieved by combining data 
from a neutrino telescope with direct detection experiments. This is, however, 
a general phenomenon which can be exploited even for larger (and more
realistic) parameter spaces.

\section{Conclusions}
\label{sec:conclusions}
\makeatletter{}
We have examined the effect of combining future direct detection and neutrino 
telescope data on reconstructions of the standard WIMP particle physics 
parameters (\mwimp, \sigmapsi and \sigmapsd) and the local speed distribution 
$f(v)$. We account for uncertainties in the DM speed distribution by using 
two parametrisations: the binned parametrisation proposed in 
Ref. \cite{Peter:2011} and the polynomial $\ln f(v)$ parametrisation from 
Ref. \cite{Kavanagh:2013a}. Direct detection data alone is only sensitive to 
speeds above a (WIMP-mass-dependent) minimum value. However the inclusion of 
neutrino telescope data allows the full range of WIMP speeds, down to zero, 
to be probed.

Our main conclusions can be summarised as follows:
\begin{itemize}
\item When only data from direct detection experiments are used, the 
polynomial speed parameterisation provides an unbiased measurement of the WIMP 
particle physics parameters. Even for benchmarks A and B, where the constraint 
contours for the particle physics parameters are not closed, the best-fit 
values are very close to the input values. This had previously been found for 
the case of SI-only interactions~\cite{Kavanagh:2013a,Kavanagh:2013b}. Here 
we have shown that it also holds when the SI and SD cross sections are both 
non-zero. We have also confirmed the bias in the WIMP mass induced by the 
binned $f(v)$ parametrisation (as found in Ref.~\cite{Kavanagh:2012}).

\item The inclusion of IceCube mock data significantly narrows the constraints 
on the WIMP mass, for both the binned and polynomial parametrisations. Most 
notably, the bias towards lower WIMP masses experienced with the binned 
parametrisation is removed.

\item For the polynomial parametrisation of $f(v)$, including mock IceCube 
data eliminates a region of parameter space where the WIMP mass and 
scattering cross-sections are all large. With only the data from direct 
detection experiments the cross sections are degenerate with the shape of 
$f(v)$, so that increasing the velocity integral at low speeds (where the 
direct detection experiments are not sensitive), balances the effect of the 
large cross sections. Including the information from neutrino telescopes 
breaks this degeneracy, since these solutions overproduce neutrinos. The net 
effect is that with the addition of IceCube data upper limits can be placed 
on the strength of the SI and SD cross sections.

\item With the combination of direct detection and neutrino data, the speed 
distribution is reconstructed to within an order of magnitude, over a range 
of speeds of $\sim 200 \, {\rm km \, s}^{-1}$, for all four benchmarks 
considered, independently of the speed parametrisation employed. For the 
binned parametrisation the accuracy achieved is better (reduced to a factor 
of 3--4 for certain speeds), over a range as wide as  
$\sim  400 \, {\rm km \, s}^{-1}$. The maximum sensitivity to the shape of 
$f(v)$ is achieved for speeds just above the threshold energies of the 
direct detection experiments. We have also demonstrated how these parametrisations can be used to make robust statistical comparisons between different speed distributions.

\item Of the two parametrisations we have used, the binned method typically 
provides tighter constraints on both WIMP particle physics parameters and 
the shape of $f(v)$. The polynomial parametrisation allows a broader range 
of speed distributions to be explored, resulting in wider uncertainties on 
the reconstructed parameters. Some of these speed distributions are probably 
not physically well-motivated, for instance those that rise or fall steeply 
at low speeds. With data only from direct detection experiments, the 
polynomial parametrisation should be used to avoid the bias in the WIMP mass 
which can arise for the binned parametrisation. Given a future signal in both 
direct detection and neutrino telescope experiments, both parametrisation 
methods should be used, as a consistency check. However, if the speed 
distribution does not contain sharp features, the binned parametrisation 
will allow a reconstruction of the WIMP particle physics parameters, and 
also the speed distribution, that is reliable and more accurate.

\item Even with the inclusion of IceCube data, it is not always possible to 
derive upper and lower limits on both the SI and SD cross sections. This is 
due to a residual degeneracy between the two. However, it is possible to 
define an effective cross section, $\sigma_\textrm{eff}$, that determines the 
total event rate and incorporates both the SI and SD cross sections. The 
combination of direct detection and neutrino telescope data allows both upper 
and lower limits to be placed on $\sigma_\textrm{eff}$.
\end{itemize}

We have shown that by combining direct detection and neutrino telescope data, 
unbiased reconstructions of not only the WIMP mass, but also the WIMP 
interaction cross sections, can be obtained without making restrictive (and
potentially unjustified) assumptions about the WIMP speed distribution. 
Furthermore, the form of the speed distribution can also be reconstructed. 
This is possible because neutrino telescopes are sensitive to the entire 
low-speed WIMP population that lies beneath the thresholds of direct detection 
experiments. The addition of neutrino telescope data thus solves a problem 
that afflicts any strategy to recover the WIMP particle physics parameters 
and to probe $f(v)$ using direct detection data without making astrophysical 
assumptions. This demonstrates, and extends, the complementarity of the 
different techniques employed in the search for DM.

\section*{Acknowledgments}
The authors thank Annika Peter and Joakim Edsj\"{o} for useful discussions. 
AMG and BJK are both supported by STFC and MF by the Leverhulme Trust. BJK is 
also supported by the European Research Council ({\sc Erc}) under the EU 
Seventh Framework Programme (FP7/2007-2013)/{\sc Erc} Starting Grant 
(agreement n.\ 278234 --- `{\sc NewDark}' project). The authors gratefully 
acknowledge access to the University of Nottingham High Performance Computing 
Facility. MF also acknowledges the support of project MultiDark CSD2009-00064.

\bibliographystyle{apsrev4-1}
\bibliography{IceCubePaper.bib}

\end{document}